\documentclass[a4paper,11pt]{article}
\pdfoutput=1

\usepackage[utf8]{inputenc}
\usepackage{jheppub}
\usepackage{amsmath,amssymb,mathtools}
\usepackage{enumerate}
\usepackage{multirow}
\usepackage{graphicx}
\usepackage{xspace}
\usepackage{subcaption}

\usepackage[export]{adjustbox}
\usepackage{pifont}

\usepackage{spverbatim}
\usepackage{listings}

\definecolor{codegreen}{rgb}{0,0.6,0}
\definecolor{codegray}{rgb}{0.5,0.5,0.5}
\definecolor{codepurple}{rgb}{0.58,0,0.82}
\definecolor{backcolour}{rgb}{0.95,0.95,0.92}

\lstdefinestyle{mystyle}{
    backgroundcolor=\color{backcolour},   
    commentstyle=\color{codegreen},
    keywordstyle=\color{magenta},
    numberstyle=\tiny\color{codegray},
    stringstyle=\color{codepurple},
    basicstyle=\ttfamily\footnotesize,
    breakatwhitespace=false,         
    breaklines=true,                 
    captionpos=b,                    
    keepspaces=true,                 
    numbers=left,                    
    numbersep=5pt,                  
    showspaces=false,                
    showstringspaces=false,
    showtabs=false,                  
    tabsize=2
}

\renewcommand{\tabcolsep}{2em}

\allowdisplaybreaks

\newcommand{\fpiegam}{\left( 4\pi e^{-\gamma_E}\right)}
\newcommand{\muomuf}{\left( \mu^2 / \mu_a^2\right)}
\def\bs{\boldsymbol}

\preprint{{\raggedleft%
ZU-TH 01/22
}}

\title{Photon Fragmentation in the Antenna Subtraction Formalism}

\author{Thomas Gehrmann,}
\author{Robin Sch\"urmann}
\affiliation{Physik-Institut, Universit\"at Z\"urich, Winterthurerstrasse 190, CH-8057 Z\"urich, Switzerland}

\emailAdd{thomas.gehrmann@uzh.ch}
\emailAdd{robins@physik.uzh.ch}

\abstract{
The theoretical description of photon production at particle colliders combines direct photon radiation and fragmentation processes,
which can not be separated from each other for definitions of photon isolation used in experimental measurements. The theoretical 
description of these processes must account for collinear parton-photon configurations, retaining the dependence  on the photon momentum
fraction, and includes the parton-to-photon fragmentation functions. 
We extend the antenna subtraction method to include photon fragmentation processes up to next-to-next-to-leading order (NNLO) in 
QCD. Collinear photon radiation is handled using newly introduced fragmentation antenna functions and associated phase space mappings. 
We derive the integrated forms of the fragmentation antenna functions and describe their interplay with the 
mass factorisation of the photon fragmentation functions. The construction principles of antenna subtraction terms up to NNLO 
for identified photons are outlined, thereby enabling the application of the method to different photon production processes at colliders. 
}

\begin{document}
\maketitle

\section{Introduction}\label{sec:intro}

The production of photons at large transverse momenta is studied for a variety of final-state configurations at particle colliders, for example in
 inclusive photon production, photon pair production or photon-plus-jet production. 
These observables probe fundamental QCD and QED dynamics, help to constrain the parton content of the colliding hadrons, and yield 
final states that are also of interest in new particle searches. 
At the LHC, measurements of single-photon~\cite{ATLAS:2017nah,CMS:2018qao,ATLAS:2019buk,ATLAS:2019iaa} and di-photon~\cite{CMS:2014mvm,ATLAS:2017cvh,ATLAS:2021mbt} observables are now reaching an experimental accuracy of a few per cent, thereby demanding a comparable level of precision for the corresponding theory predictions.

The leading-order 
parton-level production process of photons at large transverse momenta is their radiation off  quarks, 
which is also called prompt or direct production. Another source of 
final-state photons is their radiation in the hadronisation process of an ordinary jet production event, called fragmentation process. 
This photon fragmentation process is described by (non-perturbative) fragmentation functions 
of different partons into photons~\cite{Koller:1978kq,Laermann:1982jr}. The contribution of the fragmentation process to a photon production observable 
can be minimised by imposing an isolation criterion, which requires the photon to be well-separated from any final-state hadrons in the event. 
In experimental measurements, the photon isolation is formulated by allowing only a limited amount of hadronic energy  in a fixed-size cone around 
the photon. For a finite-sized cone, this hadronic energy threshold must be non-zero to ensure infrared safety of the resulting observables, 
consequently leading to a non-vanishing fragmentation contribution that must also be accounted for in the theory predictions. 
An alternative isolation procedure is to use a  dynamical cone~\cite{Frixione:1998jh}, which lowers the hadronic energy threshold towards the 
center of the cone and fully suppresses the fragmentation contribution. While theory predictions at higher orders frequently employ 
the dynamical cone isolation due to its simplicity, all experimental measurements to date are based on fixed-cone isolation. The uncertainty 
resulting from using different isolation prescriptions in theory and experiment forms a systematic source of error that is difficult to quantify. 

  For a fixed-size cone isolation, it is not even possible to disentangle the 
 prompt and fragmentation processes, since the parton-level collinear photon radiation off a final-state quark is 
kinematically indistinguishable from photon fragmentation. After renormalisation and mass factorisation of the 
incoming  parton distributions, this parton-level process yields a left-over collinear singularity, which is 
absorbed into the mass factorisation of the photon fragmentation functions~\cite{Koller:1978kq}. 
Consequently, the next-to-leading order (NLO) corrections for inclusive photon~\cite{Aurenche:1987fs,Baer:1990ra,Aurenche:1992yc,Gordon:1993qc,Gluck:1994iz,Catani:2002ny}, 
photon-plus-jet~\cite{Aurenche:2006vj}
and di-photon production~\cite{Binoth:1999qq} depend on the photon fragmentation functions. 
These fulfil DGLAP-type evolution equations~\cite{Laermann:1982jr} with an inhomogeneous term from the quark-to-photon splitting,
with a priori unknown non-perturbative boundary conditions. Parametrisations of the photon fragmentation 
functions mainly rely on models for these boundary conditions~\cite{Owens:1986mp,Gluck:1992zx,Bourhis:1997yu}.
The only measurements were performed up to now at LEP~\cite{Buskulic:1995au,Ackerstaff:1997nha}, enabling an 
determination of the photon fragmentation functions~\cite{GehrmannDeRidder:1997gf}
and a critical assessment~\cite{GehrmannDeRidder:1998ba} of the previously available models. 

Calculations of next-to-next-to-leading order (NNLO) QCD corrections for inclusive-photon~\cite{Campbell_2017,Chen_2020}, photon-plus-jet~\cite{Chen_2020,Campbell_2017a}
 di-photon~\cite{Catani:2011qz,Campbell:2016yrh,Catani:2018krb,Gehrmann:2020oec,Chawdhry:2021hkp,Badger:2021ohm}
 or tri-photon production~\cite{Chawdhry:2019bji,Kallweit:2020gcp}
have been performed up to now only for dynamical cone isolation (or variations thereof, \cite{Siegert:2016bre}). 
Theory predictions for fixed-cone isolation were not feasible at NNLO QCD up to now, since none of the available 
QCD subtraction techniques at NNLO is able to handle fragmentation processes. Most recently, first steps in this direction were taken with the calculation of heavy-hadron fragmentation in top quark decays~\cite{Czakon:2021ohs}, which incorporates the perturbative heavy-quark fragmentation at 
NNLO QCD~\cite{Melnikov:2004bm,Mitov:2004du}. 
It is the objective of this paper to 
extend the antenna subtraction method~\cite{GehrmannDeRidder:2005cm,Daleo:2006xa,Currie:2013vh} to be able to 
account for
photon fragmentation up to NNLO. 

 In section~\ref{sec:dfrag}, we review the mass factorisation of the photon fragmentation functions up to 
NNLO, which forms the basis for the compensation of collinear singularities between direct and fragmentation processes. 
The different contributions to photon production cross sections up to NNLO in QCD  are described in detail in 
section~\ref{sec:xsec}, where we construct the antenna subtraction terms that are required to handle collinear photon radiation at 
NLO and NNLO.  These antenna subtraction terms contain novel fragmentation antenna functions for double real radiation at tree level and 
single real radiation at one loop, which are differential in the final-state photon momentum fraction. The integration of 
these fragmentation antenna functions over the respective antenna phase spaces 
is described in sections~\ref{sec:X40int} and \ref{sec:X31int}. Finally, we conclude in section~\ref{sec:conc} with a discussion of 
possible applications and extensions of the newly developed formalism. Two appendices document the relevant mass factorisation 
kernels and all integrated NLO fragmentation antenna functions for identified photons or partons.

\section{Mass Factorisation of the Photon Fragmentation Functions}\label{sec:dfrag}

Collinear photon radiation off partons leads to singularities in cross sections involving identified final-state photons. These singularities
are absorbed into a redefinition (mass factorisation) of the parton-to-photon fragmentation functions. This factorisation 
is performed at a fragmentation scale $\mu_a$, and the resulting mass-factorised fragmentation functions consequently depend on $\mu_a$. 
 The relation between mass-factorised and bare fragmentation functions can be expressed as 
\begin{equation}
D_{i\to \gamma}(z,\mu_a^2) = \sum_j \mathbf{\Gamma}_{i\to j}(z,\mu_a^2) \otimes D_{j \to \gamma}^{B}(z) \, ,
\label{eq:DgamDb_component}
\end{equation}
where flavours $i,j \in \{ g, q , \bar{q} , \gamma \}$. $\mathbf{\Gamma}_{i \to j}$ are the mass factorisation kernels of the fragmentation functions. We use a bold letter to indicate that these kernels carry colour factors. For a compact notation we have introduced a photon-to-photon fragmentation function $D_{\gamma \to \gamma}$. It is given by
\begin{equation}
D_{\gamma \to \gamma}(z,\mu_a^2) = D_{\gamma \to \gamma}^B(z) = \delta(1-z) \, .
\label{eq:Dgamtogam}
\end{equation} 
In the convolution on the right-hand side of \eqref{eq:DgamDb_component}, we indicate the variable $z$ on both components with the implicit understanding that $z$ only emerges after performing the convolution. This prescription will allow us in the subsequent sections to distinguish 
convolutions in final-state momentum fractions $z$ and in initial-state momentum fractions $x$ related to the parton distribution functions (PDF), which appear simultaneously in some of the 
higher-order expressions.

 Equation \eqref{eq:DgamDb_component} can be written in matrix form, i.e.\
\begin{equation}
\mathbf{D}_{\gamma}(z,\mu_a^2) = \mathbf{\Gamma}(z,\mu_a^2) \otimes \mathbf{D}_{\gamma}^B(z).
\label{eq:DgamDb_matrix}
\end{equation}
In the equation at hand $\mathbf{D}_{\gamma}$ and $\mathbf{D}_{\gamma}^B$ are vectors in flavour space and $\mathbf{\Gamma}$ is a matrix in flavour space.  The mass factorisation kernel has a perturbative  expansion in the strong coupling constant $\alpha_s$ and 
in the electromagnetic coupling $\alpha$. 
The bare fragmentation functions can now be expressed in terms of the mass-factorised fragmentation functions by inversion of \eqref{eq:DgamDb_matrix},
\begin{equation}
\mathbf{D}_{\gamma}^B(z) = \mathbf{\Gamma}^{-1}(z,\mu_a^2) \otimes \mathbf{D}_{\gamma}(z,\mu_a^2),
\end{equation}
which can be expanded in $\alpha$ and $\alpha_s$ to obtain the bare fragmentation 
functions up to a required perturbative order. For the calculation of isolated photon production processes up to NNLO in QCD, this 
expansion is required to order $\alpha^1\alpha_s^1$. 

For the quark-to-photon fragmentation function we find
\begin{eqnarray}
D_{q \to \gamma}^B(z) &=&  D_{q \to \gamma}(z,\mu_a^2) - \frac{\alpha}{2 \pi} \mathbf{\Gamma}^{(0)}_{q \to \gamma} \nonumber \\
&&- \frac{\alpha_s}{2\pi} \left( \mathbf{\Gamma}^{(1)}_{q \to q} \otimes D_{q\to \gamma} + \mathbf{\Gamma}^{(1)}_{q \to g} \otimes D_{g\to \gamma} + \frac{\alpha}{2\pi} \mathbf{\Gamma}^{(1)}_{q \to \gamma} - \frac{\alpha}{2\pi} \mathbf{\Gamma}^{(1)}_{q \to q} \otimes \mathbf{\Gamma}^{(0)}_{q \to \gamma} \right),
\label{eq:Dqtogambar}
\end{eqnarray}
while the gluon-to-photon fragmentation function reads
\begin{eqnarray}
D_{g \to \gamma}^B(z) &=&  D_{g\to \gamma}(z,\mu_a^2) -\frac{\alpha_s}{2\pi} \bigg( \mathbf{\Gamma}^{(1)}_{g \to g} \otimes D_{g \to \gamma} + \sum_{q} \mathbf{\Gamma}^{(1)}_{g \to q} \otimes D_{q \to \gamma} \nonumber \\
&&\quad + \frac{\alpha}{2 \pi} \mathbf{\Gamma}_{g\to \gamma}^{(1)}  - \frac{\alpha}{2 \pi} \sum_q \mathbf{\Gamma}_{g \to q}^{(1)} \otimes \mathbf{\Gamma}_{q\to \gamma}^{(0)} \bigg) \, ,
\label{eq:Dgtogambar}
\end{eqnarray}
where the sum runs over all quark flavours (and also includes anti-quarks) and we have used that $\mathbf{\Gamma}^{(0)}_{g \to \gamma} = 0$.  

It will prove useful to introduce some additional notation for combinations of terms that are 
involved in the mass factorisation of the fragmentation functions. We define
\begin{eqnarray}
{\mathbf{F}}^{(0)}_{q \to \gamma} &=& \frac{\alpha}{2 \pi} \mathbf{\Gamma}^{(0)}_{q \to \gamma} \, , \nonumber \\
{\mathbf{F}}^{(1)}_{q \to \gamma} &=& \mathbf{\Gamma}^{(1)}_{q \to q} \otimes \left( D_{q\to \gamma} - \frac{\alpha}{2\pi}  \mathbf{\Gamma}^{(0)}_{q \to \gamma} \right) + \mathbf{\Gamma}^{(1)}_{q \to g} \otimes D_{g \to \gamma} + \frac{\alpha}{2\pi} \mathbf{\Gamma}^{(1)}_{q \to \gamma} \, ,  \nonumber \\
{\mathbf{F}}^{(0)}_{g \to \gamma} &= &\mathbf{\Gamma}^{(0)}_{g \to \gamma} = 0 \, , \nonumber \\
{\mathbf{F}}^{(1)}_{g \to \gamma} &= & \mathbf{\Gamma}^{(1)}_{g \to g} \otimes D_{g \to \gamma} +  \sum_{q} \mathbf{\Gamma}^{(1)}_{g \to q} \otimes \left( D_{q \to \gamma} -\frac{\alpha}{2\pi} \mathbf{\Gamma}_{q\to \gamma}^{(0)} \right) + \frac{\alpha}{2 \pi} \mathbf{\Gamma}_{g\to \gamma}^{(1)} \, ,
\label{eq:kernelbar}
\end{eqnarray}
and rewrite the relation between the bare and the mass-factorised fragmentation functions as
\begin{equation}
D_{i \to \gamma}^B(z) = D_{i \to \gamma}(z,\mu_a^2) - {\mathbf{F}}^{(0)}_{i \to \gamma}(z,\mu_a^2) - \frac{\alpha_s}{2\pi} {\mathbf{F}}^{(1)}_{i \to \gamma}(z,\mu_a^2) \, .
\label{eq:DbartoRcom}
\end{equation}
We can further decompose $\mathbf{F}^{(1)}_{i \to \gamma}$ into
\begin{equation}
{\mathbf{F}}^{(1)}_{i \to \gamma} = {\mathbf{F}}^{(1),A}_{i \to \gamma} +  {\mathbf{F}}^{(1),B}_{i \to \gamma} + {\mathbf{F}}^{(1),C}_{i \to \gamma}
\end{equation}
with
\begin{eqnarray}
{\mathbf{F}}^{(1),A}_{q \to \gamma} &= &\mathbf{\Gamma}^{(1)}_{q \to q} \otimes  D_{q\to \gamma} + \mathbf{\Gamma}^{(1)}_{q \to g} \otimes D_{g \to \gamma} \, , \nonumber\\
{\mathbf{F}}^{(1),A}_{g \to \gamma} &= &\mathbf{\Gamma}^{(1)}_{g \to g} \otimes D_{g \to \gamma} + \sum_{q} \mathbf{\Gamma}^{(1)}_{g \to q} \otimes D_{q \to \gamma} \, , \nonumber\\
{\mathbf{F}}^{(1),B}_{q \to \gamma} &= &-\frac{\alpha}{2 \pi} \mathbf{\Gamma}^{(1)}_{q \to q} \otimes  \mathbf{\Gamma}^{(0)}_{q \to \gamma} \, ,\nonumber \\
{\mathbf{F}}^{(1),B}_{g \to \gamma} &= &-\frac{\alpha}{2\pi} \sum_{q} \mathbf{\Gamma}^{(1)}_{g \to q} \otimes \mathbf{\Gamma}^{(0)}_{q \to \gamma} \, ,\nonumber \\
{\mathbf{F}}^{(1),C}_{q \to \gamma} &= &\frac{\alpha}{2\pi} \mathbf{\Gamma}^{(1)}_{q \to \gamma} \, , \nonumber\\
{\mathbf{F}}^{(1),C}_{g \to \gamma} &= &\frac{\alpha}{2\pi} \mathbf{\Gamma}^{(1)}_{g \to \gamma} \, .
\label{eq:gambarABC}
\end{eqnarray}

\section{Photon Production Cross Section}
\label{sec:xsec}

Any isolated photon production cross section at higher orders in QCD consists of a direct and a fragmentation contribution. Its general form reads:
\begin{equation}
\text{d} \hat{\sigma}^{\gamma + X} = \text{d} \hat{\sigma}_{\gamma} + \sum_{p} \text{d} \hat{\sigma}_{p} \otimes D^B_{p \to \gamma} \, ,
\label{eq:csgamgeneral}
\end{equation}
with $p \in \{ q_j, \bar{q}_j , g \}$. $\text{d} \hat{\sigma}_p$ is the cross section for the production of parton $p$ with large transverse momentum
and  $\text{d} \hat{\sigma}_{\gamma}$ describes the direct contribution to the photon production cross section. Beyond the Born approximation, it contains singularities originating from configurations where partons are collinear to the photon. The bare
fragmentation contribution in the above equation further decomposes in two parts: a piece where $\text{d} \hat{\sigma}_p$ is convoluted with the mass-factorised fragmentation functions and the mass factorisation counterterms of the fragmentation functions, which will cancel the parton-photon collinear singularities in the direct contribution. 

Genuine QCD infrared singularities that do not involve the photon are fully contained inside the direct contribution, where they compensate each other 
between partonic subprocesses of different multiplicity. By using a dynamical photon isolation, which regulates any parton-photon collinear configuration and discards the fragmentation contribution, these singularities can be handled with generic QCD subtraction methods up to NNLO. Following this
procedure, NNLO results 
have been obtained for photon-plus-jet production~\cite{Campbell_2017,Chen_2020}, di-photon 
production~\cite{Catani:2011qz,Campbell:2016yrh,Catani:2018krb,Gehrmann:2020oec}, 
di-photon-plus-jet production~\cite{Chawdhry:2021hkp,Badger:2021ohm} and tri-photon production~\cite{Chawdhry:2019bji,Kallweit:2020gcp}. 
In the following, it is assumed that the genuine QCD singularities have already been handled using antenna subtraction, 
such that only the remaining  parton-photon collinear singularities remain to be dealt with. The subtractions for infrared-singular 
genuine QCD and parton-photon collinear configurations are largely independent (except for the occurrence 
of simple collinear quark-photon singularities in  a
single type of genuine QCD subtraction terms, discussed in Section~\ref{sec:directRR} below) 
up to NNLO, such that the corresponding subtraction terms can just be
combined in an additive manner. We will thus discuss only the construction of   parton-photon collinear subtractions, their 
interplay with the mass factorisation of the parton-to-photon fragmentation functions, and generic fragmentation function contributions in the following.

The cross section $\text{d} \hat{\sigma}_i$ is expanded in powers of $\alpha_s$, i.e.
\begin{equation}
\text{d} \hat{\sigma}_i = \text{d} \hat{\sigma}^{{\rm LO}}_i + \frac{\alpha_s}{2 \pi} \text{d} \hat{\sigma}^{{\rm NLO}}_i + \left(\frac{\alpha_s}{2 \pi} \right)^2 \text{d} \hat{\sigma}^{{\rm NNLO}}_i + \mathcal{O}(\alpha_s^3) \, .
\end{equation}
With the power counting of the fragmentation functions given by  $D_{q/\bar{q}/g \to \gamma} = \mathcal{O}(\alpha)$, the different contributions to the photon cross section at the different levels of accuracy read 
\begin{eqnarray}
\text{d} \hat{\sigma}^{\gamma+ X,{\rm LO}} &= &\text{d} \hat{\sigma}_{\gamma}^{{\rm LO}}, 
\label{eq:siggampX0} \\
\text{d} \hat{\sigma}^{\gamma+ X,{\rm NLO}} &= &\text{d} \hat{\sigma}_{\gamma}^{{\rm NLO}} + {\rm d} \hat{\sigma}^{{\rm LO}}_{g} \otimes D_{g \to \gamma}  + \sum_q \text{d} \hat{\sigma}^{{\rm LO}}_{q} \otimes D_{q \to \gamma} - \sum_q \text{d} \hat{\sigma}^{{\rm LO}}_{q} \otimes \mathbf{F}^{(0)}_{q \to \gamma},\label{eq:siggampX1} 
\end{eqnarray}
\newpage
and 
\begin{eqnarray}
\text{d} \hat{\sigma}^{\gamma+ X,{\rm NNLO}} &= & \text{d} \hat{\sigma}_{\gamma}^{{\rm NNLO}} + \sum_q \text{d} \hat{\sigma}_{q}^{{\rm NLO}} \otimes D_{q \to \gamma} - \sum_q \text{d} \hat{\sigma}_{q}^{{\rm NLO}} \otimes \mathbf{F}^{(0)}_{q \to \gamma} 
\nonumber \\
&&\quad - \sum_q \text{d} \hat{\sigma}_{q}^{{\rm LO}} \otimes \frac{\alpha_s}{2 \pi}\mathbf{F}^{(1)}_{q \to \gamma} + \text{d} \hat{\sigma}_{g}^{{\rm NLO}} \otimes D_{g \to \gamma} - \text{d} \hat{\sigma}_{g}^{{\rm LO}} \otimes \frac{\alpha_s}{2\pi} \mathbf{F}^{(1)}_{g \to \gamma}.
\label{eq:siggampX2}
\end{eqnarray}
We used \eqref{eq:DbartoRcom} to express the bare fragmentation functions in terms of the mass-factorised fragmentation functions and used that $\mathbf{F}^{(0)}_{g \to \gamma} = 0$. The sums in the above equations also run over anti-quarks. In general, the fragmentation functions are flavour-sensitive while the mass factorisation kernels are flavour-blind. 

In the cross section $\text{d}\hat{\sigma}_i$ the final-state particle $i$ has to be identified. This holds not only for the case $i = \gamma$ but also when $i$ is a parton. Therefore, it is useful to rewrite the higher-order cross section as
\begin{eqnarray}
\text{d}\hat{\sigma}_i^{{\rm NLO}} &=& \int_{n+1}  \left(\text{d} \hat{\sigma}^R_i - \text{d} \hat{\sigma}_{i}^S  -\text{d} \hat{\sigma}_{j(i)}^S\right)
\nonumber  \\
&&+ \int_{n}  \left( \text{d} \hat{\sigma}^V_i - \text{d} \hat{\sigma}^T_i - \text{d} \hat{\sigma}^T_{j(i)} \right) 
\label{eq:NLOsigid}
\end{eqnarray}
and 
\begin{eqnarray}
\text{d}\hat{\sigma}_i^{{\rm NNLO}} &=& \int_{n+2}  \left(\text{d} \hat{\sigma}^{RR}_i - \text{d} \hat{\sigma}_{i}^S  -\text{d} \hat{\sigma}_{j(i)}^S\right)\nonumber  \\
&&+ \int_{n+1} \left( \text{d} \hat{\sigma}^{RV}_i - \text{d} \hat{\sigma}^T_i - \text{d} \hat{\sigma}^T_{j(i)} \right) \nonumber \\
&&+ \int_{n} \left( \text{d} \hat{\sigma}^{VV}_i - \text{d} \hat{\sigma}^U_i - \text{d} \hat{\sigma}^U_{j(i)} \right) \, ,
\label{eq:NNLOsigid}
\end{eqnarray}
where we divided the subtraction terms in a part in which the identified particle remains resolved and a part in which particle $i$ is unresolved, i.e.\ in the reduced matrix element there is no momentum corresponding to particle $i$ alone, but it is part of a cluster of identity $j \in \{q_j, \bar{q}_j, g\}$. 
The subscripts of the integral signs refer to the number of final-state particles.

All explicit poles in $\text{d}\hat{\sigma}^T_{j(\gamma)}$ and $\text{d}\hat{\sigma}^U_{j(\gamma)}$ have to cancel against the mass factorisation terms of the fragmentation functions. It should be noted that the composition of these cross sections slightly deviates from the pure 
QCD case~\cite{Currie:2013vh}, where all counterterms from the mass factorisation of the incoming parton distributions are contained in
$\text{d}\hat{\sigma}^T$ and $\text{d}\hat{\sigma}^U$.  In $\text{d}\hat{\sigma}^T_{j(\gamma)}$ and $\text{d}\hat{\sigma}^U_{j(\gamma)}$, 
only counterterms associated with the parton distributions are included, while the mass factorisation 
counterterms of the photon fragmentation functions are not included, but added 
explicitly to \eqref{eq:NLOsigid} and \eqref{eq:NNLOsigid}. This distinction will allow a more transparent identification of infrared cancellations associated with the photon fragmentation process in the following.

\subsection{Subtraction at NLO}
The form of the NLO cross section is given in \eqref{eq:siggampX1}.
Using additionally the notation of \eqref{eq:NLOsigid} for $\text{d}\hat{\sigma}^{{\rm NLO}}_{\gamma}$, we have
\newpage
\begin{eqnarray}
\text{d} \hat{\sigma}^{\gamma+ X,{\rm NLO}} &=& \int_{n+1} \left(\text{d} \hat{\sigma}^R_{\gamma} - \text{d} \hat{\sigma}_{\gamma}^S  -\sum_q \text{d} \hat{\sigma}_{q(\gamma)}^S\right) \nonumber \\
&&+ \int_{n} \left( \text{d} \hat{\sigma}^V_{\gamma} - \text{d} \hat{\sigma}^T_{\gamma} \right)  \nonumber\\
&&+  \int_{n}  \sum_q \left(-  \text{d} \hat{\sigma}^T_{q(\gamma)} -  \text{d} \hat{\sigma}^{B}_{q} \otimes \mathbf{F}^{(0)}_{q \to \gamma} \right) \nonumber \\
&&+ \int_{n}  \sum_q \text{d} \hat{\sigma}^{B}_{q} \otimes D_{q \to \gamma} + \int_{n} \text{d} \hat{\sigma}^{B}_{g} \otimes D_{g \to \gamma} \, ,
\label{eq:gampXNLO}
\end{eqnarray}
where each individual line is free of implicit and explicit divergences. 
${\rm d} \hat{\sigma}^S_{q(\gamma)}$ subtracts the quark-photon singular collinear configurations of ${\rm d} \hat{\sigma}^R_{\gamma}$. At NLO, these configurations always yield a quark as a parent cluster so that there is no contribution to ${\rm d} \hat{\sigma}^S_{g(\gamma)}$. The flavour sum runs over $\{ u,d,s,c,b \}$ and does not distinguish between quarks and anti-quarks as $D_{q \to \gamma} = D_{\bar{q} \to \gamma}$. 

The full $\text{d} \hat{\sigma}^S_{q(\gamma)}$ subtraction term is a sum of contributions of the type,
\begin{eqnarray}
\text{d} \hat{\sigma}^S_{q(\gamma)} &=& \mathcal{N}^R_{NLO} \sum_{{\rm perm.}} \text{d} \Phi_{n+1}(k_1,\, ...\,,k_n , k_{\gamma}; p_1 , p_2) \frac{1}{S_{n+1}} \nonumber \\
&&\times A^0_3(\check{k}_{\bar{q}}, k_{\gamma}^{{\rm id.}}, k_q) \, Q_q^2 \, M^0_{n+2}( ... \, , k_{(q\gamma)} , \, ...) \emph{J}^{(n)}_{m} ( \{ \tilde{k} \}_n ; z).
\label{eq:gernicSqgam}
\end{eqnarray}
The antenna function in \eqref{eq:gernicSqgam} mimics the singular $q\parallel \gamma$ limit of the real-radiation matrix element. At NLO one can always choose the $A^0_3$ antenna function in its final-final or initial-final crossing to subtract this limit. We indicate the reference momentum with a check-mark and the identified particle with a superscript (id.). The reduced matrix element is a Born-level jet matrix element and it is multiplied with the charge $Q_q$ of the quark, to which the photon becomes collinear. $z$ is the momentum fraction of the photon within the cluster momentum $k_{(q\gamma)}$. It is given by $z = z_3\left(\check{k}_{\bar{q}}, k_{\gamma}^{{\rm id.}}, k_q\right)$ with the general definition for the NLO momentum fraction
\begin{equation}
z_3\left(\check{k}_a , k_b^{{\rm id.}}, k_c\right) = \frac{s_{ab}}{s_{ab} + s_{ac}} \, .
\label{eq:derz3generic}
\end{equation}
The jet function $J^{(n)}_m$ applies the jet algorithm as well as any cuts on the photon. Consequently, it retains an explicit functional dependence on $z$.
The new class of antenna functions in \eqref{eq:gernicSqgam}, in which a final-state particle is identified, are called fragmentation antenna functions. 

The limit in which the photon becomes collinear to the reference momentum corresponds to the limit $z \to 0$, and this configuration will be vetoed by the jet function. The reference particle can be either in the final or in the initial state, i.e.\  final-final or initial-final fragmentation antenna functions can be used. In the case of an initial-final fragmentation antenna function we exclusively use the initial-state momentum as the reference direction in the definition of the momentum fraction $z$. Therefore, in this case we have $\check{k}_{\bar{q}}=\check{p}_q$ in \eqref{eq:gernicSqgam}.

To integrate the subtraction term we have to factorise the phase space in \eqref{eq:gernicSqgam} and make the integration over $z$ explicit. A different  phase space factorisation applies for the  cases of initial-final and  final-final antenna functions.

Using the initial-final phase space factorisation~\cite{Daleo:2006xa}, we obtain
\begin{equation}
  {\rm d} \Phi_{n+1}(\dots,k_q,k_\gamma,\dots;p_q,p_2) = {\rm d} \Phi_n(\dots, k_{(q \gamma)},\dots;\bar{p}_{q},p_2) \frac{{\rm d} x}{x} \frac{Q^2}{2\pi}{\rm d} \Phi_2   \delta\left( z -  \frac{s_{ \check{q} \gamma}}{s_{ \check{q} \gamma}+s_{\check{q} q}} \right) {\rm d} z \, ,
\end{equation}
where $q^2 =  (p_{q} - k_{\gamma} - k_{q})^2=-Q^2$ and ${\rm d} \Phi_2 = {\rm d} \Phi_2(q,p_q;k_{\gamma},k_q)$. We used that in the case of an initial-final antenna we have $\check{k}_{\bar{q}}=\check{p}_q$. 

Using the final-final phase space factorisation~\citep{GehrmannDeRidder:2005cm}, one can rewrite the $n+1$ particle phase space as
\begin{equation}
{\rm d} \Phi_{n+1}(\dots,k_{\bar{q}},k_\gamma,\dots;p_1,p_2) = {\rm d} \Phi_n(\dots,\tilde{k}_{\bar{q}}, k_{(q \gamma)},\dots;p_1,p_2) {\rm d} \Phi_3 P_2^{-1}   \delta\left( z -  \frac{s_{ \bar{q} \gamma}}{s_{ \bar{q} \gamma}+s_{\bar{q} q}} \right) {\rm d} z \, ,
\end{equation}
where ${\rm d} \Phi_3 = {\rm d} \Phi_3(k_{\bar{q}},k_{\gamma},k_q;\tilde{k}_{\bar{q}}+k_{(q\gamma)})$ and $P_2$ is the integrated two-body phase space, i.e.\
\begin{equation}
P_2 = \int {\rm d} \Phi_2 = 2^{-3+2\epsilon} \pi^{-1+\epsilon} \frac{\Gamma(2-2\epsilon)}{\Gamma(1-\epsilon)} s^{-\epsilon} \, .
\end{equation}

After factorising the phase space in \eqref{eq:gernicSqgam}, the integration of the subtraction term ${\rm d} \hat{\sigma}^S_{q(\gamma)}$ can be performed:
\begin{eqnarray}
\text{d}\hat{\sigma}^T_{q(\gamma)} &=& - \mathcal{N}^V_{NLO} Q_q^2 \sum_{{\rm perm.}} \int \frac{\text{d}x}{x}  \int_0^1 \text{d} z \, \text{d} \Phi_n(k_1, \, ... \, , k_{q} , \, ... \, , k_n;x p_1, p_2 ) \nonumber \\
&&\times \frac{1}{S_n} \mathcal{A}^{0, \, {\rm id.} \gamma}_{3,\bar{q}}(x , z) \, M^0_{n+2}( ... \, , k_q , \, ...) \, \emph{J}^{(n)}_m( \{ k \}_n ; z) \, ,
\label{eq:gernicTqgam}
\end{eqnarray}
where $\mathcal{N}^V_{NLO} = \mathcal{N}^R_{NLO} \, C(\epsilon)$, with $C(\epsilon) = \fpiegam^{\epsilon}/(8 \pi^2)$. In case a final-final antenna function is used there is no explicit $x$-dependence in the subtraction term. 
$\mathcal{A}^{0, \, {\rm id.} \gamma}_{3,\bar{q}}$ is the integrated fragmentation antenna function. The subscript corresponds to the reference particle used in the definition of the momentum fraction. The integration of a general $X^{0,{\rm id.} j}_{i,jk}$ fragmentation antenna function with identified particle $j$ and reference particle $i$ in the initial-final configuration reads
\begin{eqnarray}
\mathcal{X}^{0,{\rm id.} j}_{3,i}\left(x,z\right) &=& \frac{1}{C(\epsilon)}\int {\rm d} \Phi_2(q,p_{i};k_{j},k_k) \, X_{i,jk}^{0,{\rm id.} j} \, \frac{Q^2}{2\pi} \, \delta\left( z -  \frac{s_{ij}}{s_{ij}+s_{ik}} \right) \nonumber  \\
 &= &\frac{Q^2}{2} \frac{e^{\gamma_E \epsilon}}{\Gamma(1-\epsilon)} \left(Q^2\right)^{-\epsilon} \mathcal{J}(x,z) \, X_{i,jk}^{0,{\rm id.} j}(x,z) 
\label{eq:intX30IFfrag}
\end{eqnarray}
with $q^2 = (p_i-k_j-k_k)^2 = -Q^2<0$ and the Jacobian factor is given by
\begin{equation}
\mathcal{J}(x,z) = (1-x)^{-\epsilon} x^{\epsilon} z^{-\epsilon} (1-z)^{-\epsilon} \, .
\label{eq:JacPhi2}
\end{equation}

It originates from expressing the integration over the two-body phase space as a single integration over $z$. After expressing the invariants in the antenna function in terms of $x$ and $z$, all terms of the form $(1-x)^{-1-\epsilon}$ and $(1-z)^{-1-\epsilon}$ are expanded in distributions, where we use the notation
\begin{equation}
\mathcal{D}_n(u) = \left[\frac{\log^n (1-u) }{1-u} \right]_+ \, \, , \, n \in \mathbb{N}_0 \, .
\label{eq:Dndef}
\end{equation}

The integrated $A^0_3$ fragmentation antenna function in the initial-final configuration reads
\begin{eqnarray}
\mathcal{A}^{0, {\rm id.} \gamma}_{3, \hat{q}}(x,z)&=& \left(Q^2\right)^{-\epsilon} \bigg[ -\frac{1}{2\epsilon}\delta(1-x) p^{(0)}_{\gamma q}(z)  + \frac{1}{2}-\frac{x}{2}+\frac{z}{4}+\frac{x z}{4}+\frac{1}{2} z \delta(1-x) \nonumber \\
&&+\left(-\frac{1}{4}-\frac{x}{4}+\frac{1}{2} \mathcal{D}_0(x)+\frac{1}{2} \delta(1-x) \left( \log (1-z)+\log(z)\right)\right)
p^{(0)}_{\gamma q}(z) \bigg] + \mathcal{O}(\epsilon) \, , \nonumber \\
\label{eq:A30qgamIF}
\end{eqnarray}
where $p^{(0)}_{\gamma q}$ denotes the quark-photon splitting function given in \eqref{eq:LOsplittingfunc}.
In the final-final configuration the integration of a fragmentation antenna functions $X^{0,{\rm id.} j}_{ijk}$ with identified particle $j$ and reference particle $i$ takes the form
\begin{eqnarray}
\mathcal{X}_{3, i}^{0, \, {\rm id.} j}(z) &=&\frac{1}{C(\epsilon)} \int  {\rm d} \Phi_3\left(k_i,k_{j},k_k;\tilde{k}_i+k_{(kj)}\right) P_2^{-1} \delta\left( z -  \frac{s_{ij}}{s_{ij}+s_{ik}} \right) X^{0, \, {\rm id.} j}_{ijk} \nonumber \\
&= &\frac{e^{\gamma_E \epsilon}}{\Gamma(1-\epsilon)} s_{ijk}^{-1+2\epsilon} z^{-\epsilon} (1-z)^{-\epsilon} \int_0^{s_{ijk}} {\rm d} s_{jk} (s_{ijk}-s_{jk})^{1-2\epsilon} s_{jk}^{-\epsilon} X^{0, \, {\rm id.} j}_{ijk}(s_{jk},z) \, . \nonumber \\
\end{eqnarray}

As in the initial-final integration, we obtain a Jacobian factor from rewriting one of the two non-trivial integrations of the three-body phase space as an integration over $z$. The remaining integration is straightforward for the tree-level $X^0_3$ fragmentation antenna functions. The final-final fragmentation antenna function needed to subtract quark-photon collinear singularities is $A^0_3(\check{\bar{q}},\gamma^{\rm id.},q)$. Its integrated form reads
\begin{eqnarray}
\mathcal{A}^{0, {\rm id.} \gamma}_{3,\bar{q}}(z)&=& \left(s_{q\gamma \bar{q}}\right)^{-\epsilon} \bigg[ -\frac{1}{2 \epsilon } p^{(0)}_{\gamma q}(z) +\frac{1}{4}+\frac{z}{8}+\left(-\frac{3}{8}+\frac{1}{2} \log (1-z)+\frac{\log (z)}{2}\right) p^{(0)}_{\gamma q}(z) \bigg] \nonumber \\
&&+\mathcal{O}(\epsilon)
\label{eq:A30qgamFF}
\end{eqnarray}

It can be seen from \eqref{eq:A30qgamIF} and \eqref{eq:A30qgamFF} that the quark-photon collinear singularity is manifest in an $1/\epsilon$-pole at the integrated level.
This pole is cancelled by the mass factorisation contribution of the fragmentation functions, which reads
\begin{eqnarray}
\text{d} \hat{\sigma}^B_q \otimes \mathbf{F}^{(0)}_{q \to \gamma} &=& \frac{1}{2} \, \mathcal{N}^V_{NLO} \, Q_q^2 \sum_{{\rm perm.}} \int_0^1 \text{d} z \, \text{d} \Phi_n(k_1 ,  ..., k_q ,  ...  , k_n ; p_1 , p_2 ) \nonumber \\
&& \times\frac{1}{S_n} \, \mu_a^{-2\epsilon} \, \Gamma_{\gamma q}^{(0)}(z) M^0_{n+2}( ... \, , k_q , \, ...) \, \emph{J}^{(n)}_m( \{ k \}_n ; z) \, ,
\label{eq:gernicBqgam}
\end{eqnarray}
where $\mu_a$ denotes the fragmentation scale and we have expanded out the mass factorisation counterterm 
$\mathbf{F}^{(0)}_{q \to \gamma}$ in terms of coupling factors and colour-ordered coefficients $\Gamma$ as outlined
 in \eqref{eq:kernelbar} and in detailed appendix \ref{app:MFkernels}. 
 The Born cross section is
\begin{equation}
{\rm d} \hat{\sigma}^B_q = \mathcal{N}^{LO}_{{\rm jet}}\sum_{\rm perm.} {\rm d} \Phi_n(\{k\}_n;p_1,p_2)\frac{1}{S_n}M_n^0(\dots,k_q,\dots)J_m^{(n)}(\{k\}_n;z)\, .
\end{equation}
In here, the jet function depends on $z$ because the quark momentum $k_q$ denotes a quark-photon cluster containing 
a photon with momentum fraction $z$. 
The normalisation factors are related by $\mathcal{N}^B_{{\rm jet}} \alpha/({2 \pi} )\fpiegam^{\epsilon} = \ \mathcal{N}^V_{NLO}/2$. 

Adding the integrated subtraction term \eqref{eq:gernicTqgam} and the mass factorisation contribution \eqref{eq:gernicBqgam}, one has
\begin{eqnarray}
\text{d}\hat{\sigma}^T_{q(\gamma)} + \text{d} \hat{\sigma}^B_q \otimes \mathbf{F}^{(0)}_{q \to \gamma} &=& -  \, \mathcal{N}^V_{NLO} \, Q_q^2 \sum_{{\rm perm.}} \int \frac{\text{d}x}{x} \, \int_0^1 \text{d} z \frac{1}{S_n} \text{d} \Phi_n(k_1 ,  ..., k_q ,  ...  , k_n ;x p_1 ,p_2 ) \nonumber \\
&& \times  \bs{J}^{(1), \, {\rm id.} \gamma}_{2,\bar{q}}(\bar{q},q) \, M^0_{n+2}( ... \, , k_q , \, ...) \, \emph{J}^{(n)}_m( \{ k \}_n ; z) \, .
\end{eqnarray}

Combination of the integrated initial-final fragmentation antenna function \eqref{eq:A30qgamIF} with the mass factorisation kernel yield the NLO fragmentation dipole
\begin{equation}
\bs{J}^{(1), {\rm id.} \gamma}_{2,\hat{q}}(\hat{q},q) = \mathcal{A}^{0,{\rm id. \gamma}}_{3,\hat{q}}(z,x) - \frac{1}{2} \, \mu_a^{-2\epsilon} \, \Gamma^{(0)}_{ \gamma q}(z) \, \delta(1-x) \, .
\label{eq:iddipoleqgamIF}
\end{equation}
In case a final-final $A^0_3$ fragmentation antenna function is used, the dipole reads
\begin{equation}
\bs{J}^{(1), \, {\rm id.} \gamma}_{2,\bar{q}}(\bar{q},q) = \mathcal{A}^{0,{\rm id. \gamma}}_{3,\bar{q}}(z) - \frac{1}{2} \, \mu_a^{-2\epsilon} \, \Gamma^{(0)}_{\gamma q}(z) \, ,
\label{eq:iddipoleqgamFF}
\end{equation}
where the integrated fragmentation antenna function is given in \eqref{eq:A30qgamFF}. The fragmentation dipoles \eqref{eq:iddipoleqgamIF} and \eqref{eq:iddipoleqgamFF} are $\epsilon$-finite. Therefore, having expressed $\text{d}\hat{\sigma}^T_{q(\gamma)} + \text{d} \hat{\sigma}^B_q \otimes \mathbf{F}^{(0)}_{q \to \gamma}$ in terms of these dipoles the pole cancellation between the direct part and the mass factorisation contribution is guaranteed. 

The fragmentation contribution to the photon production cross section at NLO takes the form
\begin{eqnarray}
\text{d} \hat{\sigma}^{B}_{i} \otimes D_{i \to \gamma} &= &\mathcal{N}^B_{{\rm jet}}   \sum_{{\rm perm.}} \int_0^1 \text{d} z \, \text{d} \Phi_n(k_1 ,  ..., k_i ,  ...  , k_n ; p_1 , p_2 )\nonumber  \\
&& \times\frac{1}{S_n} \, D_{i \to \gamma}(z,\mu_a^2) M^0_{n+2}( ... \, , k_i , \, ...) \, \emph{J}^{(n)}_m( \{ k \}_n ; z) \, ,
\end{eqnarray}
where $i$ can be a gluon or a quark. $z$ is the momentum fraction which the photon carries away from its parent momentum $k_i$ during the process of fragmentation.

\subsection{Subtraction at NNLO}
The pole cancellation among the different pieces \eqref{eq:siggampX2} at NNLO is more involved. These pieces can be rearranged according to 
whether they contain the mass-factorised photon fragmentation functions (fragmentation contribution) or not (direct contribution). The direct 
contribution contains all photon-parton singular configurations and their associated counterterms from the mass factorisation of the 
photon fragmentation functions. 

\subsubsection{Fragmentation Contribution}

The fragmentation contribution appears in isolated photon cross sections only from NLO onwards. Consequently, its contribution at NNLO amounts 
to an  NLO-type correction to the production of an identified parton, which subsequently fragments into the photon. This contribution 
can be further divided into the generic NLO hard subprocess cross sections, 
\begin{eqnarray}
\text{d}\hat{\sigma}^{\gamma +X,{\rm NNLO}}_{f_1} &= &\sum_q \int_{n+1}  \left( \text{d} \hat{\sigma}^R_q - \text{d}\hat{\sigma}^S_q - \text{d} \hat{\sigma}^S_{q(q)} - \text{d} \hat{\sigma}^S_{g(q)} \right) \otimes D_{q \to \gamma} \nonumber \\
&&+ \sum_q \int_{n}  \left( \text{d}\hat{\sigma}^V_q - \text{d} \hat{\sigma}^T_q - \text{d}\hat{\sigma}^T_{q(q)} - \text{d} \hat{\sigma}^T_{g(q)} \right) \otimes D_{q \to \gamma} \nonumber\\
&&+ \int_{n+1}  \left( \text{d} \hat{\sigma}^R_g - \text{d} \hat{\sigma}^S_g - \text{d} \hat{\sigma}^S_{g(g)} - \text{d} \hat{\sigma}^S_{q(g)} \right) \otimes D_{g \to \gamma}\nonumber \\
&&+ \int_{n}  \left( \text{d} \hat{\sigma}^V_{g} - \text{d} \hat{\sigma}^T_g - \text{d} \hat{\sigma}^T_{g(g)} - \text{d} \hat{\sigma}^T_{q(g)} \right) \otimes D_{g \to \gamma} 
\label{eq:sigf1}
\end{eqnarray}
and terms resulting from the mass factorisation of the photon fragmentation functions, 
corresponding to 
the mass factorisation kernels $\mathbf{F}^{(1),A}_{q\to \gamma}$ and $\mathbf{F}^{(1),A}_{g\to \gamma}$ in \eqref{eq:siggampX2}:
\begin{equation}
\text{d}\hat{\sigma}^{\gamma +X,{\rm NNLO}}_{f_2} =  - \sum_q \text{d} \hat{\sigma}_{q}^{{\rm LO}} \otimes \frac{\alpha_s}{2 \pi}\mathbf{F}^{(1),A}_{q \to \gamma} -\text{d} \hat{\sigma}_{g}^{{\rm LO}} \otimes \frac{\alpha_s}{2\pi} \mathbf{F}^{(1),A}_{g \to \gamma}
\end{equation}.

Expanding these kernels yields the full fragmentation contribution: 
\begin{eqnarray}
\text{d}\hat{\sigma}^{\gamma +X,{\rm NNLO}}_{f} &= & \text{d}\hat{\sigma}^{\gamma +X,{\rm NNLO}}_{f_1} 
+\text{d}\hat{\sigma}^{\gamma +X,{\rm NNLO}}_{f_2}\nonumber \\
&=&\sum_q \int_{n+1} \left( \text{d} \hat{\sigma}^R_q - \text{d}\hat{\sigma}^S_q - \text{d} \hat{\sigma}^S_{q(q)} - \text{d} \hat{\sigma}^S_{g(q)} \right) \otimes D_{q \to \gamma} \nonumber\\
&&+ \sum_q \int_{n} \left( \text{d}\hat{\sigma}^V_q - \text{d} \hat{\sigma}^T_q - \text{d}\hat{\sigma}^T_{q(q)} - \text{d} \hat{\sigma}^B_q \otimes \frac{\alpha_s}{2 \pi} \mathbf{\Gamma}^{(1)}_{q \to q} \right) \otimes D_{q \to \gamma} \nonumber\\
&&+ \sum_q \int_{n} \left( - \text{d} \hat{\sigma}^T_{g(q)} - \text{d} \hat{\sigma}^B_g \otimes \frac{\alpha_s}{2\pi} \mathbf{\Gamma}^{(1)}_{g \to q} \right) \otimes D_{q\to \gamma} \nonumber\\
&&+ \int_{n+1} \left( \text{d} \hat{\sigma}^R_g - \text{d} \hat{\sigma}^S_g - \text{d} \hat{\sigma}^S_{g(g)} - \text{d} \hat{\sigma}^S_{q(g)} \right) \otimes D_{g \to \gamma} \nonumber\\
&&+ \int_{n} \left( \text{d} \hat{\sigma}^V_g - \text{d} \hat{\sigma}^T_g - \text{d} \hat{\sigma}^T_{g(g)} - \text{d} \hat{\sigma}^B_{g} \otimes \frac{\alpha_s}{2 \pi} \mathbf{\Gamma}^{(1)}_{g\to g} \right) \otimes D_{g \to \gamma}\nonumber \\
&&+ \int_{n}  \left( - \text{d} \hat{\sigma}^T_{q(g)} - \text{d} \hat{\sigma}^B_{q} \otimes \frac{\alpha_s}{2 \pi} \mathbf{\Gamma}^{(1)}_{q\to g} \right) \otimes D_{g \to \gamma} \, ,
\label{eq:fconNNLO}
\end{eqnarray}
where we used the expression for the kernels given in \eqref{eq:gambarABC}. Each line in \eqref{eq:fconNNLO} is free of explicit and implicit divergences. 

The only singularity which is subtracted in $\text{d}\hat{\sigma}^S_{q(q)}$ corresponds to the $q \parallel g$ limit, in which the two partons form a parent quark and the original quark momentum within the quark-gluon cluster is identified. The subtraction term takes the form
\begin{eqnarray}
\text{d}\hat{\sigma}^S_{q(q)} &=& \mathcal{N}^R_{{\rm jet}} \sum_{{\rm perm.}} \text{d} \Phi_{n+1} (k_1,  ... , k_q, ... , k_g, ..., k_{n+1}; p_1,p_2) \frac{1}{S_{n+1}} \nonumber \\
&& \times X^0_3(\check{k}_s,k_g,k_q^{{\rm id.}}) M^0_{n+2}(... , k_{(gq)}, ... ) \emph{J}^{(n)}_m ( \{ \tilde{k} \}_n ; u) \, ,
\label{eq:sigSqqid}
\end{eqnarray}
where we used the same notation as in \eqref{eq:gernicSqgam} to indicate the reference parton with momentum $k_s$ and the identified quark momentum $k_q$. The momentum $k_g$ is the momentum of the gluon which is colour connected to the quark for the specific colour ordering. $u$ is the momentum fraction of the quark in the cluster momentum $k_{(gq)}$. It reads
\begin{equation}
u = z_3\left(\check{k}_s, k_q^{{\rm id.}} , k_g\right) = \frac{s_{sq}}{s_{sq}+s_{sg}} \, .
\end{equation}

For the subtraction of a quark-gluon collinear limit either a $D^0_3$ or  $A^0_3$ antenna function can be used and the reference particle can be in the initial or final state. The convolution of the subtraction term with the quark-to-photon fragmentation functions is given by
\begin{eqnarray}
\text{d}\hat{\sigma}^S_{q(q)} \otimes D_{q \to \gamma} &&=\mathcal{N}^R_{{\rm jet}}  \sum_{{\rm perm.}} \int_0^1  \text{d} v \, \text{d} \Phi_{n+1} (k_1,  ... , k_q, ... , k_g, ..., k_{n+1}; p_1,p_2) \frac{1}{S_{n+1}} \nonumber \\
&&\hspace{-6mm} \times X^0_3(\check{k}_s,k_g,k_q^{{\rm id.}}) M^0_{n+2}(... , k_{(gq)}, ... ) \emph{J}^{(n)}_m( \{ \tilde{k} \}_n ; z = uv) \, D_{q \to \gamma}(v,\mu_a^2) \, .
\label{eq:sigSqqidconD}
\end{eqnarray}
In here, we have two momentum fractions:  $v$ denotes the fraction of the photon momentum in the quark-photon cluster. The quark itself is part of a 
quark-gluon cluster, in which it carries a momentum fraction $u$. 
 The jet function $\emph{J}^{(n)}_m$ has to reconstruct the photon momentum from the mapped momentum $ k_{(gq)}$.  The momentum fraction of the photon in the quark-gluon-photon cluster momentum $ k_{(gq)}$ is given by $z=uv$, which is entering the jet function. 

The subtraction term $\text{d}\hat{\sigma}^S_{g(q)}$ subtracts the $q \parallel \bar{q}$ limit. In this case the cluster identity is a gluon. This subtraction term reads
\begin{eqnarray}
\text{d}\hat{\sigma}^S_{g(q)} &= &\mathcal{N}^R_{{\rm jet}}  \bigg\lbrace \sum_{{\rm perm.}} \text{d} \Phi_{n+1} (k_1,  ... , k_{q}, ... , k_{\bar{q}}, ..., k_{n+1}; p_1,p_2) \frac{1}{S_{n+1}}\nonumber  \\
&& \times X^0_3(\check{k}_s,k_{q}^{{\rm id.}},k_{\bar{q}}) M^0_{n+2}(... , k_{(q\bar{q})}, ... ) \emph{J}^{(n)}_m( \{ \tilde{k} \}_n ; u) + ( q \leftrightarrow \bar{q}) \bigg\rbrace \, ,
\label{eq:sigSgqid}
\end{eqnarray}
where we have indicated that there is an identical term in which the anti-quark is identified. Both of these terms are convoluted with the same fragmentation function since $D_{q \to \gamma}=D_{\bar{q} \to \gamma}$. The antenna function in \eqref{eq:sigSgqid} can be either a $E^0_3$ or $G^0_3$ antenna and they can be used in the initial-final or final-final configuration. The convolution of \eqref{eq:sigSgqid}  with the fragmentation function takes the same form as \eqref{eq:sigSqqidconD}. 

The integration of the subtraction terms \eqref{eq:sigSqqid} and \eqref{eq:sigSgqid} proceeds in the same manner as the integration of ${\rm d} \hat{\sigma}^S_{q(\gamma)}$. After integrating over the antenna phase space the fragmentation antenna function retains an explicit dependence on the momentum fraction.
Performing the convolution with the fragmentation functions, we find
\begin{eqnarray}
\text{d} \hat{\sigma}^T_{q(q)} \otimes D_{q \to \gamma} &=& 
- \mathcal{N}^V_{{\rm jet}} \sum_{{\rm perm.}} \int \frac{\text{d}x}{x}\, \int_0^1 \text{d} z \int_z^1 \frac{\text{d} u}{u} \, \, \text{d} \Phi_n(k_1,...,k_q,...,k_n;x p_1 ,  p_2) \nonumber \\
&&\times \frac{1}{S_n} \mathcal{X}^{0, \, {\rm id.} q}_{3,s}(x , u) \, D_{q \to \gamma}\left(\frac{z}{u},\mu_a^2\right)  M^0_{n+2}(...,k_q,...) \emph{J}^{(n)}_m(\{k\}_n;z) \, .
\label{eq:sigTqqid}
\end{eqnarray}
The jet function again only depends on   $z=uv$. In case a final-final antenna function is used to subtract the $q \parallel g$ limit,  there is no explicit dependence on and integration over $x$. 

To obtain a fragmentation dipole from the integrated fragmentation antenna function it has to be combined with the mass factorisation contribution from the fragmentation functions. For the case at hand the corresponding mass factorisation contribution is $\mathbf{F}^{(1),A}_{q \to \gamma}$. It reads
\begin{eqnarray}
\lefteqn{\text{d}\hat{\sigma}^B_q \otimes \frac{\alpha_s}{2\pi} \mathbf{\Gamma}^{(1)}_{q \to q} \otimes D_{q\to \gamma} =}\nonumber \\
&& \mathcal{N}^V_{{\rm jet}}  \sum_{{\rm perm.}}  \int_0^1 \text{d} z \int_z^1 \frac{\text{d} u}{u} \frac{1}{S_n}  \text{d} \Phi_n(k_1,...,k_q,...,k_n;p_1 , p_2) 
\nonumber  \\
&&\times \mu_a^{-2\epsilon} \, \Gamma^{(1)}_{q q}(u)  D_{q \to \gamma}\left(\frac{z}{u},\mu_a^2\right) \, M^0_{n+2}(...,k_q,...) \, \emph{J}^{(n)}_m(\{ k \}_n ; z) \, ,
\label{eq:sigBqcongam}
\end{eqnarray}
where we used $\mathcal{N}^V_{{\rm jet}} = \mathcal{N}^B_{{\rm jet}} \fpiegam^{\epsilon} {\alpha_s N}/({2\pi})$
and where the mass factorisation kernel is expanded according to \eqref{eq:allkernels}.   
Combining \eqref{eq:sigTqqid} and \eqref{eq:sigBqcongam}, we find
\begin{eqnarray}
\lefteqn{\text{d}\hat{\sigma}^T_{q(q)} \otimes D_{q \to \gamma} + \text{d} \hat{\sigma}^B_q \otimes \frac{\alpha_s}{2\pi} \mathbf{\Gamma}^{(1)}_{q \to q} \otimes D_{q \to \gamma} }\nonumber  \\
&=&- \mathcal{N}^V_{{\rm jet}} \sum_{{\rm perm.}} \int \frac{\text{d}x}{x}\, \int_0^1 \text{d} z \int_z^1 \frac{\text{d} u}{u} \, \text{d} \Phi_n(k_1,...,k_q,...,k_n; x p_1 , p_2) \frac{1}{S_n} D_{q \to \gamma}\left(\frac{z}{u},\mu_a^2\right) \nonumber \\
&& \times \left( \mathcal{X}^{0, \, {\rm id.} q}_{3,s}(x,u) - \mu_a^{-2\epsilon} \delta(1-x) \Gamma^{(1)}_{qq}(u) \right)   M^0_{n+2}(..., k_q, ...) \emph{J}^{(n)}_m(\{ k\}_n ; z) \, .
\label{eq:sigTqidq}
\end{eqnarray}
If the reference particle in the fragmentation antenna function $X^0_3$ is in the initial state there is an additional contribution from the mass factorisation terms of the initial-state parton distributions. In this case we would have to replace the integrated fragmentation antenna function by the dipole which includes this contribution. There are two final-final fragmentation dipoles for the limit under consideration, 
\begin{eqnarray}
\bs{J}^{(1), \, {\rm id.} q}_{2,\bar{q}}(\bar{q},q) &=& \mathcal{A}^{0,{\rm id.} \, q}_{3,\bar{q}}(u) - \mu_a^{-2\epsilon} \, \Gamma^{(1)}_{q  q}(u) \, , 
\nonumber \\
\bs{J}^{(1), \, {\rm id.} q}_{2,g}(g,q) &=& \mathcal{D}^{0,{\rm id.} \, q}_{3,g}(u) - \mu_a^{-2\epsilon} \, \Gamma^{(1)}_{q  q}(u) \, .
\label{eq:iddipolesqqclusterFF}
\end{eqnarray}

In the initial-final configuration we have
\begin{eqnarray}
\bs{J}^{(1) , \, {\rm id.} q}_{2,\hat{q}}(\hat{q},q) &= &\mathcal{A}^{0,{\rm id.} \, q}_{3,\hat{q}}(u,x) - \mu_F^{-2\epsilon} \, \Gamma^{(1)}_{qq}(x) \delta(1-u) - \mu_a^{-2 \epsilon} \, \Gamma^{(1)}_{qq}(u) \delta(1-x) \, , \nonumber \\
\bs{J}^{(1), \, {\rm id.} q}_{2,\hat{g}}(\hat{g},q) &= &\mathcal{D}^{0,{\rm id.} \, q}_{3,\hat{g}}(u,x) - \frac{1}{2} \mu_F^{-2\epsilon} \, \Gamma^{(1)}_{gg}(x) \delta(1-u) - \mu_a^{-2 \epsilon} \, \Gamma^{(1)}_{q q}(u) \delta(1-x) \, ,
\label{eq:iddipolesqqclusterIF}
\end{eqnarray}
where we also included the contribution from the mass factorisation terms of the initial-state parton distributions, which are mass-factorised at the factorisation scale $\mu_F$. The remaining poles of the identified dipoles are all proportional to $\delta(1-x) \, \delta(1-u)$ so that a cancellation with the virtual contribution  can take place. The expressions for the $\mathcal{A}^{0,{\rm id.} \, q}_{3,q}$ and 
$\mathcal{D}^{0,{\rm id.} \, q}_{3,g}$
integrated antenna functions are given in appendix~\ref{app:X30integration}.

In case the identified quark is part of a quark-anti-quark cluster resulting from a gluon splitting, we have 
\newpage
\begin{eqnarray}
\text{d} \hat{\sigma}^T_{g(q)} \otimes D_{q \to \gamma}&=& -2 \, \mathcal{N}^V_{{\rm jet}}  \sum_{{\rm perm.}} \int \frac{\text{d}x}{x}\int_0^1 \text{d} z \int_z^1 \frac{\text{d} u}{u}  \, \text{d} \Phi_n(k_1,...,k_g,...,k_n;x p_1 , p_2)\nonumber  \\
&&\times \frac{1}{S_n} \mathcal{X}^{0, \, {\rm id.} q}_{3,s}(x,u) \, D_{q \to \gamma}\left(\frac{z}{u}, \mu_a^2\right)  M^0_{n+2}(...,k_g,...) \emph{J}^{(n)}_m(\{k\}_n;z)\, ,
\label{eq:sigTgqid}
\end{eqnarray}
where $k_g$ is the momentum of the $q\bar{q}$-cluster. The overall factor 2 accounts results from summation over quark and anti-quark of flavour $q$, which have identical fragmentation functions to photons. 
The additional contribution coming from $\mathbf{F}^{(1),A}_{g \to \gamma}$ in \eqref{eq:fconNNLO} takes the form
\begin{eqnarray}
\lefteqn{\text{d}\hat{\sigma}^B_g \otimes \frac{\alpha_s}{2\pi} \mathbf{\Gamma}^{(1)}_{g \to q} \otimes D_{q\to \gamma}}  \nonumber \\
&=&2 \, \mathcal{N}^V_{{\rm jet}} \sum_{{\rm perm.}} \int_0^1 \text{d} z \int_z^1 \frac{\text{d} u}{u}  \text{d} \Phi_n(k_1,...,k_g,...,k_n;xp_1 ,p_2) \, \frac{1}{S_n} \nonumber  \\
&&\times \mu_a^{-2\epsilon} \,  \Gamma^{(1)}_{qg}(u)  D_{q \to \gamma}\left(\frac{z}{u},\mu_a^2\right) \, M^0_{n+2}(...,k_g,...) \, \emph{J}^{(n)}_n(\{ k \} ; z = uv) \, ,
\label{eq:sigBgcongam}
\end{eqnarray}
where we used $\mathcal{N}^V_{{\rm jet}} = \fpiegam^{\epsilon}{\alpha_s }/({2\pi}) \tilde{\mathcal{N}}^B_{{\rm jet}}$. Usually there is an additional factor of $N$ when going from the Born normalisation to the virtual normalisation factor. This factor is absent in \eqref{eq:sigBgcongam} since
  the normalisation in \eqref{eq:sigTgqid} refers to a four-quark matrix element, while the Born normalisation factor in \eqref{eq:sigBgcongam} refers to a two-quark matrix element. 
  
Combination of both contributions yields
\begin{eqnarray}
\lefteqn{\text{d}\hat{\sigma}^T_{g(q)} \otimes D_{q \to \gamma} + \text{d} \hat{\sigma}^B_g \otimes \frac{\alpha_s}{2\pi} \mathbf{\Gamma}^{(1)}_{g \to \gamma} \otimes D_{q \to \gamma} } \nonumber \\
&=&- 2 \mathcal{N}^V_{{\rm jet}} \sum_{{\rm perm.}} \int \frac{\text{d}x}{x} \int_0^1 \text{d} z \int_z^1 \frac{\text{d} u}{u}\, \text{d} \Phi_n(k_1,...,k_g,...,k_n;x p_1 ,p_2) \frac{1}{S_n} D_{q \to \gamma}\left(\frac{z}{u},\mu_a^2\right)  \nonumber \\
&& \times  \left( \mathcal{X}^{0, {\rm id.} q}_{3,s}(x,u) - \mu_a^{-2\epsilon} \delta(1-x) \Gamma^{(1)}_{qg} (u)\right) \, M^0_{n+2}(..., k_g, ...) \emph{J}^{(n)}_m(\{ k\}_n ; z) \, .
\end{eqnarray}

The term which is convoluted with the fragmentation function defines another fragmentation dipole, 
containing  a $G^0_3$ or $E^0_3$ antenna function for the $q \parallel \bar{q}$ limit. For the reference particle being in the final state the corresponding dipoles read
\begin{eqnarray}
\bs{J}^{(1), \, {\rm id.} q'}_{2,q}(q,g) &=& \mathcal{E}^{0,{\rm id.} \, q'}_{3,q}(u) - \mu_a^{-2 \epsilon} \, \Gamma^{(1)}_{qg}(u) \, ,\nonumber \\
\bs{J}^{(1), \, {\rm id.} q'}_{2,g}(g,g) &=& \mathcal{G}^{0,{\rm id.} \, q'}_{3,g}(u) - \mu_a^{-2 \epsilon} \, \Gamma^{(1)}_{qg}(u)\, ,\
\label{eq:J21qpidFF}
\end{eqnarray}
and for an initial-state reference particle we have 
\begin{eqnarray}
\bs{J}^{(1), \, {\rm id.} q'}_{2,\hat{q}}(\hat{q},g) = \mathcal{E}^{0,{\rm id.} \, q'}_{3,\hat{q}}(u,x) - \mu_a^{-2 \epsilon} \, \Gamma^{(1)}_{qg}(u) \delta(1-x) \, , \nonumber\\ 
\bs{J}^{(1), \, {\rm id.} q'}_{2,\hat{g}}(\hat{g},g) = \mathcal{G}^{0,{\rm id.} \, q'}_{3,\hat{g}}(u,x) - \mu_a^{-2 \epsilon} \, \Gamma^{(1)}_{qg}(u) \delta(1-x) \, .
\label{eq:J21qpidIF}
\end{eqnarray}
The integrated antenna functions $\mathcal{E}^{0,{\rm id.} \, q'}_{3,q}=\mathcal{E}^{0,{\rm id.} \, \bar{q}'}_{3,q}$ and $\mathcal{G}^{0,{\rm id.} \, q'}_{3,g}=\mathcal{G}^{0,{\rm id.} \, \bar{q}'}_{3,g}$ are documented in appendix~\ref{app:X30integration}. As the
dipoles \eqref{eq:J21qpidFF} and \eqref{eq:J21qpidIF} correspond to a flavour-changing limit, they
are $\epsilon$-finite. 

Using these fragmentation dipoles the NNLO fragmentation contribution \eqref{eq:fconNNLO} can be implemented using the antenna subtraction formalism.

\subsubsection{Direct Contribution: Structure and Final-State Mass Factorisation}

The direct contribution to the NNLO photon production cross section reads according to \eqref{eq:siggampX2}:
\begin{eqnarray}
\text{d}\hat{\sigma}_{\gamma}^{{\rm NNLO}} &=& \int_{n+2} \left(\text{d} \hat{\sigma}^{RR}_{\gamma} - \text{d} \hat{\sigma}_{\gamma}^S  - \sum_q \text{d} \hat{\sigma}_{q(\gamma)}^S - \text{d} \hat{\sigma}_{g(\gamma)}^S \right)\nonumber \\
&&+ \int_{n+1} \left( \text{d} \hat{\sigma}^{RV}_{\gamma} - \text{d} \hat{\sigma}^T_{\gamma} -  \sum_q \text{d} \hat{\sigma}^T_{q(\gamma)} - \text{d} \hat{\sigma}^T_{g(\gamma)} \right) \nonumber\\
&&+ \int_{n} \left( \text{d} \hat{\sigma}^{VV}_{\gamma} - \text{d} \hat{\sigma}^U_{\gamma} -  \sum_q \text{d} \hat{\sigma}^U_{q(\gamma)} - \text{d} \hat{\sigma}^U_{g(\gamma)} \right).
\label{eq:sigNNLOdir}
\end{eqnarray}
This contribution contains final-state parton-photon collinear singularities that are cancelled by the mass 
factorisation terms of the fragmentation functions. The relevant terms at NNLO are as follows:
\begin{eqnarray}
\text{d}\hat{\sigma}^{\gamma +X,{\rm NNLO}}_{{\rm MF}} &=& - \sum_q \left( \text{d} \hat{\sigma}^{{\rm NLO}}_q \otimes \mathbf{F}^{(0)}_{q \to \gamma} + \text{d} \hat{\sigma}^{{\rm LO}}_q \otimes \frac{\alpha_s}{2\pi} \mathbf{F}^{(1),B}_{q \to \gamma} \right)  - \text{d} \hat{\sigma}^{{\rm LO}}_g \otimes \frac{\alpha_s}{2\pi} \mathbf{F}^{(1),B}_{g\to \gamma} \nonumber \\
&& 
- \sum_q \text{d} \hat{\sigma}^{{\rm LO}}_q \otimes \frac{\alpha_s}{2\pi}\mathbf{F}^{(1),C}_{q \to \gamma} - \text{d} \hat{\sigma}^{{\rm LO}}_g \otimes \frac{\alpha_s}{2\pi} \mathbf{F}^{(1),C}_{g\to \gamma} \nonumber \\
&=&\sum_q \int_{n+1}  \left( \text{d} \hat{\sigma}^R_q - \text{d}\hat{\sigma}^S_q - \text{d} \hat{\sigma}^S_{q(q)} - \text{d} \hat{\sigma}^S_{g(q)} \right) \otimes \left( -  \frac{\alpha}{2\pi} \mathbf{\Gamma}^{(0)}_{q \to \gamma} \right) \nonumber \\
&&+ \sum_q \int_{n}  \left( \text{d}\hat{\sigma}^V_q - \text{d} \hat{\sigma}^T_q - \text{d}\hat{\sigma}^T_{q(q)} - \text{d} \hat{\sigma}^B_q \otimes \frac{\alpha_s}{2 \pi} \mathbf{\Gamma}^{(1)}_{q \to q} \right) \otimes \left( -  \frac{\alpha}{2\pi} \mathbf{\Gamma}^{(0)}_{q \to \gamma} \right)\nonumber \\
&&+ \sum_q \int_{n}  \left( - \text{d} \hat{\sigma}^T_{g(q)} - \text{d} \hat{\sigma}^B_g \otimes \frac{\alpha_s}{2\pi} \mathbf{\Gamma}^{(1)}_{g \to q} \right) \otimes \left( - \frac{\alpha}{2\pi} \mathbf{\Gamma}^{(0)}_{q \to \gamma} \right) \nonumber \\
 &&- \sum_q \int_{n}  \text{d} \hat{\sigma}^{B}_q \otimes \frac{\alpha_s}{2 \pi} \frac{\alpha}{2\pi} \mathbf{\Gamma}^{(1)}_{q \to \gamma} - \int_{n}  \text{d} \hat{\sigma}^{B}_g \otimes \frac{\alpha_s}{2\pi} \frac{\alpha}{2\pi} \mathbf{\Gamma}^{(1)}_{g \to \gamma}.
\label{eq:sigMF}
\end{eqnarray}

The individual lines in the above expressions \eqref{eq:sigNNLOdir} and \eqref{eq:sigMF} are free of implicit singularities. 
 However, each term contains explicit poles in $\epsilon$, which eventually have to cancel among the different contributions. To guarantee the cancellation we also have to include additional mass factorisation terms of the initial-state parton distributions at different levels of the calculation. These terms 
 are contained inside the $\text{d} \hat{\sigma}^{T,U}_{i}$ in the above expression and  take the general form
\begin{eqnarray}  
\text{d} \hat{\sigma}^{\gamma + X,{\rm NNLO}}_{{\rm ISMF1}} &=& -\int_{n+1}  \mathbf{\Gamma}^{(1)}_{{\rm PDF}} \otimes \left(- \text{d} \hat{\sigma}^{S,{\rm NLO}}_{q(\gamma)} \right) \, ,
\label{eq:sigISMF1} \\
\text{d} \hat{\sigma}^{\gamma + X,{\rm NNLO}}_{{\rm ISMF2}} &= &-\int_n  \mathbf{\Gamma}^{(1)}_{{\rm PDF}} \otimes \left(- \text{d} \hat{\sigma}^{T,{\rm NLO}}_{q(\gamma)} \right) \, .
\label{eq:sigISMF2}
\end{eqnarray}

Finally, products of mass factorisation terms of initial-state parton distributions and final-state fragmentation functions appear at the
double virtual level. These mixed terms read: 
\begin{eqnarray}
\text{d} \hat{\sigma}^{\gamma +X,{\rm NNLO}}_{{\rm MF3}} &=&  
 \int_n  \mathbf{\Gamma}^{(1)}_{{\rm PDF}}  \otimes 
\left(  \frac{\alpha}{2\pi} \mathbf{\Gamma}^{(0)}_{q \to \gamma} \right) \otimes
 \left( \text{d} \hat{\sigma}^{B}_{q} \right) \, .
\label{eq:MF3mix}
\end{eqnarray}
 We discuss the cancellation of the implicit and explicit singularities  at each level of final-state multiplicity. 

\subsubsection{Direct Contribution: Double Real Level}
\label{sec:directRR}

The double-real subtraction terms in which the photon becomes unresolved can be decomposed into the parts $\text{d}\hat{\sigma}^{S,a}, \text{d}\hat{\sigma}^{S,b},\text{d}\hat{\sigma}^{S,c}$ and $\text{d}\hat{\sigma}^{S,d}$ as it is done for genuine QCD subtraction terms~\cite{Currie:2013vh}. 
The notion of colour connection is however slightly different here, since the final-state photon can not become soft. It is thus only colour connected to one of the hard emitters in its antenna function. The second hard emitter is only used as reference momentum to define the collinear momentum fraction, 
and does not play a role in any unresolved limit. If this reference momentum is shared with another antenna function, the configuration is still viewed 
as colour unconnected and thus part of $\text{d}\hat{\sigma}^{S,d}$.

The single unresolved subtraction term $\text{d}\hat{\sigma}^{S,a}$ follows the same construction pattern as the NLO real subtraction term. Moreover, in a single unresolved limit the photon can only become part of quark-photon cluster. Consequently, we have $\text{d} \hat{\sigma}^{S,a}_{g(\gamma)} =0$. The subtraction term for a single collinear $q \parallel \gamma $ limit reads
\begin{eqnarray}
\text{d} \hat{\sigma}^{S,a}_{q(\gamma)} &=& \mathcal{N}^{RR}  \sum_{{\rm perm.}} \text{d} \Phi_{n+2}(k_1,\, ...\,,k_q, \, ...\,,k_{n+1} , k_{\gamma}; p_{\hat{q}} , p_2) \frac{1}{S_{n+2}}\nonumber \\
&&\times A^0_3(\check{p}_{\hat{q}}, k_{\gamma}^{{\rm id.}}, k_q) \, Q_q^2 \, M^0_{n+3}( ... \, , k_{(q\gamma)} , \, ...) \emph{J}^{(n+1)}_{n} \left( \{ \tilde{k} \}_{n+1} ; z\right) \, ,
\label{eq:gernicSaqgam}
\end{eqnarray}
where the momentum fraction $z=z_3(\check{p}_{\hat{q}},k_{\gamma}^{\rm id.},k_q)$ is given in \eqref{eq:derz3generic}. The subtraction term \eqref{eq:gernicSaqgam} subtracts the single collinear quark-photon limits of the corresponding double real radiation matrix element. In the construction of the double real subtraction term we use the $A^0_3$ fragmentation antenna function in its initial-final configuration, where it contains  
only the final-state quark-photon collinear limit.  In case there is no initial-state quark in the corresponding double real matrix element 
an initial-state gluon momentum is used as reference momentum in the $A^0_3$ antenna function. 

The subtraction term at hand introduces spurious additional singularities in almost colour connected and colour disconnected limits as the jet function allows an additional parton to become unresolved. Likewise, the genuine QCD double real subtraction term of type $\text{d}\hat{\sigma}^{S,a}$
(which were originally constructed for 
a dynamical photon isolation) 
contains reduced matrix elements that can develop collinear quark-photon singularities once a fixed-cone photon isolation is applied.
One has to account for both these types of 
spurious singularities when constructing the subtraction terms $\text{d}\hat{\sigma}^{S,c}$ and $\text{d}\hat{\sigma}^{S,d}$.
Terms of the form of \eqref{eq:gernicSaqgam} are reintroduced at the real-virtual level upon integration over the antenna phase space. They combine with the term $\text{d} \hat{\sigma}^R_q \otimes (- \mathbf{F}^{(0)}_{q \to \gamma})$ in \eqref{eq:sigMF} and form fragmentation dipoles, which were introduced in \eqref{eq:iddipoleqgamIF}. 

The subtraction terms for colour connected double unresolved limits including the photon are $\text{d} \hat{\sigma}^{S,b_1}_{q(\gamma)}$ and $\text{d} \hat{\sigma}^{S,b_1}_{g(\gamma)}$. In both cases the unresolved limits correspond to triple collinear configurations. The singular limit which is subtracted by $\text{d} \hat{\sigma}^{S,b_1}_{q(\gamma)}$ is the triple collinear $q \parallel g \parallel \gamma$ limit, while the limit subtracted by $\text{d} \hat{\sigma}^{S,b_1}_{g(\gamma)}$ is the triple collinear $q \parallel \gamma \parallel \bar{q}$ limit. Therefore, the only $X^0_4$ antenna functions needed are $\tilde{A}^0_4(q,g,\gamma,\bar{q})$ and $\tilde{E}^0_4(q,q',\gamma,\bar{q}')$~\cite{GehrmannDeRidder:2005cm}. We use these antenna functions exclusively in their initial-final configuration. 

The subtraction term for the limit where the photon and a gluon simultaneously become unresolved reads
\begin{eqnarray}
\text{d} \hat{\sigma}^{S,b_1}_{q(\gamma)} &=&  \mathcal{N}^{RR} \sum_{{\rm perm.}} \text{d} \Phi_{n+2}(k_1, ...,k_q  , ... , \, k_g ,  ...  ,k_{n+1} , k_{\gamma}; p_{\hat{q}} , p_2) \frac{1}{S_{n+2}}\nonumber \\
&&\times \tilde{A}^0_4(\check{p}_{\hat{q}},k_g, k_{\gamma}^{{\rm id.}}, k_q) \, Q_q^2 \, M^0_{n+2}( ... \, , k_{(q\gamma g)} , \, ...), \emph{J}^{(n)}_{n} \left( \{ \tilde{k} \}_n ; z\right) \, ,
\label{eq:sigSb1qgam}
\end{eqnarray}
where the momentum $k_g$ is the momentum of the gluon to which the final-state quark is colour connected. The momentum fraction of the photon in the cluster momentum $k_{(q\gamma g)}$ is  $z=z_4\left( \check{p}_{\hat{q}}, k_{\gamma}^{{\rm id.}},k_g,k_q\right)$ with the general definition of the NNLO momentum fraction 
\begin{equation}
z_4\left(\check{k}_a,k_b^{{\rm id.}},k_c,k_d\right) = \frac{s_{ab}}{s_{ab} + s_{ac} +s_{ad}} \, .
\end{equation}

The subtraction term for the $q \parallel \gamma \parallel \bar{q}$ limit reads
\begin{eqnarray}
\text{d} \hat{\sigma}^{S,b_1}_{g(\gamma)} &=&  \mathcal{N}^{RR} \sum_{{\rm perm.}} \text{d} \Phi_{n+2}(k_1, ...,k_{q'}  , ... , \, k_{\bar{q}'} ,  ...  ,k_{n+1} , k_{\gamma}; p_{\hat{q}} , p_2) \frac{1}{S_{n+2}} \nonumber \\
&&\times \tilde{E}^0_4(\check{p}_{\hat{q}},k_{q'}, k_{\gamma}^{{\rm id.}}, k_{\bar{q}'}) \, Q_{q'}^2 \, M^0_{n+2}( ... \, , k_{(q'\gamma \bar{q}')} , \, ...), \emph{J}^{(n)}_{n} \left( \{ \tilde{k} \}_n ; z\right) \, ,
\label{eq:sigSb1ggam}
\end{eqnarray}
where the momentum fraction $z$ is given by $z=z_4\left(\check{p}_q,k_{\gamma}^{{\rm id.}}, k_{q'}, k_{\bar{q}'}\right)$. 

In case there is no initial-state quark in the corresponding double real matrix element, we use an initial-state gluon momentum 
in the $X^0_4$ antenna functions in \eqref{eq:sigSb1qgam} and \eqref{eq:sigSb1ggam} as the reference momentum. The integration of these two $X^0_4$ fragmentation antenna functions is explained in detail in 
section~\ref{sec:X40int}.  After integrating over the antenna phase space the contributions \eqref{eq:sigSb1qgam} and \eqref{eq:sigSb1ggam} are added back at the double virtual level. 

The antenna functions in $\text{d} \hat{\sigma}^{S,b_1}_{j(\gamma)}$ 
contain single unresolved singular limits which have to be subtracted to guarantee an overall successful subtraction (see~\cite{Currie:2013vh} for details). 
The single unresolved limits of $\text{d} \hat{\sigma}^{S,b_1}_{q(\gamma)}$ are subtracted by
\begin{eqnarray}
\text{d} \hat{\sigma}^{S,b_2}_{q(\gamma)} &=&  -\mathcal{N}^{RR} \sum_{{\rm perm.}} \text{d} \Phi_{n+2}(k_1, ...,k_q  , ... , \, k_g ,  ...  ,k_{n+1} , k_{\gamma}; p_{\hat{q}} , p_2) \frac{1}{S_{n+2}} Q_q^2 \nonumber \\
&&\times \bigg( A^0_3(p_{\hat{q}},k_g,  k_q) A^0_3(\check{\bar{p}}_{\tilde{\hat{q}}},k_{\gamma}^{{\rm id.}}, k_{(gq)})  \, M^0_{n+2}( ... \, , k_{((gq) \gamma)} , \, ...)  \emph{J}^{(n)}_{n} \left( \{ \tilde{\tilde{k}} \}_n ; z\right) \nonumber \\
&&\quad + A^0_3(\check{p}_{\hat{q}},k_{\gamma}^{{\rm id.}},  k_q) A^0_3(\check{\bar{p}}_{\tilde{\hat{q}}},k_{g}, k^{{\rm id.}}_{(\gamma q)}) \, M^0_{n+2}( ... \, , k_{((\gamma q) g)} , \, ...) \emph{J}^{(n)}_{n} \left( \{ \tilde{\tilde{k}} \}_n ; z= u v\right) \bigg) \, .\nonumber \\
\label{eq:sigSb2qgam}
\end{eqnarray}

The first term in the above equation subtracts the single unresolved gluon limit while the second term subtracts the single unresolved photon limit of $\tilde{A}^0_4$. In the first term the momentum fraction $z$ of the photon in the cluster momentum is calculated from the first mapped momentum set, i.e.\
\begin{equation}
z = z_3\left(\check{\bar{p}}_{\tilde{\hat{q}}}, k_{\gamma}^{{\rm id.}},k_{(gq)}\right) = \frac{s_{\tilde{\hat{q}} \gamma}}{s_{\tilde{\hat{q}} \gamma } + s_{\tilde{\hat{q}} (gq) }} = \frac{s_{\hat{q} \gamma}}{s_{\hat{q} \gamma} + s_{\hat{q} q} + s_{\hat{q} g}} = z_4\left(\check{p}_{\hat{q}}, k_{\gamma}^{{\rm id.}} ,k_g , k_q\right),
\label{eq:defzwithmap}
\end{equation}
using an initial-final mapping:
\begin{eqnarray}
k_{(gq)} &=& k_g + k_q - (1-x) p_{\hat{q}} \, ,\nonumber  \\
\bar{p}_{\tilde{\hat{q}}} &=& x p_{\hat{q}} \, ,
\label{eq:mappingIF}
\end{eqnarray} 
with $x$ being the initial-state momentum fraction. It is crucial that the momentum fractions in \eqref{eq:sigSb1qgam} and in \eqref{eq:sigSb2qgam} coincide in the single unresolved limits to guarantee the cancellation of the single unresolved singularities. 

For the second term in \eqref{eq:sigSb2qgam} two momentum fractions have to be calculated. In the first antenna function we identify the photon, i.e.\ we calculate its momentum fraction in the $(q\gamma)$-cluster, which reads
\begin{equation}
u = z_3\left(\check{p}_{\hat{q}}, k_{\gamma}^{{\rm id.}},k_q\right) = \frac{s_{\hat{q} \gamma}}{s_{\hat{q} \gamma} +s_{\hat{q} q}} \, .
\end{equation}

In the second antenna function we identify the $(q\gamma)$-cluster within the $((q\gamma)g)$-cluster. The corresponding momentum fraction reads
\begin{equation}
v = z_3\left(\check{\bar{p}}_{\tilde{\hat{q}}},k_{(\gamma q)}^{{\rm id.}}, k_g\right) = \frac{s_{\tilde{\hat{q}} (\gamma q)}}{s_{\tilde{\hat{q}} (\gamma q)} + s_{\tilde{\hat{q}} g}} = \frac{s_{\hat{q} \gamma} + s_{\hat{q} q}}{s_{\hat{q} \gamma} + s_{\hat{q} q} + s_{\hat{q} g}} \, ,
\end{equation}
where we again used \eqref{eq:mappingIF} to rewrite the momentum fraction in terms of the original momentum set. The momentum fraction of the photon within the $(qg\gamma)$-cluster is then given by
\begin{equation}
z = u \, v = \frac{s_{\hat{q} \gamma}}{s_{\hat{q} \gamma} + s_{\hat{q} g} + s_{\hat{q} q}} = z_4\left(\check{p}_{\hat{q}}, k_{\gamma}^{{\rm id.}} ,k_g , k_q\right),
\end{equation}
which coincides with the NNLO momentum fraction in \eqref{eq:sigSb1qgam}. Note that the two terms in \eqref{eq:sigSb2qgam} are added back at the real-virtual level after integration over the primary antenna phase space. The term in which the photon is in the primary antenna will combine with the contribution $\text{d} \hat{\sigma}^S_{q(q)} \otimes (-\mathbf{F}^{(0)}_{q \to \gamma})$ in $\text{d} \hat{\sigma}^{\gamma +X,{\rm NNLO}}_{{\rm MF}}$ to form the fragmentation dipole of \eqref{eq:iddipoleqgamIF}. The term in which the photon is part of the secondary antenna will contribute to $\text{d} \hat{\sigma}^{T,b}_{q(\gamma)}$ below and combine with the newly introduced one-loop fragmentation antenna functions. 

The subtraction of the single unresolved limits of the $\tilde{E}^0_4$ antenna function in \eqref{eq:sigSb1ggam} takes a similar form. However, this antenna function only contains single unresolved limits involving the photon. Therefore, we only obtain terms in which the photon is part of the primary antenna function, i.e.\
\begin{eqnarray}
\text{d} \hat{\sigma}^{S,b_2}_{g(\gamma)} &=&  -\mathcal{N}^{RR} \sum_{{\rm perm.}} \text{d} \Phi_{n+2}(k_1, ...,k_{q'}  , ... , \, k_{\bar{q}'} ,  ...  ,k_{n+1} , k_{\gamma}; p_{\hat{q}} , p_2) \frac{1}{S_{n+2}} Q_{q'}^2\nonumber  \\
&&\times \bigg( A^0_3(\check{p}_{\hat{q}},k_{\gamma}^{{\rm id.}},  k_{\bar{q}'}) E^0_3(\check{\bar{p}}_{\tilde{\hat{q}}}, k_{q'},k_{(\bar{q}'\gamma)}^{{\rm id.}}) \,  M^0_{n+2}( ... \, , k_{((\bar{q}'\gamma) q')} , \, ...) \emph{J}^{(n)}_{n} \left( \{ \tilde{\tilde{k}} \}_n ; z = u\,v\right) \nonumber \\
&&+ A^0_3(\check{p}_{\hat{q}},k_{\gamma}^{{\rm id.}},  k_{q'}) E^0_3(\check{\bar{p}}_{\tilde{\hat{q}}},k_{(q'\gamma)}^{{\rm id.}}, k_{\bar{q}'}) \,  M^0_{n+2}( ... \, , k_{((q'\gamma) \bar{q}')} , \, ...) \emph{J}^{(n)}_{n} \left( \{ \tilde{\tilde{k}} \}_n ; z= u \, v\right) \bigg) \, .\nonumber \\
\label{eq:sigSb2ggam}
\end{eqnarray}

The first term subtracts the $\gamma \parallel \bar{q}'$ limit and the second term the $\gamma \parallel q'$ limit. In both cases two momentum fractions are calculated. In the first term we have $u = z_3\left(\check{p}_{\hat{q}}, k_{\gamma}^{{\rm id.}},k_{\bar{q}'}\right)$ and $v = z_3\left(\check{\bar{p}}_{\tilde{\hat{q}}},k_{(\bar{q}'\gamma)}^{{\rm id.}}, k_{q'}\right)$. The momentum fraction of the photon in the $(q'\gamma \bar{q}')$-cluster is given by the product of the two momentum fractions, i.e.\
\begin{equation}
z = u \, v = \frac{s_{\hat{q}\gamma}}{s_{\hat{q} \gamma} + s_{\hat{q} q'} + s_{\hat{q} \bar{q}'}} = z_4\left( \check{p}_{\hat{q}} , k_{\gamma}^{{\rm id.}}, k_{q'} , k_{\bar{q}'} \right)\, ,
\end{equation}
where we expressed the mapped momenta from the first initial-final mapping in terms of the original momentum set. For the second term the construction of the momentum fraction follows the same steps with the replacement $q' \leftrightarrow \bar{q}'$. Since both terms in \eqref{eq:sigSb2ggam} have the photon in the primary antenna function, they both combine with the contribution $\text{d} \hat{\sigma}^S_{g(q)} \otimes (-\mathbf{F}^{(0)}_{q \to \gamma})$ upon integration and form a fragmentation dipole of the form of \eqref{eq:iddipoleqgamIF}. 

To successfully compensate all oversubtractions of photonic limits in $\text{d} \hat{\sigma}^{S,a}$ and $\text{d} \hat{\sigma}^{S,b}$, 
as well as in the genuine QCD $\text{d} \hat{\sigma}^{S,a}$, 
one also has to introduce the $\text{d} \hat{\sigma}^{S,c}_{q(\gamma)}$ and $\text{d} \hat{\sigma}^{S,d}_{q(\gamma)}$ subtraction terms. 
The subtraction terms in $\text{d} \hat{\sigma}^{S,c}_{q(\gamma)}$ consist of a primary QCD antenna function, in which the final-state quark to which the photon becomes collinear acts as a hard radiator and the fragmentation antenna function $A^0_3$, i.e.
\begin{eqnarray}
\text{d} \hat{\sigma}^{S,c}_{q(\gamma)} &= &\mathcal{N}^{RR} \sum_{{\rm perm.}} \text{d} \Phi_{n+2}(k_1, ...,k_q  , ... , \, k_g ,  ...  ,k_{n+1} , k_{\gamma}; p_{\hat{q}} , p_2)  \frac{1}{S_{n+2}} \nonumber \\
&&\times X^0_3(k_{m}, k_g , k_{q}) A^0_3(\check{p}_{\hat{q}}, k_{\gamma}^{{\rm id.}},k_{(qg)})\, Q_q^2\, M^0_{n+2}(... , k_{((q g)\gamma)} , ... ) \emph{J}^{(n)}_n\left( \{ \tilde{\tilde{k}} \}_n ; z \right) \, , \nonumber \\
\label{eq:sigScsqgam}
\end{eqnarray}
where parton $m$ is colour connected to the gluon. The subtraction terms in $\text{d} \hat{\sigma}^{S,c}_{q(\gamma)}$ are needed to reproduce the correct soft-collinear limits of the real radiation matrix element.

The last contribution to the double real subtraction term is $\text{d} \hat{\sigma}^{S,d}_{q(\gamma)}$. It  takes care of colour disconnected double unresolved limits.  It reads
\begin{eqnarray}
\text{d} \hat{\sigma}^{S,d}_{q(\gamma)} &=& \mathcal{N}^{RR}  \sum_l \sum_{{\rm perm.}} \text{d} \Phi_{n+2}(k_1, ...,k_q  , ... , \, k_g ,  ...  ,k_{n+1} , k_{\gamma}; p_{\hat{q}} , p_2) \frac{1}{S_{n+2}} \nonumber \\
&&\times  A^0_3(\check{p}_{\hat{q}}, k_{\gamma}^{{\rm id.}},k_q) X^0_3(k_i, k_l , k_m) Q_q^2 M^0_{n+2}(... , k_{(\gamma q)} , ... ) \emph{J}^{(n)}_n\left( \{ \tilde{k} \}_n ; z\right) \, .
\label{eq:sigSdqggam}
\end{eqnarray}
Here one of the radiators $i$ or $m$ can correspond to the initial-state quark used in the $A^0_3$ antenna function. Recall that the reconstructed momentum fraction $z_3$ vanishes in the initial-state collinear limit. Therefore, the above subtraction term does only subtract colour disconnected unresolved limits even if they share the same initial-state radiator. Terms in $\text{d} \hat{\sigma}^{S,d}_{q(\gamma)}$ sharing the same initial-state radiator are added back at the real-virtual level while terms with distinct radiators are added back at the double virtual level. 

There are no contributions of the form $\text{d} \hat{\sigma}^{S,c}_{g(\gamma)}$ or $\text{d} \hat{\sigma}^{S,d}_{g(\gamma)}$.

\subsubsection{Direct Contribution: Real-Virtual Level}
\label{sec:directRV}

The real-virtual subtraction terms in which the photon becomes unresolved can be decomposed into the parts $\text{d}\hat{\sigma}^{T,a}, \text{d}\hat{\sigma}^{T,b}$ and 
$\text{d}\hat{\sigma}^{T,c}$, following the structure used for genuine QCD subtraction terms~\cite{Currie:2013vh}.

The first contribution to the real-virtual subtraction term $\text{d} \hat{\sigma}^{T}_{q(\gamma)}$ is given by integrating $\text{d} \hat{\sigma}^{S,a}_{q(\gamma)}$ in \eqref{eq:gernicSaqgam} over the antenna phase space. One finds
\begin{eqnarray}
-\text{d} \hat{\sigma}^{T,a}_{q(\gamma)} &=& \mathcal{N}^{RV} \int \frac{\text{d} x}{x} \int_0^1 \text{d}z \sum_{{\rm perm.}} \text{d} \Phi_{n+1}(k_1, ... , k_q, ... , k_{n+1} ; x p_{\hat{q}} ,p_2) \nonumber  \\
&&\times \frac{1}{S_{n+1}} \mathcal{A}^{0, {\rm id.} \gamma}_{3,\hat{q}}(x,z) \, Q_q^2 \, M^0_{n+3} ( ... , k_q, ...) \, J^{(n+1)}_n(\{k\}_{n+1} ;z) \, ,
\end{eqnarray}
where the integrated fragmentation antenna function is given in \eqref{eq:A30qgamIF}.
The reduced matrix element in $\text{d} \hat{\sigma}^{T,a}_{q(\gamma)}$ is a real radiation jet matrix element. Therefore,
there is a corresponding contribution in $\text{d} \hat{\sigma}^R_q \otimes (- \mathbf{F}^{(0)}_{q \to \gamma} )$, which is part of $\text{d}\hat{\sigma}^{\gamma +X,{\rm NNLO}}_{{\rm MF}}$. It reads
\begin{eqnarray}
-\text{d} \hat{\sigma}^R_q \otimes \mathbf{F}^{(0)}_{q \to \gamma} &= & -\frac{1}{2} \, \mathcal{N}^{RV} \, Q_q^2 \sum_{{\rm perm.}} \int_0^1 \text{d} z \, \text{d} \Phi_{n+1}(k_1 ,  ..., k_q ,  ...  , k_{n+1} ; p_{\hat{q}} , p_2 ) \nonumber \\
&&  \times\frac{1}{S_{n+1}} \, \mu_a^{-2\epsilon} \, \Gamma_{\gamma q}^{(0)}(z) M^0_{n+3}( ... \, , k_q , \, ...) \, \emph{J}^{(n+1)}_n( \{ k \}_{n+1} ; z) \, ,
\end{eqnarray}
where, as at NLO, the factor 1/2 is due to the different normalisations of the photon and jet matrix elements. Combining both contributions, we find
\begin{eqnarray}
-\text{d} \hat{\sigma}^{T,a}_{q(\gamma)}  - \text{d} \hat{\sigma}^R_q \otimes  \mathbf{F}^{(0)}_{q \to \gamma}
&=& \mathcal{N}^{RV} \, Q_q^2  \sum_{{\rm perm.}} \int \frac{\text{d}x}{x} \int_0^1 \text{d} z \, \text{d} \Phi_{n+1}(k_1 ,  ..., k_q ,  ...  , k_{n+1} ;x p_{\hat{q}} ,p_2 ) 
\nonumber \\
&& \times \frac{1}{S_{n+1}} \, \bs{J}^{(1), {\rm id.} \gamma}_{2,\hat{q}}(\hat{q},q) \, M^0_{n+3}( ... \, , k_q , \, ...) \, \emph{J}^{(n+1)}_n( \{ k \}_{n+1} ; z) \, ,\nonumber \\
\label{eq:sigTacomsigRoGam0}
\end{eqnarray}
where the initial-final fragmentation dipole is given in \eqref{eq:iddipoleqgamIF}. Terms of the form of \eqref{eq:sigTacomsigRoGam0} are $\epsilon$-finite but contain single unresolved limits, which have to be subtracted to guarantee an overall successful cancellation of the singularities.

To subtract the $q \parallel \gamma$ limit of the one-loop matrix elements, the following term is required:
\begin{eqnarray}
\text{d} \hat{\sigma}^{T,b}_{q(\gamma)} &=&\phantom{+}\mathcal{N}^{RV} \sum_{\rm {perm.}} \text{d} \Phi_{n+1}(k_1, ... , k_q, ...,k_n, k_{\gamma};p_{\hat{q}} , p_2) \frac{1}{S_{n+1}} \nonumber \\
&&\qquad \times A^0_3(\check{p}_{\hat{q}}, k_{\gamma}^{\rm id.}, k_q) \, Q_q^2 \, M^1_{n+2}(..., k_{(q\gamma)} ,...) J^{(n)}_n\left(\{ \tilde{k} \}_n ; z\right) \, .
\nonumber \\
&& +  \mathcal{N}^{RV} \sum_{\rm {perm.}} \int \frac{\text{d} x}{x} \text{d} \Phi_{n+1}(k_1, ... , k_q , ...,k_n, k_{\gamma};x p_{\hat{q}} , p_2) \frac{1}{S_{n+1}} Q_q^2  \nonumber \\
&&\qquad\times \left( \tilde{A}^1_3(\check{\bar{p}}_{\hat{q}}, k_{\gamma}^{\rm id.}, k_q) \delta(1-x) + \bs{J}^{(1)}_2(\hat{q}(\bar{p}_{\hat{q}}),q(k_q))(x) \, A^0_3(\check{\bar{p}}_{\hat{q}}, k_{\gamma}^{\rm id.}, k_q) \right) \nonumber \\
&&\qquad\times \, M^0_{n+2}(..., k_{(q\gamma)} ,...) J^{(n)}_n\left(\{ \tilde{k} \}_n ; z\right) \, ,
\label{eq:sigTb}
\end{eqnarray}
where we used that the $q \parallel \gamma$ limit of a one-loop matrix element can be subtracted using the one-loop colour-subleading antenna function $\tilde{A}^1_3$~\cite{GehrmannDeRidder:2005cm}. For photon production it is sufficient to use this one-loop fragmentation antenna function in 
 the initial-final configuration, with the initial-state momentum as a reference momentum in the definition of the momentum fraction. The integration of this class of fragmentation antenna functions is discussed in section \ref{sec:X31int}. Note that in case there is no quark in the initial-state we use an initial-state gluon momentum as the reference momentum.

The $\bs{J}^{(1)}_2$ term in \eqref{eq:sigTb} is a QCD dipole, unrelated to the photon. It 
contains the integral of the first contribution in \eqref{eq:sigSb2qgam}, 
where the primary antenna is a QCD antenna function, as well as contributions from the mass factorisation of the incoming parton distributions from 
 \eqref{eq:sigISMF1}. Combining these two contributions yields the integrated inclusive QCD dipole factor which can be found in~\cite{Currie:2013vh}. 
 The momentum fraction entering the jet function is the same as at NLO, i.e.\ $z=z_3\left(\check{p}_{\hat{q}}, k_{\gamma}^{{\rm id.}},k_{q}\right)$. 

The integration over the contribution $\text{d} \hat{\sigma}^{S,c}_{q(\gamma)}$ takes the same form as the last term in \eqref{eq:sigTb}, i.e.\
\begin{eqnarray}
\text{d} \hat{\sigma}^{T,c\, (s)}_{q(\gamma)} &=& \mathcal{N}^{RV} \sum_{\rm {perm.}} \int \frac{\text{d} x}{x} \text{d} \Phi_{n+1}(k_1, ... , k_q , ...,k_n, k_{\gamma};x p_{\hat{q}} ,p_2) \frac{1}{S_{n+1}}  \nonumber \\
&&\times \bs{J}^{(1)}_2(m(k_m),q(k_q))(x) \, A^0_3(\check{\bar{p}}_{\hat{q}}, k_{\gamma}^{\rm id.}, k_q) \, Q_q^2 \, M^0_{n+2}(..., k_{(q\gamma)} ,...) J^{(n)}_n\left(\{ \tilde{k} \}_n ; z\right) \, ,\nonumber \\ &&
\label{eq:sigTcs}
\end{eqnarray}
where $\bs{J}^{(1)}_2$ is the inclusive dipole corresponding to the primary antenna function used in $\text{d} \hat{\sigma}^{S,c}_{q(\gamma)}$ in \eqref{eq:sigScsqgam}. The superscript $s$ indicates that the photon enters the secondary, unintegrated antenna function. In case parton $m$ is in the initial state, $\bs{J}^{(1)}_2$  also contains the mass factorisation countertems for the incoming parton distribution. 

In general, it is necessary to include additional terms to correctly subtract the $q \parallel \gamma$ limit of the corresponding real-virtual matrix element 
without introducing spurious poles in $\epsilon$. 
These terms take the same form as \eqref{eq:sigTcs}, but in this case the integrated dipole contains momenta from the mapped momentum set $\{ \tilde{k} \}$. These extra terms also form part of  $\text{d} \hat{\sigma}^{T,c\, (s)}_{q(\gamma)} $. 

Combining the contributions \eqref{eq:sigTb} and \eqref{eq:sigTcs}, the expression
\begin{equation}
\int_{n+1} \left( \text{d} \hat{\sigma}^{RV}_{\gamma} - \text{d} \hat{\sigma}^T_{\gamma} -  \sum_q \left( \text{d} \hat{\sigma}^{T,b}_{q(\gamma)}  - \text{d} \hat{\sigma}^{T,c\, (s)}_{q(\gamma)} \right)  \right) 
\end{equation}
is free of explicit and implicit singularities. 

At this point we have not added back the contribution in $\text{d} \hat{\sigma}^{S,b_2}_{q(\gamma)}$ in which the photon is part of the primary antenna function. To distinguish this contribution from the contribution which is included in \eqref{eq:sigTb} we denote it as $\text{d} \hat{\sigma}^{S,b_2(p)}_{q(\gamma)}$, where the superscript $p$ indicates that this contribution originates from the piece of double real subtraction term $\text{d} \hat{\sigma}^{S,b_2}_{q(\gamma)}$ in which the photon is in the primary antenna. 

After integrating over the phase space of the primary antenna, this contribution takes the form
\newpage
\begin{eqnarray}
\text{d} \hat{\sigma}^{T,b_2(p)}_{q(\gamma)} &=& \mathcal{N}^{RV} Q_q^2 \sum_{{\rm perm.}}  \int_0^1 \text{d} v \int \frac{\text{d} x}{x} \text{d} \Phi_{n+1}(k_1, ... , k_q , ... , k_{n+1}; x p_{\hat{q}} ,p_2) \nonumber \\
&&\times \frac{1}{S_{n+1}} \mathcal{A}^{0, {\rm id.} \gamma}_{3,\hat{q}}(x,v) A^0_3(\check{\bar{p}}_{\hat{q}}, k_g , k_q^{{\rm id.}}) M^0_{n+2} ( ... , k_{(qg)}, ...) J^{(n)}_n\left(\{ \tilde{k} \}_n ; z=uv\right) \nonumber \\
\label{eq:sigTb2qgampri}
\end{eqnarray}
This expression is very similar to  \eqref{eq:sigSqqidconD}. The momentum fraction $v$ is the momentum fraction of the photon in the $(q\gamma)$-cluster. It is an external convolution variable. $u=z_3(\check{\bar{p}}_{\hat{q}},k_q^{{\rm id.}},k_g)$ is the momentum fraction of the quark in the $(qg)$-cluster, which is calculated during the mapping. Therefore, $z=uv$ describes the momentum fraction of the photon within the $(qg\gamma)$-cluster. Note that the momenta entering the integrated antenna function are unmapped momenta. 
Equation 
\eqref{eq:sigTb2qgampri}  combines with 
 the counterterm contribution 
\begin{eqnarray}
\text{d}\hat{\sigma}^S_{q(q)} \otimes  \mathbf{F}^{(0)}_{q \to \gamma} &=& \mathcal{N}^R_{{\rm jet}}\frac{\alpha}{2 \pi} \fpiegam^{\epsilon} Q_q^2 \sum_{{\rm perm.}} \int_0^1  \text{d} v \nonumber \\
&&\times \text{d} \Phi_{n+1} (k_1,  ... , k_q, ... , k_g, ..., k_{n+1}; p_{\hat{q}},p_2) \frac{1}{S_{n+1}} \, \mu_a^{-2\epsilon} \, \Gamma^{(0)}_{\gamma q}(v)
\nonumber \\
&&\times A^0_3(\check{p}_{\hat{q}},k_g,k_q^{{\rm id.}}) M^0_{n+2}(... , k_{(qg)}, ... ) \emph{J}^{(n)}_n\left( \{ \tilde{k} \}_n ; z=uv\right) \, ,
\label{eq:sigSqqcongam0}
\end{eqnarray}
such that
\begin{eqnarray}
\lefteqn{-\text{d} \hat{\sigma}^{T,b_2(p)}_{q(\gamma)} + \text{d} \hat{\sigma}^S_{q(q)} \otimes \mathbf{F}^{(0)}_{q \to \gamma}} \nonumber \\
&=& -\mathcal{N}^{RV} Q_q^2  \sum_{{\rm perm.}} \int_0^1 \text{d} v \int \frac{\text{d}x}{x} \text{d} \Phi_{n+1}(k_1, ... , k_q ,...,k_g, ... , k_{n+1}; x p_{\hat{q}} ,p_2) \frac{1}{S_{n+1}}\nonumber  \\
&&\times \bs{J}^{(1),\, {\rm id.} \gamma}_{2,\hat{q}}(\hat{q}(\bar{p}_{\hat{q}}),q(k_{q}))\left(x,v\right) A^0_3(\check{\bar{p}}_{\hat{q}}, k_g , k_q^{{\rm id.}}) M^0_{n+2} ( ... , k_{(qg)}, ...) J^{(n)}_n\left(\{ \tilde{k} \}_n ; z=uv\right) \, .
\nonumber \\
\label{eq:sigTb2posigSqq}
\end{eqnarray}

Compared to \eqref{eq:sigTacomsigRoGam0} the fragmentation dipole does not multiply a real radiation matrix element but an unintegrated antenna function and a reduced matrix element. 
In the case in which the photon becomes unresolved in a gluon type cluster we have
\begin{eqnarray}
\lefteqn{-\text{d} \hat{\sigma}^{T,b_2(p)}_{g(\gamma)} + \text{d} \hat{\sigma}^S_{g(q)} \otimes  \mathbf{F}^{(0)}_{q \to \gamma}} 
\nonumber \\
&=& -\mathcal{N}^{RV} Q_{q'}^2 \sum_{{\rm perm.}} \int_0^1 \text{d} v \int \frac{\text{d}x}{x}  \text{d} \Phi_{n+1}(k_1, ... , k_{\bar{q}'}, ...,k_{q'} , ... , k_{n+1}; x p_{\hat{q}} ,p_2) \frac{1}{S_{n+1}}\nonumber \\
&&\quad \quad \times \bigg( \bs{J}^{(1),\, {\rm id.} \gamma}_{2,\hat{q}}(\hat{q}(\bar{p}_{\hat{q}}),\bar{q}'(k_{\bar{q}'}))\left(x,v\right) E^0_3(\check{\bar{p}}_{\hat{q}}, k_{q'}, k_{\bar{q}'}^{{\rm id.}} ) \nonumber\\
&&\quad \quad + \bs{J}^{(1),\, {\rm id.} \gamma}_{2,\hat{q}}(\hat{q}(\bar{p}_{\hat{q}}),q'(k_{q'}))\left(x,v\right) E^0_3(\check{\bar{p}}_{\hat{q}}, k_{q'}^{{\rm id.}},  k_{\bar{q}'}) \bigg) \nonumber \\
&&\quad \quad \times M^0_{n+2} ( ... , k_{(q'\bar{q}')}, ...) J^{(n)}_n\left(\{ \tilde{k} \}_n ; z=uv\right) \, .
\label{eq:sigb2pggam}
\end{eqnarray}
Contributions of the form of \eqref{eq:sigTb2posigSqq} and \eqref{eq:sigb2pggam} subtract parts of the single unresolved limits of \eqref{eq:sigTacomsigRoGam0}. 

In general it is necessary to include additional terms to achieve an overall subtraction of the unresolved limits in \eqref{eq:sigTacomsigRoGam0}. Two classes of these additional terms are distinguished. The first class consists of all subtraction terms in which the unintegrated antenna function is a fragmentation antenna function. We call this contribution $\text{d} \hat{\sigma}^{T,c_1\, (p)}_{i(\gamma)}$. It has the form
\begin{eqnarray}
\lefteqn{-\text{d} \hat{\sigma}^{T,c_1 \, (p)}_{q(\gamma)} + \text{d} \hat{\sigma}^S_{q(q)} \otimes \mathbf{F}^{(0)}_{q \to \gamma}}
\nonumber \\
&= &- \mathcal{N}^{RV} Q_q^2  \sum_{{\rm perm.}} \int_0^1 \text{d} v \int \frac{\text{d}x}{x} \text{d} \Phi_{n+1}(k_1, ... , k_q, ... , k_{n+1}; x p_{\hat{q}} , p_2) \frac{1}{S_{n+1}}\nonumber \\
&&  \times \bs{J}^{(1),\,  {\rm id.} \gamma}_{2,\hat{q}}(\hat{q}(\bar{p}_{\hat{q}}),q(k_{(qg)}))\left(x,v \right) X^0_3(\check{k}_{l}, k_g , k_q^{{\rm id.}})  M^0_{n+2} ( ... , k_{(qg)}, ...) J^{(n)}_n\left(\{ \tilde{k} \}_n ; z =uv\right) \, .\nonumber \\ &&
\end{eqnarray}
In contrast to \eqref{eq:sigTb2posigSqq} the momenta entering the identified dipole belong to the mapped momentum set $\{ \tilde{k} \}$. $-\text{d} \hat{\sigma}^{T,c_1 \, (p)}_{g(\gamma)} + \text{d} \hat{\sigma}^S_{g(q)} \otimes  \mathbf{F}^{(0)}_{q \to \gamma} $ takes a similar form. In this case the unintegrated antenna function subtracts a $q\parallel\bar{q}$ limit. 

The second class of newly introduced subtraction terms does not contain an unintegrated fragmentation antenna function. We denote this contribution by  $\text{d} \hat{\sigma}^{T,c_2 \, (p)}_{q(\gamma)}$ and it has the form 
\begin{eqnarray}
\lefteqn{-\text{d} \hat{\sigma}^{T,c_2\, (p)}_{q(\gamma)} + \text{d} \hat{\sigma}^S_{q} \otimes  \mathbf{F}^{(0)}_{q \to \gamma} } \nonumber \\
&=& -\mathcal{N}^{RV} Q_q^2 \sum_{l} \sum_{{\rm perm.}} \int_0^1 \text{d} z \int \frac{\text{d}x}{x} \text{d} \Phi_{n+1}(k_1, ... , k_q , ... , k_{n+1}; x p_{\hat{q}} ,p_2)  \frac{1}{S_{n+1}}\nonumber \\
&&\times \bs{J}^{(1), \, {\rm id.} \gamma}_{2,\hat{q}}(\hat{q}(\bar{p}_{\hat{q}}),q(k_q))\left(x,z\right) X^0_3(k_i, k_l , k_m)  M^0_{n+2} ( ... k_I, k_M, ..., k_q, ...) J^{(n)}_n\left(\{ \tilde{k} \}_n ; z\right) \, ,\nonumber \\ &&
\end{eqnarray}
where $i$ and $m$ can be any hard radiator but not the final-state quark entering the integrated dipole. 
This contribution also contains those terms from  $\text{d} \hat{\sigma}^{S,d}_{q(\gamma)}$, in which the two antenna functions share the 
same initial-state radiator. 
Note that there is no contribution of the type $\text{d} \hat{\sigma}^{T,c_2\, (p)}_{g(\gamma)}$. 

Combining all terms in which the photon is part of the primary (integrated) antenna function, we find
\begin{eqnarray}
\lefteqn{ \left( \text{d} \hat{\sigma}^R_q - \text{d}\hat{\sigma}^S_q - \text{d} \hat{\sigma}^S_{q(q)} - \text{d} \hat{\sigma}^S_{g(q)} \right) \otimes \bs{J}^{(1), \, {\rm id.} \gamma}_{2,\hat{q}} } \nonumber \\
& = &\left(-\text{d} \hat{\sigma}^{T,a}_{q(\gamma)}  - \text{d} \hat{\sigma}^R_q \otimes  \mathbf{F}^{(0)}_{q \to \gamma} \right) + \left(- \text{d} \hat{\sigma}^{T,b_2\, (p)}_{q(\gamma)} + \text{d} \hat{\sigma}^S_{q(q)} \otimes  \mathbf{F}^{(0)}_{q \to \gamma} \right) \nonumber \\
&&+\left( -\text{d} \hat{\sigma}^{T,c_1}_{q(\gamma)} + \text{d} \hat{\sigma}^S_{q(q)} \otimes  \mathbf{F}^{(0)}_{q \to \gamma} \right) + \left(-\text{d} \hat{\sigma}^{T,c_1}_{g(\gamma)} + \text{d} \hat{\sigma}^S_{g(q)} \otimes  \mathbf{F}^{(0)}_{q \to \gamma} \right) \nonumber \\
&&+ \left( -\text{d} \hat{\sigma}^{T,c_2}_{q(\gamma)} + \text{d} \hat{\sigma}^S_{q} \otimes  \mathbf{F}^{(0)}_{q \to \gamma} \right)
\end{eqnarray}
where we have absorbed the contributions from $\text{d} \hat{\sigma}^{T,c_i \, (p)}_{q(\gamma)}$ into $\text{d} \hat{\sigma}^{T,c_i}_{q(\gamma)}$. 

None of these terms subtracts explicit poles or unresolved limits of $\text{d} \hat{\sigma}^{RV}_{\gamma}$, thus decoupling from the remaining subtraction at the real-virtual level.

\subsubsection{Direct Contribution: Double Virtual Level}
\label{sec:directVV}

At the double virtual level all subtraction terms which have not yet 
been added back are combined. The terms in $\text{d} \hat{\sigma}^U_{j(\gamma)}$ include integrals of subtraction terms in which the photon becomes unresolved. All explicit poles in $\text{d} \hat{\sigma}^{U}_{j(\gamma)}$ cancel against the mass factorisation terms of the fragmentation functions.

The first contribution in $\text{d} \hat{\sigma}^U_{q(\gamma)}$, $\text{d} \hat{\sigma}^{U,a}_{q(\gamma)}$ is given by the first term of $\text{d} \hat{\sigma}^{T,b}_{q(\gamma)}$ after integrating over the antenna phase space. It is combined with the corresponding contribution $\text{d} \hat{\sigma}^V_q \otimes ( - \mathbf{F}^{(0)}_{q \to \gamma})$ in \eqref{eq:sigMF}: 
\begin{eqnarray}
\lefteqn{-\text{d} \hat{\sigma}^{U,a}_{q(\gamma)} - \text{d} \hat{\sigma}^V_q \otimes  \mathbf{F}^{(0)}_{q \to \gamma}}  \nonumber \\
&= &\mathcal{N}^{VV} Q_q^2 \sum_{\rm {perm.}} \int_0^1 \text{d} z \int \frac{\text{d} x}{x}  \text{d} \Phi_{n}(k_1, ... , k_q , ...,k_n;x p_{\hat{q}} , p_2)  \nonumber \\
&&\times  \frac{1}{S_{n}}  \bs{J}^{(1), \, {\rm id.} \gamma}_{2,\hat{q}}(\hat{q}(\bar{p}_{\hat{q}}),q(k_q))(x,z) M^1_{n+2}(..., k_{q} ,...) J^{(n)}_n(\{ k \}_n ; z) \, .
\label{eq:sigUAcomqgam}
\end{eqnarray} 
This expression still exhibits explicit poles in the dimensional regulator $\epsilon$ coming from the one-loop matrix element. The poles of the one-loop matrix element cancel with one-loop  dipoles, resulting from integrated antenna functions in inclusive or fragmentation kinematics.   

To organise the cancellation of the terms, it is helpful to collect the contributions from $\text{d} \hat{\sigma}^{S,b_1}$ and $\text{d} \hat{\sigma}^{T,b}$ and combine them with mass factorisation contributions from the fragmentation functions alone \eqref{eq:sigMF} as well as with mixed initial-final mass factorisation contributions \eqref{eq:MF3mix}. For the case in which the photon becomes unresolved in a quark-type cluster this combination yields
\begin{eqnarray}
\lefteqn{-\text{d} \hat{\sigma}^{U,b}_{q(\gamma)} + \text{d} \hat{\sigma}^T_{q(q)} \otimes \mathbf{F}^{(0)}_{q \to \gamma} - \frac{\alpha_s}{2\pi} \text{d} \hat{\sigma}^B_q \otimes ( \mathbf{F}^{(1),B}_{q \to \gamma} + \mathbf{F}^{(1),C}_{q \to \gamma}) + \text{d} \hat{\sigma}^B_{q} \otimes \mathbf{\Gamma}_{{\rm PDF}} \otimes \mathbf{F}^{(0)}_{q \to \gamma}}  \nonumber \\
&= &\mathcal{N}^{VV} Q_q^2 \sum_{\rm {perm.}} \int_0^1 \text{d} z \int \frac{\text{d} x}{x}  \text{d} \Phi_{n}(k_1, ... , k_q , ...,k_n;x p_{\hat{q}} , p_2) \frac{1}{S_{n}} \nonumber \\
&&\times   \left( \tilde{\mathcal{A}}^{0,{\rm id.} \gamma}_{4,\hat{q}}(x,z) + \tilde{\mathcal{A}}^{1,{\rm id.} \, \gamma}_{3,\hat{q}}(x,z) - \mu_F^{-2\epsilon} \mathcal{A}^{0,{\rm id.} \gamma}_{3,\hat{q}}(x,z) \otimes \Gamma^{(1)}_{qq}(x) \right. \nonumber\\
&&\left. -\frac{1}{2} \mu_a^{-2\epsilon}  \mathcal{A}^{0,{\rm id.} \, q}_{3,\hat{q}}(x,z) \otimes \Gamma^{(0)}_{\gamma q}(z)+ \frac{1}{2} \left( \mu_F \, \mu_a \right)^{-2\epsilon} \Gamma^{(0)}_{\gamma q}(z) \, \Gamma^{(1)}_{qq}(x) \right. \nonumber \\
&&\left. +\frac{1}{2}  \left(\mu_a^2\right)^{-2\epsilon} \left(  \Gamma^{(0)}_{\gamma q}(z) \otimes \Gamma^{(1)}_{qq}(z) - \Gamma^{(1)}_{\gamma q}(z) \right) \right)  M^0_{n+2}(..., k_{q} ,...) J^{(n)}_n(\{ k \}_n ; z) \, . \nonumber \\
\label{eq:sigUBqgam}
\end{eqnarray}
Note that this combination of antenna functions and mass factorisation kernels can be related to a combination of NNLO coefficient functions for semi-inclusive deep inelastic scattering~\cite{Gehrmann:2021lwb}. 
It is useful to rewrite \eqref{eq:sigUBqgam} as a sum of a finite two-loop fragmentation dipole and a convolution of two fragmentation dipoles, i.e.\
\newpage
\begin{eqnarray}
\lefteqn{-\text{d} \hat{\sigma}^{U,b}_{q(\gamma)} + \text{d} \hat{\sigma}^T_{q(q)} \otimes \mathbf{F}^{(0)}_{q \to \gamma} - \frac{\alpha_s}{2\pi} \text{d} \hat{\sigma}^B_q \otimes \left( \mathbf{F}^{(1),B}_{q \to \gamma} + \mathbf{F}^{(1),C}_{q \to \gamma} \right) + \text{d} \hat{\sigma}^B_{q} \otimes \mathbf{\Gamma}_{{\rm PDF}} \otimes \mathbf{F}^{(0)}_{q \to \gamma}}  \nonumber \\
&=&\mathcal{N}^{VV} Q_q^2 \sum_{\rm {perm.}} \int_0^1 \text{d} z \int \frac{\text{d} x}{x}  \text{d} \Phi_{n}(k_1, ... , k_q , ...,k_n;x p_{\hat{q}} , p_2) \frac{1}{S_{n}} \nonumber \\
&&\times \left( \bs{J}^{(2),{\rm id.} \gamma}_{2,\hat{q}}(\hat{q},q) + \bs{J}^{(1),{\rm id.} \gamma}_{2,\hat{q}}(\hat{q},q) \otimes \bs{J}^{(1),{\rm id.} q}_{2,\hat{q}}(\hat{q},q) \right) M^0_{n+2}(..., k_{q} ,...) J^{(n)}_n(\{ k \}_n ; z) \, ,\nonumber \\
\label{eq:sigUBqgamre}
\end{eqnarray}
where the one-loop dipoles are given in \eqref{eq:iddipoleqgamIF} and in \eqref{eq:iddipolesqqclusterIF} respectively. The two-loop quark-to-photon
dipole $\bs{J}^{(2),{\rm id.} \gamma}_{2,\hat{q}}(\hat{q},q) $ expressed in terms of fragmentation antenna functions and mass factorisation terms reads
\begin{eqnarray}
\bs{J}^{(2), {\rm id.} \gamma}_{2,\hat{q}}(\hat{q},q) &= &\tilde{\mathcal{A}}^{0, {\rm id.} \gamma}_{4,\hat{q}}(x,z) + \tilde{\mathcal{A}}^{1, {\rm id.} \gamma}_{3,\hat{q}}(x,z) \nonumber \\
&&- \left(\mathcal{A}^{0,{\rm id.}\gamma}_{3,\hat{q}}(x,z) \otimes ( \mathcal{A}^{0,{\rm id.} q}_{3,\hat{q}}(x,z) - \mu_a^{-2\epsilon} \, \Gamma^{(1)}_{q q}(z)) \right) \nonumber \\
&&- \frac{1}{2} \left(\mu_a^2 \right)^{-2\epsilon} \, \Gamma^{(1)}_{\gamma q}(z) \, .
\label{eq:J22qgam}
\end{eqnarray}
Since this two-loop dipole is $\epsilon$-finite, the poles in \eqref{eq:sigUBqgamre} are all contained in the 
fragmentation dipoles for identified partons $\bs{J}^{(1),{\rm id.} q}_{2,\hat{q}}(\hat{q},q)$. These poles partly cancel the poles in \eqref{eq:sigUAcomqgam}. 

For the case in which the photon is clustered into a gluon, we find
\begin{eqnarray}
\lefteqn{-\text{d} \hat{\sigma}^{U,b}_{g(\gamma)} + \text{d} \hat{\sigma}^T_{g(q)} \otimes \mathbf{F}^{(0)}_{q \to \gamma} - \frac{\alpha_s}{2\pi} \text{d} \hat{\sigma}^B_g \otimes \left( \mathbf{F}^{(1),B}_{g \to \gamma} + \mathbf{F}^{(1),C}_{g \to \gamma} \right)} \nonumber \\
&=& \mathcal{N}^{VV} N_{q'} Q_{q'}^2 \sum_{\rm {perm.}} \int_0^1 \text{d} z \int \frac{\text{d} x}{x}  \text{d} \Phi_{n}(k_1, ... , k_g , ...,k_n;x p_{\hat{q}} , p_2) \frac{1}{S_{n}} \nonumber \\
&&\times   \left( \tilde{\mathcal{E}}^{0,{\rm id.} \gamma}_{4,\hat{q}}(x,z) - \mu_a^{-2\epsilon} \mathcal{E}^{0,{\rm id.} \, q'}_{3,\hat{q}}(x,z) \otimes \Gamma^{(0)}_{\gamma q}(z) \right. \nonumber\\
&&  \left. + \left( \mu_a^2 \right)^{-2\epsilon} \left( \Gamma^{(0)}_{\gamma q}(z) \otimes \Gamma^{(1)}_{qg}(z) - \tilde{\Gamma}^{(1)}_{\gamma g}(z) \right) \right)  M^0_{n+2}(..., k_{g} ,...) J^{(n)}_n(\{ k \} ; z) \, ,
\label{eq:sigUBggam}
\end{eqnarray}
 with 
\begin{equation}
\tilde{\Gamma}^{(1)}_{\gamma g} = \frac{1}{2} \left( \frac{1}{2\epsilon^2} p^{(0)}_{q g} \otimes p^{(0)}_{\gamma q} - \frac{1}{2 \epsilon} p^{(1)}_{\gamma g} \right) \,.
\end{equation}
We can rewrite \eqref{eq:sigUBggam} as a sum of a two-loop dipole and the convolution of two fragmentation dipoles, 
\begin{eqnarray}
\lefteqn{-\text{d} \hat{\sigma}^{U,b}_{g(\gamma)} + \text{d} \hat{\sigma}^T_{g(q)} \otimes \mathbf{F}^{(0)}_{q \to \gamma} - \frac{\alpha_s}{2\pi} \text{d} \hat{\sigma}^B_g \otimes \left( \mathbf{F}^{(1),B}_{g \to \gamma} + \mathbf{F}^{(1),C}_{g \to \gamma} \right)} \nonumber \\
&=& \mathcal{N}^{VV} N_{q'} Q_{q'}^2 \sum_{\rm {perm.}} \int_0^1 \text{d} z \int \frac{\text{d} x}{x}  \text{d} \Phi_{n}(k_1, ... , k_g , ...,k_n;x p_{\hat{q}} , p_2) \frac{1}{S_{n}}  \nonumber \\
&&\times   \left( \bs{J}^{(2), {\rm id.} \gamma}_{2,\hat{q}}(\hat{q},g) + 2 \, \bs{J}^{(1), {\rm id.} \gamma}_{2,\hat{q}}(\hat{q},q) \otimes \bs{J}^{(1), {\rm id.} q'}_{2,\hat{q}}(\hat{q},g) \right)  M^0_{n+2}(..., k_{g} ,...) J^{(n)}_n(\{ k \} ; z) \, ,\nonumber\\
\label{eq:sigUBggamre}
\end{eqnarray}
where the one-loop dipoles are given in \eqref{eq:iddipoleqgamIF} and in \eqref{eq:J21qpidIF} respectively. The two-loop 
gluon-to-photon dipole reads
\begin{eqnarray}
\bs{J}^{(2), \, {\rm id.} \gamma}_{2, \hat{q}}(\hat{q},g) &= &\tilde{\mathcal{E}}^{0,{\rm id.} \gamma}_{4,\hat{q}}(x,z) - 2 \left( \mathcal{A}^{0,{\rm id.} \gamma}_{3,\hat{q}}(x,z) \otimes ( \mathcal{E}^{0, {\rm id.} q'}_{3,\hat{q}}(x,z) - \mu_a^{-2\epsilon} \Gamma^{(1)}_{qg}(z) ) \right) \nonumber\\
&&- \left( \mu_a^2 \right)^{-2\epsilon} \tilde{\Gamma}^{(1)}_{\gamma g}(z) \, .
\label{eq:J22ggam}
\end{eqnarray}
Note that all three dipoles in \eqref{eq:sigUBggamre} correspond to flavour-changing limits. Therefore, all three of them are by themselves $\epsilon$-finite. 

Two more contributions from the double real subtraction terms and real-virtual subtraction terms have to be added back, $\text{d} \hat{\sigma}^{S,d}_{q(\gamma)}$ and $\text{d} \hat{\sigma}^{T,c}_{q(\gamma)}$. Among these contributions only the terms in $\text{d} \hat{\sigma}^{T,c_1\, (p)}_{q(\gamma)}$ consist of two fragmentation antenna functions. 
Integration over the antenna phase space and combination with mass factorisation contributions from \eqref{eq:sigMF} and \eqref{eq:MF3mix} yields
\begin{eqnarray}
\lefteqn{-\text{d} \hat{\sigma}^{U,c_1}_{q(\gamma)} + \text{d} \hat{\sigma}^T_{q(q)} \otimes \mathbf{F}^{(0)}_{q \to \gamma} - \frac{\alpha_s}{2\pi} \text{d} \hat{\sigma}^B_q \otimes \mathbf{F}^{(1),B}_{q \to \gamma}  + \text{d} \hat{\sigma}^B_{q} \otimes \mathbf{\Gamma}_{{\rm PDF}} \otimes \mathbf{F}^{(0)}_{q \to \gamma}} \nonumber \\
&=& \mathcal{N}^{VV} Q_q^2 \sum_{\rm {perm.}} \int_0^1 \text{d} z \int \frac{\text{d} x_1}{x_1} \frac{\text{d}x_2}{x_2}  \text{d} \Phi_{n}(k_1, ... , k_q , ...,k_n;x_1 p_{\hat{q}} , x_2 p_2) \frac{1}{S_{n}} \nonumber \\
&&\times \left( \bs{J}^{(1),{\rm id.} \gamma}_{2,\hat{q}}(\hat{q},q) \otimes \bs{J}^{(1),{\rm id.} q}_{2,l}(l,q) \right) M^0_{n+2}(..., k_{q} ,...) J^{(n)}_n(\{ k \}_n ; z) \, ,
\label{eq:sigUC1qgamre}
\end{eqnarray}
where $l$ can either be a quark or a gluon. If $l$ is in the final state there is no contribution from the mixed initial-final mass factorisation contribution \eqref{eq:MF3mix}.
 
Integrating the subtraction terms of $\text{d} \hat{\sigma}^{T,c\, (s)}_{q(\gamma)}$ over the antenna phase space, we obtain
\begin{eqnarray}
\lefteqn{-\text{d} \hat{\sigma}^{U,c_2}_{q(\gamma)} + \text{d} \hat{\sigma}^B_{q} \otimes \mathbf{\Gamma}_{{\rm PDF}} \otimes \mathbf{F}^{(0)}_{q \to \gamma}}\nonumber \\
&= &\mathcal{N}^{VV} Q_q^2 \sum_{\rm {perm.}} \int_0^1 \text{d} z \int \frac{\text{d} x_1}{x_1} \frac{\text{d}x_2}{x_2}  \text{d} \Phi_{n}(k_1, ... , k_q , ...,k_n;x_1 p_{\hat{q}} , x_2 p_2) \frac{1}{S_{n}} \nonumber \\
&&\times \left( \bs{J}^{(1),{\rm id.} \gamma}_{2,\hat{q}}(\hat{q},q) \otimes \bs{J}^{(1)}_{2}(l,q) \right) M^0_{n+2}(..., k_{q} ,...) J^{(n)}_n(\{ k \}_n ; z) \, ,
\label{eq:sigUC2qgamre}
\end{eqnarray}
where $l$ is either a quark or a gluon in the initial or final state. In contrast to \eqref{eq:sigUC1qgamre}, the secondary dipole in the equation at hand is an inclusive dipole, i.e.\ it has no explicit $z$ dependence so that the convolution in the final-state momentum fraction is trivial. In the convolution of the two dipoles we have included terms from the mixed initial-final mass factorisation contribution \eqref{eq:MF3mix}. To complete the fragmentation dipoles in \eqref{eq:sigUC2qgamre} also pure final-state mass factorisation terms are needed. Since these terms cancel after summation of the different terms in \eqref{eq:sigUC2qgamre}, they do not appear on the left-hand side of the equation.

The last contribution to the double virtual subtraction takes into account the terms from $\text{d} \hat{\sigma}^{S,d}_{q(\gamma)}$ and $\text{d} \hat{\sigma}^{T,c_2\, (p)}_{q(\gamma)}$ and we denote it $\text{d} \hat{\sigma}^{U,d}_{q(\gamma)}$. Adding  the corresponding mass factorisation contributions yields
\newpage
\begin{eqnarray}
\lefteqn{-\text{d} \hat{\sigma}^{U,d}_{q(\gamma)} + \text{d} \hat{\sigma}^T_{q} \otimes \mathbf{F}^{(0)}_{q \to \gamma} + \text{d} \hat{\sigma}^B_{q} \otimes \mathbf{\Gamma}_{{\rm PDF}} \otimes \mathbf{F}^{(0)}_{q \to \gamma}}\nonumber \\
&=& \mathcal{N}^{VV} Q_q^2 \sum_{\rm {perm.}} \int_0^1 \text{d} z \int \frac{\text{d} x_1}{x_1} \frac{\text{d}x_2}{x_2}  \text{d} \Phi_{n}(k_1, ... , k_q , ...,k_n;x_1 p_{\hat{q}} , x_2 p_2) \frac{1}{S_{n}}\nonumber  \\
&&\times \left( \bs{J}^{(1),{\rm id.} \gamma}_{2,\hat{q}}(\hat{q},q) \otimes \bs{J}^{(1)}_{2}(i,m) \right) M^0_{n+2}(..., k_{q} ,...) J^{(n)}_n(\{ k \}_n ; z) \, ,
\label{eq:sigUDqgamre}
\end{eqnarray}
where $i$ and $m$ can be any partons in the process but not the identified quark in the final state.

We have rewritten all double virtual subtraction terms in which the photon becomes unresolved in terms of  two newly introduced two-loop fragmentation dipoles $\bs{J}^{(2),{\rm id.} \gamma}_{2,\hat{q}}(\hat{q},q)$ and $\bs{J}^{(2),{\rm id.} \gamma}_{2,\hat{q}}(\hat{q},g)$ and convolutions of two dipoles in which one dipole is always given by $\bs{J}^{(1),{\rm id.} \gamma}_{2,\hat{q}}(\hat{q},q)$. All two-loop parton-to-photon 
dipoles are by themselves $\epsilon$-finite. The poles in the convolution terms cancel the explicit poles in $\text{d} \hat{\sigma}^{U,a}_{q(\gamma)}$.

\section{Integration of $X^0_4$ Fragmentation Antenna Functions}\label{sec:X40int}

$X^0_4$ initial-final antenna functions are kinematically described by a scattering process of the form
\begin{equation}
q+p \rightarrow k_j + k_l + k_k \, .
\end{equation}
The final-state momenta and the initial-state momentum $p$ are massless $p^2=k_j^2=k_l^2=k_k^2=0$ and we have $q^2= -Q^2 <0$. The fully inclusive integrated $\mathcal{X}^0_4$ antenna functions are obtained by integration over the corresponding three-body phase space~\cite{Daleo:2009yj}:
\begin{equation}
\mathcal{X}^0_{i,jkl}(x) = \frac{1}{C(\epsilon)^2} \int \text{d} \Phi_3(k_j,k_k,k_l;p,q) \frac{Q^2}{2 \pi} X^0_{i,jkl} \, ,
\label{eq:def_ifantenna_qcd}
\end{equation}
with $x= {Q^2}/({2p \cdot q})$ and the normalisation factor 
\begin{equation}
C(\epsilon) = \frac{\left(4\pi e^{-\gamma_E}\right)^{\epsilon}}{8 \pi^2} \, .
\end{equation}

For initial-final  fragmentation antenna functions the same normalisation as in \eqref{eq:def_ifantenna_qcd} is used but the integration remains differential in the final-state momentum fraction $z$, i.e.
\begin{equation}
\mathcal{X}^{0, \, {\rm id.}  j}_{i,j k l}(x,z) = \frac{1}{C(\epsilon)^2} \int \text{d} \Phi_3(k_j,k_k,k_l;p,q) \, \delta\left(z -x \frac{(p+k_j)^2}{Q^2} \right) \frac{Q^2}{2\pi} X^0_{i,j k l} \, .
\label{eq:def_intphotonicantenna}
\end{equation}
The final-state momentum fraction is fixed by the additional $\delta$-distribution and it describes the fraction of energy carried by particle $j$ in the unresolved limit. In the definition of the momentum fraction the initial-state momentum $p$ is used as a reference momentum, which can be seen by rewriting its definition
\begin{equation}
z = x\frac{(k_j + p)^2}{Q^2} = \frac{s_{jp}}{s_{jp} + s_{kp} + s_{lp}} \, .
\end{equation}

For an identified photon i.e.\ $j=\gamma$ there are two fragmentation antenna functions: $\tilde{A}^0_4(\hat{q},g,\gamma^{{\rm id.}},q)$ 
containing the triple-collinear $q\to qg\gamma$ configuration and 
 $\tilde{E}^0_4(\hat{q},q',\gamma^{{\rm id.}},\bar{q}')$ containing the triple collinear $g\to q'\bar{q}'\gamma$ configuration.
To integrate these fragmentation antenna functions, we use the reduction to master integrals technique. Using
\begin{equation}
2 \pi i \delta(k^2) = \frac{1}{k^2+ i \epsilon} - \frac{1}{k^2-i \epsilon} \, ,
\end{equation}
we rewrite the phase space integrals as $2\to 2$  three-loop-integrals with  forward scattering kinematics. 

The reduction is performed with the program \texttt{Reduze2}~\cite{vonManteuffel:2012np}. For the integration of the two photonic fragmentation antenna functions we find nine master integrals. The master integrals are calculated using their differential equations in the two kinematic variables $x$ and $z$. The boundary conditions are fixed by integrating the solution of the differential equations over $z$ and comparing the result with the inclusive master integrals calculated in~\cite{Daleo:2009yj}.

The master integrals take the general form
\begin{equation}
I(x,z) = (1-x)^{a - 2  \epsilon} ( z^{-\epsilon} A(x,z) + z^{-2\epsilon} B(x,z))
\end{equation}
with $a \in \{-1,0,1\}$. After being inserted into the antenna functions, the factor $(1-x)^{a - 2  \epsilon}$ can give rise to factors of the form $(1-x)^{-1 -2  \epsilon}$, whose expansion reads
\begin{equation}
(1-x)^{-1-2 \epsilon} = -\frac{1}{2 \, \epsilon} \delta(1-x) + \sum_n \frac{(- 2 \, \epsilon)^n}{n!} \mathcal{D}_n(x) \, ,
\label{eq:distexp}
\end{equation}
where we used the notation introduced in \eqref{eq:Dndef}.

Potential factors of the form $z^{-1 - a \epsilon}$ do not have to be expanded in terms of distributions, since the endpoint $z=0$ corresponds to a soft photon singularity. This singularity will be regulated by the jet function, which requires a minimum $p_T$ of the photon so that the endpoint $z=0$ does not contribute to any observable with a photon in the final state. 
However, to check the result of the integrated fragmentation antenna functions we also derive the master integrals with the exact scaling in $z=0$ in the limit of $z \to 0$. 

In the scattering 
\begin{equation}
q + p \to p_1(k_{\gamma}) + p_2(k_2) + p_3(k_3)
\end{equation}
12 different propagators appear from which four are cut propagators. Using four-momentum conservation $k_3 = q + p - k_{\gamma} - k_2$, they read
\begin{eqnarray}
D_1 &=& (q-k_{\gamma})^2 \, , \nonumber \\
D_2 &=& (p+q-k_{\gamma})^2 \, ,\nonumber  \\
D_3 &=& (p-k_2)^2 \, ,\nonumber  \\
D_4 &=& (q- k_2)^2 \, ,\nonumber  \\
D_5 &=& (p+q-k_2)^2 \, ,\nonumber  \\
D_6 &=& (q-k_{\gamma} -k_2)^2\, ,\nonumber  \\
D_7 &=& (p-k_{\gamma}-k_2)^2 \, , \nonumber \\
D_8 &=& (k_{\gamma} + k_2)^2\, , \nonumber \\
D_9 &=& k_{\gamma}^2\, ,\nonumber  \\
D_{10} &=& k_2^2 \, , \nonumber \\
D_{11} &=& (q+p-k_{\gamma}-k_2)^2 \, , \nonumber \\
D_{12} &=& (p-k_{\gamma})^2 + Q^2 \frac{z}{x} \, ,
\end{eqnarray}
where the cut propagators are $D_9-D_{12}$. We label the master integrals by the propagators in the corresponding integral (omitting the cut propagators, which we require in each integral), for example:
\begin{equation}
I[-3,7] = \frac{Q^2(2\pi)^{-2d+3}}{x}  \int \text{d}^d k_{\gamma} \, \text{d}^d k_2  \, \delta\left(D_9\right) \, \delta\left(D_{10}\right) \, \delta\left(D_{11}\right)  \delta\left(D_{12}\right) \frac{D_3}{D_7}.
\end{equation}
The factor $Q^2/x$ originates from rewriting the $\delta$-distribution fixing the momentum fraction $z$ in \eqref{eq:def_intphotonicantenna} in terms of $\delta(D_{12})$.
As there are seven linearly independent scalar products the integration families consist of the four cut propagators and three additional propagators. We find three integral families and in total nine master integrals which are summarised in Table\,\ref{tabMI}.
\begin{table}[t]
\centering
\resizebox{\columnwidth}{!}{%
\begin{tabular}{c|c|c|c|c}
family                & master     & deepest pole    & behaviour at $x=1$     & known to order                     \\ \hline
\multicolumn{1}{l|}{} & $I[0]$     & $\epsilon^0$    & $(1-x)^{1-2\epsilon}$  & all                                \\ \hline
\multirow{2}{*}{A}    & $I[5]$     & $\epsilon^{-1}$ & $(1-x)^{-2 \epsilon}$  & all                                \\
                      & $I[2,3,5]$ & $\epsilon^{-2}$ & $(1-x)^{-1-2\epsilon}$ & $\epsilon^1$                         \\ \hline
\multirow{4}{*}{B}    & $I[7]$     & $\epsilon^0$    & $(1-x)^{1-2\epsilon}$  & $\epsilon^2$                       \\
                      & $I[-2,7]$  & $\epsilon^0$    & $(1-x)^{1-2\epsilon}$  & $\epsilon^2$                       \\
                      & $I[-3,7]$  & $\epsilon^0$    & $(1-x)^{1-2\epsilon}$  & $\epsilon^2$                       \\
                      & $I[2,3,7]$ & $\epsilon^{-2}$ & $(1-x)^{-2\epsilon}$   & $\epsilon^0$ ($\epsilon^1$ at $x=1$) \\ \hline
\multirow{2}{*}{C}    & $I[5,7]$   & $\epsilon^{-1}$ & $(1-x)^{-2 \epsilon}$  & $\epsilon^0$ ($\epsilon^1$ at $x=1$) \\
                      & $I[3,5,7]$ & $\epsilon^{-2}$ & $(1-x)^{- 2\epsilon}$  & $\epsilon^0$ ($\epsilon^1$ at $x=1$)
\end{tabular}%
}
\caption{\label{tabMI} Summary of the double real radiation master integrals.}
\end{table}

The phase space integral $I[0]$ has been calculated directly by carrying out the three-body phase space integral and by solving the differential equation in the kinematic variable $z$ and fixing the boundary condition by comparing to the inclusive three-body phase space. It reads
\begin{equation}
I[0] = N_{\Gamma} \left(Q^2\right)^{1-2\epsilon} (1-x)^{1-2\epsilon} x^{-1+2\epsilon} z^{-\epsilon} (1-z)^{1-2\epsilon} \, , 
\end{equation}
with the normalisation factor
\begin{equation}
N_{\Gamma} = \frac{2^{-5+4\epsilon} \pi^{-3+2\epsilon}\, \Gamma^2(2-\epsilon)}{\Gamma^2\left(3- 2\epsilon\right)}\, .
\end{equation}

The only other master integral which admits a simple closed form solution is the master integral $I[5]$. We find
\begin{eqnarray}
I[5] &=& N_{\Gamma} \left(\frac{1-2\epsilon}{\epsilon} \right)^2 \left(Q^2 \right)^{-2\epsilon} (1-x)^{-2\epsilon} x^{2\epsilon} \nonumber \\
&&\times \left( z^{-\epsilon}  {}_2F_1(\epsilon,2\epsilon,1+\epsilon;z) - z^{-2\epsilon} \frac{\Gamma\left(1-2\epsilon\right)\Gamma(1+\epsilon)}{\Gamma(1-\epsilon)} \right)\, .
\end{eqnarray}
All other master integrals have been calculated in terms of a Laurent expansion in $\epsilon$.

The integrated antenna functions are then obtained by reducing the integrand in \eqref{eq:def_intphotonicantenna} to these master integrals, 
and applying \eqref{eq:distexp} to extract the end-point contributions in $x=1$. The results for
 $\tilde{\mathcal{A}}^{0, \, {\rm id.}  \gamma}_{q,\gamma qg}(x,z)$ and $\tilde{\mathcal{E}}^{0, \, {\rm id.}  \gamma}_{q,\gamma q' \bar{q}'}(x,z)$
 are too lengthy to be expressed in the text here, and are included as ancillary files.  

\section{Integration of $X^1_3$ Fragmentation Antenna Functions}
\label{sec:X31int}

The inclusive integrated one-loop antenna functions in the initial-final configuration are defined as~\cite{Daleo:2009yj}
\begin{equation}
\mathcal{X}^1_{i,jk}(x) = \frac{1}{C(\epsilon)} \int \text{d} \Phi_2(k_j,k_k;p_i,q) \frac{Q^2}{2 \pi} X^1_{i,jk} \, ,
\label{eq:def_X31_QCD}
\end{equation}
where $X^1_{i,jk}$ is the unintegrated one-loop antenna function and $\text{d} \Phi_2$ the two-particle phase space. We define the integrated  initial-final one-loop fragmentation antenna functions in line with \eqref{eq:def_X31_QCD} as 
\begin{eqnarray}
\mathcal{X}^{1, {\rm id.} j}_{i,j k}(x,z) &=& \frac{1}{C(\epsilon)} \int \text{d} \Phi_2(k_j,k_k;p_i,q) \, \delta \left( z - \frac{s_{i j}}{s_{i j} + s_{i k}} \right) \frac{Q^2}{2 \pi} X^1_{i,j k} \nonumber  \\
&=& \frac{Q^2}{2} \frac{e^{\gamma_E \epsilon}}{\Gamma(1-\epsilon)} \left(Q^2\right)^{-\epsilon} \mathcal{J}(x,z) \, X^1_{i,j k} \, .
\label{eq:def_X31_integrated_photonic}
\end{eqnarray}

The integration takes the same form as for the $X^0_3$ initial-final fragmentation antenna functions, see \eqref{eq:intX30IFfrag} above.
 The Jacobian factor $\mathcal{J}$ is given in \eqref{eq:JacPhi2}.
As can be seen from \eqref{eq:def_X31_integrated_photonic}, no actual integration has to be performed to obtain the integrated fragmentation antenna functions $\mathcal{X}^{1, {\rm id.} j}_{i,j k}$. However, to express the integrated fragmentation antenna functions in terms of distributions in $(1-x)$ and in $z$ we first have to cast the unintegrated antenna functions in a form suitable for this expansion. Therefore, deriving the integrated initial-final one-loop fragmentation antenna functions follows the steps of the derivation of the integrated initial-initial one-loop antenna functions presented in~\cite{Gehrmann:2011wi}. In contrast to the NLO $X^0_3$ antenna functions which only contain rational terms in the invariants, the one-loop antenna functions $X^1_3$ also contain logarithms and polylogarithms in the invariants. These functions have branch cuts in the limits $x \to 1$ and $z \to 0$. Therefore, the expansion in distributions in $z=0$ and $x=1$ cannot be performed directly. We follow the strategy of~\cite{Gehrmann:2011wi} and express the one-loop antenna functions in terms of one-loop master integrals. 

The one-loop master integrals appearing in the expressions for the one-loop antenna functions are the one-loop bubble ${\rm Bub}(s_{ij})$ and the one-loop ${\rm Box}(s_{ij},s_{ik})$ in all kinematic crossings. The expression for the one-loop bubble reads
\begin{equation}
\text{Bub}(s_{ij}) = \left[ \frac{(4\pi)^{\epsilon}}{16 \pi^2} \frac{\Gamma(1+\epsilon) \Gamma^2(1-\epsilon)}{\Gamma(1-2\epsilon)} \right] \frac{i}{\epsilon(1-2\epsilon)} \left( - s_{ij} \right)^{-\epsilon} \equiv A_{2,LO} \left( - s_{ij} \right)^{-\epsilon} \, , 
\label{eq:definition_bubble}
\end{equation}
and the expression for the one-loop box is
\begin{eqnarray}
\lefteqn{\text{Box}(s_{ij},s_{ik})} \nonumber  \\
&=& \frac{2 (1-2 \epsilon)}{\epsilon} A_{2,LO} \frac{1}{s_{ij}s_{ik}} \nonumber \\
&&\times \bigg[ \left(\frac{s_{ij} s_{ik}}{s_{ij}-s_{ijk}}\right)^{-\epsilon}  {}_2 F_1\left( - \epsilon, -\epsilon; 1- \epsilon ; \frac{s_{ijk} - s_{ij} - s_{ik}}{s_{ijk} - s_{ij}}\right) \nonumber  \\
&&+ \left(\frac{s_{ij} s_{ik}}{s_{ik}-s_{ijk}}\right)^{-\epsilon}  {}_2 F_1\left( - \epsilon, -\epsilon; 1- \epsilon ; \frac{s_{ijk} - s_{ij} - s_{ik}}{s_{ijk} - s_{ik}}\right) \nonumber \\
&&- \left(\frac{- s_{ijk} s_{ij} s_{ik}}{(s_{ij}-s_{ijk})(s_{ik}- s_{ijk})}\right)^{-\epsilon}  {}_2 F_1\left( - \epsilon, -\epsilon; 1- \epsilon ; \frac{s_{ijk}(s_{ijk} - s_{ij} - s_{ik})}{(s_{ijk}- s_{ij}) (s_{ijk} - s_{ik})}\right) \bigg] \, .
\label{eq:definition_box}
\end{eqnarray}

For the following discussion we adopt the labelling to $p_i \to p_1$, $k_j \to k_3$ and $k_k \to k_2$ in 
\eqref{eq:def_X31_integrated_photonic}, so that the particle with momentum $k_3$ is identified and the momentum $p_1$ is the reference momentum. Using this convention, the invariants expressed in terms of $x$, $z$ and $Q^2$ read
\begin{eqnarray}
s_{12} &=& (p_1 - k_2)^2 = - Q^2 \frac{(1-z)}{x} \, , \nonumber \\
s_{13} &=& (p_1- k_3)^2  = - Q^2 \frac{z}{x} \, , \nonumber \\
s_{23} &=& (k_2 + k_3)^2= - Q^2 \frac{(x-1)}{x} \, , \nonumber \\
s_{123} &=& (k_1 + k_2 - p_1)^2 = - Q^2 \, .
\end{eqnarray}

Both master integrals are well-defined in the Euclidean region, in which all invariants are smaller than 0. The master integrals have to be analytically continued from this kinematic region, to the kinematic region under consideration given by
\begin{equation}
s_{12} < 0 \quad , \quad s_{13} < 0 \quad , \quad s_{23} >0 \quad , \quad s_{123} = - Q^2 < 0 \, .
\label{eq:kinregion}
\end{equation}
The analytic continuation of the bubble master integral is straightforward, taking into account $s_{ij} \to s_{ij} + i\delta$ in 
\eqref{eq:definition_bubble}.

In the analytic continuation of the box integrals, the prefactors in front of the hypergeometric functions as well as the hypergeometric functions themselves have to be considered. 
In particular branch cuts of the hypergeometric functions in the kinematic endpoints $x=1$ and $z=0$ have to be avoided:
  for these values the arguments of the hypergeometric function must not be unity or $+\infty$. To further avoid explicit imaginary parts from 
  the hypergeometric functions, their arguments are moreover
   transformed to be less than $+1$ using their well-known transformation rules~\cite{bateman}.
  It is noted that this will typically require to partition the kinematic region defined by~\eqref{eq:kinregion} into 
  up to four segments~\cite{Graudenz:1993tg,Gehrmann:2002zr}, see Figure~\ref{fig:kinematic_regions_com} below. In the following, 
  we discuss the transformations of the arguments for the different hypergeometric functions appearing in the box master integrals for all kinematic crossings: 
  ${\rm Box}(s_{12},s_{23})$, ${\rm Box}(s_{13},s_{23})$ and  ${\rm Box}(s_{12},s_{13})$. 
     
In ${\rm Box}(s_{12},s_{23})$ the arguments of the hypergeometric functions read
\begin{eqnarray}
a_1(s_{12},s_{23}) &=& \frac{s_{123}-s_{12}-s_{23}}{s_{123}-s_{12}} = -\frac{z}{1-x-z} \, ,\nonumber \\ 
a_2(s_{12},s_{23}) &=& \frac{s_{123}-s_{12}- s_{23}}{s_{123}-s_{23}} = z \, , \nonumber \\ 
a_3(s_{12},s_{23}) &=& \frac{s_{123} \, s_{13} }{(s_{13}+s_{23}) (s_{12} + s_{13})} = - \frac{x \, z}{1-x-z} \, .
\label{eq:a_boss12s23}
\end{eqnarray}
All arguments vanish in the kinematic endpoint $z=0$. However, $a_1$ and $a_3$ are equal to unity in the kinematic endpoint $x=1$. 
Therefore, the analytic continuation of the corresponding hypergeometric functions proceeds by expressing these functions as hypergeometric functions in terms of  new arguments:
\begin{eqnarray}
\tilde{a}_1(s_{12},s_{23}) &=& 1-\frac{1}{a_1(s_{12},s_{23})} = \frac{1-x}{z} \, , \nonumber \\
\tilde{a}_3(s_{12},s_{23}) &=& 1-\frac{1}{a_3(s_{12},s_{23})} = \frac{(1-x)(1-z)}{xz} \, .
\end{eqnarray}
The arguments $\tilde{a}_1$ and $\tilde{a}_3$ vanish in the endpoint $x=1$ but yield unity in the endpoint $z=0$. 

Therefore, to obtain an expression for ${\rm Box}(s_{12},s_{23})$ which does not contain hypergeometric functions with branch cuts in $z=0$ and $x=1$, it is necessary to distinguish the two regions
\begin{eqnarray}
R_1 &=&\{ s_{13} , s_{23} : s_{13} + s_{23} > 0 \Leftrightarrow z < 1-x \} \, ,\nonumber \\
R_2 &=& \{ s_{13} , s_{23} : s_{13} + s_{23} < 0 \Leftrightarrow z > 1-x \} \, .
\label{eq:def_R1_R2}
\end{eqnarray}
The regions are depicted in Figure~\ref{fig:kinematic_regions_com}. In region $R_1$ which contains the endpoint $z=0$ we use the hypergeometric functions with the arguments given in \eqref{eq:a_boss12s23}, while in region $R_2$, which contains the endpoint $x=1$ we express ${\rm Box}(s_{12},s_{23})$ in terms of hypergeometric functions with arguments $\tilde{a}_1, a_2$ and $\tilde{a}_3$. 
\begin{figure}[t]
\centering
\begin{subfigure}{.28\textwidth}
  \centering
  \includegraphics[width=\linewidth]{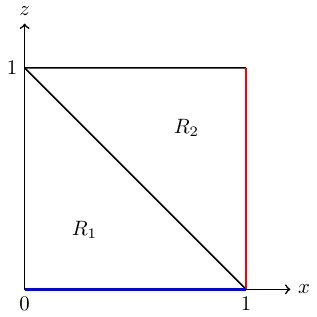}
\end{subfigure}%
\begin{subfigure}{0.28\textwidth}
  \centering
  \includegraphics[width=\linewidth]{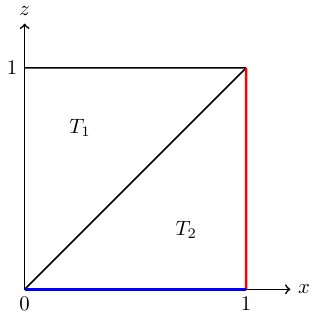}
\end{subfigure}%
\begin{subfigure}{.28\textwidth}
  \centering
  \includegraphics[width=\linewidth]{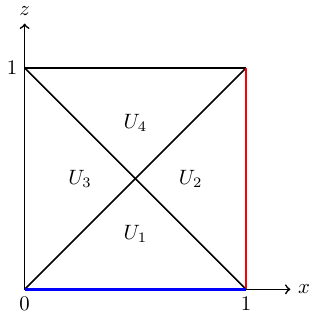}
\end{subfigure}
\caption{Kinematic regions in the $(x,z)$-plane relevant for the analytic continuation of the box master integrals . The kinematic endpoints are  $z=0$ (blue line) and $x=1$ (red line).}
\label{fig:kinematic_regions_com}
\end{figure}

For ${\rm Box}(s_{13},s_{23})$ the arguments of the hypergeometric functions read
\begin{eqnarray}
a_1(s_{13},s_{23}) &=& \frac{s_{123}-s_{13}-s_{23}}{s_{123}-s_{13}} = -\frac{1-z}{z-x} \, ,\nonumber  \\ 
a_2(s_{13},s_{23}) &=& \frac{s_{123}-s_{13}- s_{23}}{s_{123}-s_{23}} = 1-z \, , \nonumber \\ 
a_3(s_{13},s_{23}) &=& \frac{s_{123} \, s_{12} }{(s_{12}+s_{23}) (s_{13} + s_{12})} = -  \frac{x(1-z)}{z-x} \, .
\end{eqnarray}
The arguments $a_1$ and $a_3$ are equal to unity for the endpoint $x=1$. Moreover, arguments $a_2$ and $a_3$ are unity for $z=0$. 
After expressing ${\rm Box}(s_{13},s_{23})$ in terms of hypergeometric functions with arguments
\begin{eqnarray}
\tilde{a}_1(s_{13},s_{23}) &=& 1-\frac{1}{a_1(s_{13},s_{23})} = \frac{1-x}{1-z} \, , \nonumber\\ 
\tilde{a}_2(s_{13},s_{23}) &=& 1-a_2(s_{13},s_{23})= z \, , \nonumber\\ 
\tilde{a}_3(s_{13},s_{23}) &=&1-\frac{1}{a_3(s_{13},s_{23})} = \frac{z (1-x)}{x (1-z)} \, ,
\label{eq:atildes13s23}
\end{eqnarray}
none of these functions contains branch cuts in the kinematic endpoints. However, the arguments $\tilde{a}_1$ and $\tilde{a}_3$ are larger than unity for $z>x$. Therefore, the hypergeometric functions with arguments $\tilde{a}_1$ and $\tilde{a}_3$ yield a non-vanishing imaginary part in this region. To separate the imaginary part of ${\rm Box}(s_{13},s_{23})$ from the hypergeometric functions we distinguish the regions
\begin{eqnarray}
T_1 &=& \{ s_{12}, s_{23}: s_{12} + s_{23} > 0 \Leftrightarrow z > x \} \, , \nonumber \\
T_2 &=& \{ s_{12}, s_{23}: s_{12} + s_{23} < 0 \Leftrightarrow z < x \}  
\end{eqnarray}
and apply the transformations of argument $a_1$ and $a_3$ in \eqref{eq:atildes13s23} only in region $T_2$ and not in region $T_1$.

For ${\rm Box}(s_{12},s_{13})$ the arguments of the hypergeometric functions read
\begin{eqnarray}
a_1(s_{12},s_{13}) &=& \frac{s_{123}-s_{12}-s_{13}}{s_{123}-s_{12}} = \frac{1-x}{1-x-z} \, ,\nonumber \\
a_2(s_{12},s_{13}) &=& \frac{s_{123}-s_{12}-s_{13}}{s_{123}-s_{13}} = \frac{1-x}{z-x} \, , \nonumber\\
a_3(s_{12},s_{13}) &=& \frac{s_{123} \, s_{23}}{(s_{12}+s_{23}) \, (s_{13} +s_{23})} = -\frac{(1-x) \, x}{(z-x) \, (1-x-z)} \, .
\end{eqnarray}
The argument $a_1$ is equal to unity in the kinematic endpoint $z=0$. To avoid the corresponding branch cut of the hypergeometric function we map this argument to
\begin{equation}
\tilde{a}_1(s_{12},s_{13}) = 1-\frac{1}{a_1(s_{12},s_{13})} = \frac{z}{1-x} \, .
\label{eq:tildea1s12s13}
\end{equation}
The endpoint $z=0$ is mapped to $\tilde{a}_1=0$. However, we have $ \tilde{a}_1 \to + \infty$ as $x$ approaches 1. To avoid this other branch cut of the hypergeometric function we apply \eqref{eq:tildea1s12s13} only in the region $R_1$ and keep the argument $a_1$ in region $R_2$. 

The hypergeometric function with argument $a_2$ does not yield any branch cuts in the kinematic endpoints. However, the argument $a_2$ is larger than unity in region $T_1$. Therefore, we apply the following mapping in region $T_1$:
\begin{equation}
\tilde{a}_2(s_{12},s_{13}) = 1-\frac{1}{a_2(s_{12},s_{13})} = \frac{1-z}{1-x} \, .
\end{equation}
For the analytic continuation of the hypergeometric function in $a_3$ we have to distinguish the regions
\begin{eqnarray}
U_1 &=& \{s_{12}, s_{13}, s_{23}:  s_{13} + s_{23} > 0 \,  \wedge \, s_{12} + s_{23} < 0 \Leftrightarrow z < 1-x \, \wedge \, z < x \} \, ,\nonumber  \\
U_2 &=& \{s_{12}, s_{13}, s_{23}:  s_{13} + s_{23} < 0 \,  \wedge \, s_{12} + s_{23} < 0 \Leftrightarrow z > 1-x \, \wedge \, z < x \} \, ,\nonumber  \\
U_3 &=& \{s_{12}, s_{13}, s_{23}:  s_{13} + s_{23} > 0 \,  \wedge \, s_{12} + s_{23} > 0 \Leftrightarrow z < 1-x \, \wedge \, z > x \} \, , \nonumber \\
U_4 &=& \{s_{12}, s_{13}, s_{23}:  s_{13} + s_{23} < 0 \,  \wedge \, s_{12} + s_{23} > 0 \Leftrightarrow z > 1-x \, \wedge \, z > x \} \, .
\end{eqnarray}
The different regions are shown in Figure\,\ref{fig:kinematic_regions_com}. For $a_3$, we have
\begin{eqnarray}
a_3(s_{12}, s_{13} ) \geq 1 \, \,  &\text{in}& \, \, U_1 \cup U_4 \, , \\
a_3(s_{12},s_{13} ) \leq 0 \, \, &\text{in}& \, \, U_2 \cup U_3 \, .
\end{eqnarray}
Moreover, $a_3=1$ for $z=0$. Therefore, we map the argument $a_3$ of the hypergeometric function in region $U_1 \cup U_4$ to
\begin{equation}
\tilde{a}_3(s_{12},s_{13}) = 1-\frac{1}{a_3(s_{12},s_{13})} = \frac{z(1-z)}{x(1-x)} 
\end{equation}
by means of an appropriate identity for the hypergeometric function. In the region $U_2$ the hypergeometric function with argument $a_3$ does not have a branch cut in $x=1$ and no mapping of the argument is required. To obtain the analytic continuation of the third hypergeometric function in $U_3$, we take the result in region $U_1 \cup U_4$ and apply the transformation
\begin{equation}
\tilde{\tilde{a}}_3(s_{12},s_{13}) = \frac{1}{1-\tilde{a}_3(s_{12},s_{13})} = a_3(s_{12},s_{13}) \, ,
\end{equation}
to the argument of the hypergeometric function. Note that even though the arguments of the hypergeometric functions in region $U_2$ and $U_3$ are the same, the result in $U_3$ contains additional terms originating from the analytic continuation from region $U_1 \cup U_4$ to region $U_3$.

After having performed the analytic continuation of the master integrals in the different parts of the physical region the expansion in terms of distribution can safely be performed. We have checked that the expressions in the different regions are continuous at the boundaries.

 We have cast the hypergeometric functions in the box master integrals in a form that an expansion in terms of distributions in $z=0$ can be performed. The same does not hold for the endpoint $z=1$. However, at the level of the integrated fragmentation antenna functions we are able to recover any distributions in $z=1$ by exchanging particles 2 and 3 which corresponds to exchanging $z$ with $1-z$. To this end, factors of the form $1/(s_{12}s_{13})$ have to be rewritten using partial fractions.   
 
After inserting   these bubble and box master integrals in the $X_3^1$ antenna functions, the expansions of factors $z^{-1-\epsilon}$ and $(1-x)^{-1-\epsilon}$ in terms of distributions can be performed. The results in the different segments of the physical region, 
Figure~\ref{fig:kinematic_regions_com}, can then 
be recast in a form that ensures that the pole terms and the coefficients of the distributions in $z$ and $(1-x)$ take the same form in all segments. 
 
 The $z$-integration of the resulting expressions recovers
the known real-virtual initial-final master integrals~\cite{Daleo:2009yj} and enabled us to identify an error in their numerical implementation for 
jet production in deep-inelastic scattering~\cite{Currie:2017tpe}.
 
 The relevant one-loop integrated fragmentation antenna function for photon production is  $\mathcal{\tilde{A}}^{1, \, {\rm id.} \gamma}_{3,\hat{q}}(x,z)$. Its expression is very lengthy and is enclosed as ancillary file together with the expressions for the other integrated one-loop fragmentation antenna functions.

\section{Conclusions}
\label{sec:conc}

In this paper, we extended the antenna subtraction method to account for identified photons in the final state, and 
derived all required ingredients for the computation of photonic cross sections up to NNLO. This extension required to introduce 
novel fragmentation antenna functions, which are differential in the momentum fraction of the final-state photon. The unintegrated forms 
of the fragmentation antenna 
functions could  be inferred from their inclusive QCD counterparts. They come with novel forms of phase space factorisation at NLO and NNLO, allowing 
to retain the photon momentum fraction as a variable in all stages of the event reconstruction. The corresponding integrated 
fragmentation antenna functions were newly computed for all photon and parton fragmentation processes at NLO and for photon 
fragmentation up to NNLO. 

The developments in this paper allow to compute the NNLO corrections to processes involving final-state photons (also in association
with jets), with a realistic fixed-cone based isolation prescription for the photon. 
The new subtraction terms are largely separate from previously derived subtraction terms obtained for
an idealised dynamical-cone isolation, and can be added to existing NNLO implementations.  
First applications could be photon-plus-jet or di-photon production, where NNLO corrections for fixed-cone based isolation will allow 
to accurately quantify the effects of the photon isolation procedure. Moreover, it will then also become possible to compute NNLO-accurate
 cross sections
for alternative photon isolation prescriptions~\cite{Glover:1993xc,Hall:2018jub} (or even without any photon isolation) 
and to investigate observables that could allow for direct determinations 
of the photon fragmentation functions at hadron colliders~\cite{Kaufmann:2016nux}. 

The formalism derived in this paper for fragmentation antenna functions can be further generalised from photons to identified hadrons. Cross sections 
for identified hadrons are obtained by convoluting cross sections for the production of specific partons with parton-to-hadron fragmentation functions. Their description at higher orders  requires fragmentation antenna functions, differential in the momentum fraction of a final-state quark or gluon. The full set 
of these functions at NLO is already given in appendix \ref{app:X30integration}. An extension to NNLO will require the integration of all 
double real and real-virtual fragmentation 
antenna functions, each in initial-final and final-final kinematics.  In the initial-final case, no integration is required for the 
real-virtual functions and the results are 
obtained directly along the lines of the section~\ref{sec:X31int}; they are included as ancillary files. More conceptual work and 
new master integrals are needed 
for integrated fragmentation antenna functions for identified partons in the double real case, as well as in final-final kinematics.

\section*{Acknowledgements}
We would like to thank Alexander Huss and Marius H\"ofer for multiple discussions and comments that helped shaping 
and testing the formulation of the 
method that is presented in this paper. In the course of this project, we also benefitted from numerous discussions with Xuan Chen, 
Jonathan Mo and Giovanni Stagnitto, whom we would like to thank for their input.
This work has received funding from the Swiss National Science Foundation (SNF) under contract 200020-204200 and from the European Research Council (ERC) under the European Union's Horizon 2020 research and innovation programme grant agreement 101019620 (ERC Advanced Grant TOPUP). 

\begin{appendix}

\section{Mass Factorisation Kernels}
\label{app:MFkernels}
The components of the mass factorisation kernels $\mathbf{\Gamma}$ are given in~\cite{GehrmannDeRidder:1997gf}. Adopted to our notation they read
\begin{eqnarray}
\mathbf{\Gamma}^{(0)}_{q \to \gamma} &=& Q_q^2 \fpiegam^{\epsilon} \muomuf^{\epsilon} \Gamma^{(0)}_{\gamma q}(z) \, , \nonumber \\
\mathbf{\Gamma}^{(0)}_{g \to \gamma} &=&0 \, , \nonumber \\
\mathbf{\Gamma}^{(0)}_{\gamma \to \gamma} &=& \delta(1-z) \, , \nonumber \\
\mathbf{\Gamma}^{(0)}_{\gamma \to p} &=& \mathbf{\Gamma}^{(1)}_{\gamma \to p} = 0 \quad \text{for } p \in \{q, \bar{q}, g\} \,  , \nonumber \\
\mathbf{\Gamma}^{(0)}_{q \to q} &=& \delta(1-z) \, , \nonumber \\
\mathbf{\Gamma}^{(0)}_{q \to q'} &=& \mathbf{\Gamma}^{(1)}_{q \to q'} = 0 \quad \text{for } q \neq q' \, , \nonumber \\
\mathbf{\Gamma}^{(0)}_{q \to g} &=& 0 \, , \nonumber \\
\mathbf{\Gamma}^{(0)}_{g \to g} &=& \delta(1-z) \, , \nonumber \\
\mathbf{\Gamma}^{(0)}_{g \to q} &=& 0 \, , \nonumber \\
\mathbf{\Gamma}^{(1)}_{q \to \gamma} &=& \left( \frac{N^2-1}{N} \right) Q_q^2 \fpiegam^{2\epsilon} \muomuf^{2\epsilon} \Gamma^{(1)}_{\gamma q}(z) \, , \nonumber \\
\mathbf{\Gamma}^{(1)}_{g \to \gamma} &=& \fpiegam^{2\epsilon} \muomuf^{2\epsilon} \Gamma^{(1)}_{\gamma g}(z) \, , \nonumber \\
\mathbf{\Gamma}^{(1)}_{\gamma \to \gamma} &=& 0 \, , \nonumber \\
\mathbf{\Gamma}^{(1)}_{q \to q} &=& \left( \frac{N^2-1}{N} \right) \fpiegam^{\epsilon} \muomuf^{\epsilon} \Gamma^{(1)}_{qq}(z) \, , \nonumber \\
\mathbf{\Gamma}^{(1)}_{g \to g} &=& \fpiegam^{\epsilon} \muomuf^{\epsilon} \left( N \, \Gamma^{(1)}_{gg}(z) + N_f \, \Gamma^{(1)}_{gg,F}(z) \right) \, , \nonumber \\
\mathbf{\Gamma}^{(1)}_{g \to q} &=& \fpiegam^{\epsilon} \muomuf^{\epsilon} \Gamma^{(1)}_{qg}(z) \, , \nonumber \\
\mathbf{\Gamma}^{(1)}_{q \to g} &=& \left( \frac{N^2-1}{N} \right) \fpiegam^{\epsilon} \muomuf^{\epsilon} \Gamma^{(1)}_{gq}(z) \, . 
\label{eq:allkernels}
\end{eqnarray}
Since we set $D_{g \to \gamma}=\mathcal{O}(\alpha)$, the mass factorisation kernels $\mathbf{\Gamma}^{(1)}_{q\to g}$ and $\mathbf{\Gamma}^{(1)}_{g\to g}$ are non-zero. Moreover, we decomposed the kernels by factors $N$ and $N_f$.
The factorisation kernels can be expressed in terms of leading order and next-to-leading order splitting functions, i.e.\
\begin{eqnarray}
\Gamma^{(0)}_{\gamma q}(z) &=& - \frac{1}{\epsilon} p^{(0)}_{\gamma q}(z) \, , \nonumber \\
\Gamma^{(1)}_{\gamma q}(z) &=& \frac{1}{2}  \left[ \frac{1}{2 \epsilon^2} (p^{(0)}_{q q} \otimes p^{(0)}_{\gamma q })(z) - \frac{1}{2\epsilon} p^{(1)}_{\gamma q}(z) \right] \, , \nonumber \\
\Gamma^{(1)}_{\gamma g}(z) &=& \frac{1}{2} \sum_{q} Q_q^2 \left( \frac{1}{2\epsilon^2} (p^{(0)}_{q g} \otimes p^{(0)}_{\gamma q})(z) - \frac{1}{2 \epsilon} p^{(1)}_{\gamma g}(z) \right) \, , \nonumber \\
\Gamma^{(1)}_{qq}(z) &=&  - \frac{1}{2\epsilon} p^{(0)}_{q q}(z) \,  , \nonumber \\
\Gamma^{(1)}_{q g}(z) &=&  - \frac{1}{2\epsilon} p^{(0)}_{q g}(z) \, , \nonumber \\
\Gamma^{(1)}_{g q}(z) &=& - \frac{1}{2\epsilon} p^{(0)}_{gq}(z) \,  , \nonumber \\
\Gamma^{(1)}_{g g,F}(z) &=& - \frac{1}{\epsilon} p^{(0)}_{gg,F}(z) \, , \nonumber \\
\Gamma^{(1)}_{g g}(z) &=& - \frac{1}{\epsilon} p^{(0)}_{gg}(z) \, . 
\label{eq:gamnonvanishpco}
\end{eqnarray}
The factors of $1/2$ appearing in \eqref{eq:gamnonvanishpco} originate from decomposing the colour factors $C_F=(N^2-1)/(2N)$ and $T_R=1/2$ in~\cite{GehrmannDeRidder:1997gf}. 

The lowest order splitting functions are given by 
\begin{eqnarray}
p^{(0)}_{qq}(z) &=& \frac{3}{2} \delta(1-z) + 2 \mathcal{D}_0(z) -1 -z \, , \nonumber \\
p^{(0)}_{qg}(z) &=& 1- 2z +2 z^2 \, , \nonumber \\
p^{(0)}_{gq}(z) &=& \frac{2}{z} - 2 + z \, , \nonumber  \\
p^{(0)}_{gg}(z) &=& \frac{11}{6} \delta(1-z) + 2 \mathcal{D}_0(z) + \frac{2}{z} - 4 + 2z - 2z^2 \, , \nonumber  \\
p^{(0)}_{gg,F}(z) &=& -\frac{1}{3} \delta(1-z) \, , \nonumber  \\
p^{(0)}_{\gamma q}(z) &=& \frac{2}{z} - 2 + z \, ,
\label{eq:LOsplittingfunc}
\end{eqnarray}
and the next-to-leading quark-to-photon and gluon-to-photon splitting functions read
\begin{eqnarray}
p^{(1)}_{\gamma q}(z) &=& -\frac{1}{2} + \frac{9}{2} z + \left( -8 + \frac{1}{2} z \right) \log z + 2 z \log (1-z) + \left(1 - \frac{1}{2}z\right) \log^2 z \nonumber \\
&&+ \left[\log^2(1-z) + 4 \log z \log (1-z) + 8 \text{Li}_2(1-z) - \frac{4}{3} z \right] p^{(0)}_{\gamma q}(z) \, , \nonumber \\
p^{(1)}_{\gamma g}(z) &=& -2 + 6z - \frac{82}{9} z^2 + \frac{46}{9z} + \left( 5 + 7z + \frac{8}{3} z^2 + \frac{8}{3z} \right) \log z \nonumber \\
&&+ (1+z) \log^2 z \, .
\end{eqnarray}

\section{Integrated $X^0_3$ Fragmentation Antenna Functions}
\label{app:X30integration}
We express the integrated fragmentation antenna functions in terms of splitting functions \eqref{eq:LOsplittingfunc} and colour-ordered infrared singularity operators, which read
\begin{eqnarray}
\mathbf{I}^{(1)}_{q\bar{q}}(\epsilon,s_{q\bar{q}}) &=& - \frac{e^{\epsilon \gamma_E}}{2\Gamma(1-\epsilon)} \left[ \frac{1}{\epsilon^2} + \frac{3}{2\epsilon} \right] {\mathcal R}(-s_{q\bar{q}})^{-\epsilon} \, , \nonumber \\
\mathbf{I}^{(1)}_{qg}(\epsilon,s_{qg}) &=& - \frac{e^{\epsilon \gamma_E}}{2\Gamma(1-\epsilon)} \left[ \frac{1}{\epsilon^2} + \frac{5}{3\epsilon} \right] {\mathcal R}(-s_{qg})^{-\epsilon} \, , \nonumber \\
\mathbf{I}^{(1)}_{gg}(\epsilon,s_{gg}) &=& - \frac{e^{\epsilon \gamma_E}}{2\Gamma(1-\epsilon)} \left[ \frac{1}{\epsilon^2} + \frac{11}{6\epsilon} \right] {\mathcal R}(-s_{gg})^{-\epsilon} \, , \nonumber \\
\mathbf{I}^{(1)}_{q\bar{q},F}(\epsilon,s_{q\bar{q}})  &=& 0 \, , \nonumber \\
\mathbf{I}^{(1)}_{qg,F}(\epsilon,s_{qg}) &=& \frac{e^{\epsilon \gamma_E}}{2\Gamma(1-\epsilon)}  \frac{1}{6\epsilon}  {\mathcal R}(-s_{qg})^{-\epsilon} \, , \nonumber \\
\mathbf{I}^{(1)}_{gg,F}(\epsilon,s_{gg}) &=& \frac{e^{\epsilon \gamma_E}}{2\Gamma(1-\epsilon)}  \frac{1}{3\epsilon}  {\mathcal R}(-s_{gg})^{-\epsilon} \, . 
\end{eqnarray}
The invariant masses that appear in these pole terms and in the normalisation factors of the integrated antenna functions are 
always constructed from three-parton invariants as 
$q^2=s_{12}+s_{13}+s_{23}$ and $Q^2=-q^2$.

\subsection{Initial-Final Configuration}
The unintegrated $X^0_3$ antenna functions in the initial-final configuration were introduced in~\cite{Daleo:2006xa}. We recall their expressions here and give the results for their integrated form differential in the final-state momentum fraction. 

The quark-initiated quark-quark antenna function in the initial-final configuration reads
\begin{equation}
A^0_3(\hat{1}_q,3_g,2_q) =\frac{1}{s_{123}} \left(\frac{2 s_{12}^2}{s_{13} s_{23}}+\frac{2
   s_{12}}{s_{13}}+\frac{2
   s_{12}}{s_{23}}+\frac{s_{23}}{s_{13}}+\frac{s_{13}}{s_{23}} \right) + \mathcal{O}(\epsilon) \, .
\label{eq:A30IFunint}
\end{equation}
For the integration of the fragmentation antenna function we need to specify which parton in the final state is identified. In case the final-state gluon is identified we find
\begin{eqnarray}
\mathcal{A}^{0, {\rm id.} g}_{3, \hat{q}}(x,z)&=& \left(Q^2\right)^{-\epsilon} \bigg[ -\frac{1}{2\epsilon}\delta(1-x) p^{(0)}_{gq}(z)  + \frac{1}{2}-\frac{x}{2}+\frac{z}{4}+\frac{x z}{4}+\frac{1}{2} z \delta(1-x) \nonumber \\
&&+\left(-\frac{1}{4}-\frac{x}{4}+\frac{1}{2} \mathcal{D}_0(x)+\frac{1}{2} \delta(1-x) \left( \log (1-z)+\log(z)\right)\right)
p^{(0)}_{gq}(z) \bigg] + \mathcal{O}(\epsilon) \, , \nonumber \\
\end{eqnarray}
and for the case of an identified final-state quark we have
\begin{eqnarray}
\mathcal{A}^{0, {\rm id.} q}_{3, \hat{q}}(x,z) &=& -2 \mathbf{I}_{q\bar{q}}^{(1)}(\epsilon,-Q^2) \delta(1-z)  \delta(1-x) \nonumber \\
&&+\left(Q^2\right)^{-\epsilon} \bigg[    -\frac{1}{2\epsilon} \left(\delta(1-z) p^{(0)}_{qq}(x)+\delta(1-x)
p^{(0)}_{qq}(z) \right) +\frac{9}{16}\delta(1-z) \delta(1-x) \nonumber \\
&&+\delta(1-z) \left(\frac{1}{2}-\frac{x}{2}+\mathcal{D}_1(x)-\frac{1}{2}(1+x) \log (1-x)-\frac{1+x^2}{2(1-x)} \log(x)\right) \nonumber \\
&&+\delta(1-x) \left(\frac{1}{2}-\frac{z}{2} +\mathcal{D}_1(z)-\frac{1}{2}(1+z) \log(1-z)+ \frac{1+z^2}{2(1-z)} \log(z) \right) \nonumber \\
&&- \frac{3}{8} \left( \delta(1-z) p^{(0)}_{qq}(x) +
\delta(1-x) p^{(0)}_{qq}(z) \right) + \frac{1}{4} p^{(0)}_{qq}(x)p^{(0)}_{qq}(z)  \nonumber \\
&&+ \frac{3}{4}-\frac{x}{4}-\frac{z}{4}-\frac{x z}{4} \bigg] +\mathcal{O}(\epsilon) \, .
\label{eq:A30IFqq}
\end{eqnarray}
In the subtraction of quark-photon collinear limits the antenna function $A^0_3(\hat{1}_q,3_{\gamma},2_q)$ is used. Its unintegrated form coincides with \eqref{eq:A30IFunint} and we have $\mathcal{A}^{0, {\rm id.} \gamma}_{3, \hat{q}} = \mathcal{A}^{0, {\rm id.} g}_{3, \hat{q}}$.

The unintegrated $D$-type quark-initiated quark-gluon antenna function is
\begin{eqnarray}
D^0_3(\hat{1}_q,2_g,3_g) &=& \frac{1}{s_{123}^2} \bigg(\frac{2 s_{123}^2 s_{13}}{s_{12} s_{23}}+\frac{2 s_{12}
   s_{123}^2}{s_{13} s_{23}}+\frac{s_{123}
   s_{23}}{s_{12}}+\frac{2 s_{12}
   s_{13}}{s_{23}}  \nonumber \\
&&+ s_{12}+\frac{s_{123} s_{23}}{s_{13}}+4
   s_{123}+s_{13} \bigg) + \mathcal{O}
(\epsilon)\, .
\end{eqnarray}
It is symmetric under the exchange of gluons 2 and 3. Therefore, there is only one corresponding integrated fragmentation antenna function, i.e.\
\begin{eqnarray}
\mathcal{D}^{0, {\rm id.} g}_{3, \hat{q}}(x,z)&=& -2 \mathbf{I}_{qg}^{(1)}(\epsilon,-Q^2) \delta(1-x)\delta(1-z) \nonumber \\
&&+ \left(Q^2\right)^{-\epsilon} \bigg[ -\frac{1}{2\epsilon} \left(\delta(1-x) p^{(0)}_{gg}(z)+\delta(1-z) p^{(0)}_{qq}(x) \right) + \frac{11}{16}\delta(1-x)\delta(1-z) \nonumber \\
&&+\delta(1-z) \left(\frac{1}{2}-\frac{x}{2}+\mathcal{D}_1(x)-\frac{1}{2} (1+x) \log (1-x)- \frac{1+x^2}{2(1-x)} \log(x)\right) \nonumber \\
&&+\delta(1-x) \left(\mathcal{D}_1(z) +\left(-2 + \frac{1}{z} +z - z^2\right) \log (1-z) + \frac{(1-z+z^2)^2}{(1-z)z} \log(z) \right) \nonumber \\
&&-\frac{3}{8} \delta(1-x) p^{(0)}_{gg}(z)-\frac{11}{24}  p^{(0)}_{qq}(x)  \delta(1-z)+ \frac{1}{4} p^{(0)}_{gg}(z) p^{(0)}_{qq}(x) -1-\frac{1}{2 x}-x+\frac{z}{2} \nonumber \\
&&+\frac{z}{x}+\frac{x z}{2}-\frac{z^2}{2}-\frac{z^2}{x}-\frac{x z^2}{2} \bigg] + \mathcal{O}(\epsilon) \, .
\end{eqnarray}
The three-quark quark-gluon antenna functions have the form
\begin{eqnarray}
E^0_3(\hat{1}_q,2_{q'},3_{\bar{q}'}) &=& \frac{1}{s_{123}^2} \left( \frac{(s_{12}+s_{13})^2}{s_{23}}-\frac{2 s_{12}s_{13}}{s_{23}}+ (s_{12}+s_{13}) \right) + \mathcal{O}(\epsilon) \, , \\ 
E^0_3(\hat{1}_{q'},2_{q'},3_{q}) &=& -\frac{1}{s_{123}^2} \left(\frac{(s_{13}+s_{23})^2}{s_{12}}-\frac{2 s_{13}
   s_{23}}{s_{12}}+ (s_{13}+s_{23}) \right) \, + \mathcal{O}(\epsilon) \label{eq:E30qpqpq} \, . 
\end{eqnarray}
The first antenna function is symmetric in the final-state quark pair. There is one corresponding integrated fragmentation antenna function, i.e.\
\begin{eqnarray}
\mathcal{E}^{0, {\rm id.} q'}_{3, \hat{q}}(x,z) &=&  \left(Q^2\right)^{-\epsilon} \bigg[ -\frac{1}{2\epsilon} \delta(1-x) p^{(0)}_{qg}(z)  -\frac{1}{2 x}+\frac{p^{(0)}_{qg}(z)}{2 x} +\frac{1}{2}
\mathcal{D}_0(x) p^{(0)}_{qg}(z) \nonumber \\
&&+\delta(1-x) \left(\frac{1}{2}-\frac{1}{2} p^{(0)}_{qg}(z)+\frac{1}{2} \left( \log (1-z) +\log (z)\right) p^{(0)}_{qg}(z)\right) \bigg] + \mathcal{O}(\epsilon) \, . \nonumber \\
\label{eq:E30IFqqp}
\end{eqnarray}
The only unresolved limit of the  antenna function in \eqref{eq:E30qpqpq} is the flavour-changing initial-final collinear limit. However, identifying the final-state parton $q'$ prevents it from becoming collinear to the initial-state since any jet function will require a minimum transverse momentum of the identified particle. Therefore, the only integrated fragmentation antenna function corresponding to \eqref{eq:E30qpqpq} identifies the final-state quark $q$. We find
\begin{eqnarray}
\mathcal{E}^{0, {\rm id.} q}_{3, \hat{q}'}(x,z) &=& \left(Q^2\right)^{-\epsilon} \bigg[ -\frac{1}{2\epsilon} \delta(1-z) p^{(0)}_{gq}(x)- \frac{1}{2}+\frac{x}{2}+\frac{1}{2} x \delta(1-z) \nonumber \\
&&-\left(\frac{1}{2}-\frac{1}{2} \mathcal{D}_0(z) +\delta(1-z) \left(\frac{1}{2}-\frac{1}{2} \log (1-x)+\frac{\log (x)}{2}\right)\right)
p^{(0)}_{gq}(x) \bigg] + \mathcal{O}(\epsilon) \, . \nonumber \\
\end{eqnarray}
The remaining quark-initiated antenna function is
\begin{equation}
G^0_3(\hat{1}_{q'},2_{q'},3_{g}) = -\frac{1}{s_{123}^2} \left( \frac{(s_{13}+s_{23})^2}{s_{12}}-\frac{2 s_{13}
  s_{23}}{s_{12}} \right) + \mathcal{O}(\epsilon) \, .
\end{equation}
The $G^0_3$ antenna function at hand only contains the flavour-changing initial-final limit. Using the same reasoning as for the $E^0_3$ antenna function, we find only one integrated fragmentation antenna function:
\begin{eqnarray}
\mathcal{G}^{0, {\rm id.} g}_{3, \hat{q}'}(x,z) &=& \left(Q^2\right)^{-\epsilon} \bigg[- \frac{1}{2\epsilon} \delta(1-z) p^{(0)}_{gq}(x) -  \frac{1}{2}+\frac{x}{4}-\frac{z}{2}+\frac{x z}{4}+\frac{1}{2} x \delta(1-z) \nonumber \\
&&-\left(\frac{1}{4}+\frac{z}{4}-\frac{1}{2} \mathcal{D}_0(z)+\delta(1-z) \left(\frac{1}{2}-\frac{1}{2} \log (1-x)+\frac{\log (x)}{2}\right)\right)
p^{(0)}_{gq}(x) \bigg] \nonumber \\
&&+ \mathcal{O}(\epsilon) \, .
\end{eqnarray}

The gluon-initiated quark-anti-quark antenna function is given by
\begin{equation}
A^0_3(2_{\bar{q}},\hat{1}_g,3_q) = -\frac{1}{s_{123}} \left(\frac{2 s_{23}^2}{s_{12} s_{13}}+\frac{s_{13}}{s_{12}}+\frac{s_{12}}{s_{13}}+\frac{2 s_{23}}{s_{12}}+\frac{2s_{23}}{s_{13}} \right) + \mathcal{O}(\epsilon)
\end{equation}
and its integrated form with an identified quark reads
\begin{eqnarray}
\mathcal{A}^{0, {\rm id.} \, q}_{3, \hat{g}}(x,z) &=& \left(Q^2\right)^{-\epsilon} \bigg[- \frac{1}{2\epsilon}\delta(1-z) p^{(0)}_{qg}(x) - 1 + \frac{p^{(0)}_{qg}(x)}{2 z}  + \frac{1}{2} \mathcal{D}_0(z) p^{(0)}_{qg}(x) \nonumber  \\
&&- \frac{1}{2}  \delta(1-z) \left( -1 +p^{(0)}_{qg}(x) ( \log(x) - \log(1-x) ) \right) \bigg] + \mathcal{O}(\epsilon)\, .
\end{eqnarray}
As explained in~\cite{Daleo:2006xa}, the gluon-initiated $D^0_3$ antenna function has to be decomposed into a flavour-preserving and flavour-changing piece. The two resulting antenna functions are
\begin{eqnarray}
D^0_3(\hat{1}_g,2_g,3_q) &=& \frac{1}{s_{123}^2} \left( \frac{s_{12}^2}{s_{23}}+\frac{2 s_{13}^3}{s_{12} s_{23}}+\frac{4
   s_{13}^2}{s_{12}}+\frac{2 s_{23}^3}{s_{12}
   (s_{12}+s_{13})} +\frac{6 s_{13} s_{23}}{s_{12}} \right. \nonumber \\
   &&\left.+\frac{3
s_{12} s_{13}}{s_{23}}+\frac{4 s_{23}^2}{s_{12}}+6
s_{12}+\frac{4 s_{13}^2}{s_{23}}+9 s_{13}+9 s_{23}\right) + \mathcal{O}(\epsilon) \ , \\
D^0_{3, g \to q}(\hat{1}_g, 2_{q}, 3_g) &=& -\frac{1}{s_{123}^2} \left(  \frac{s_{13}^2}{s_{12}}+\frac{2 s_{23}^3}{s_{12}
   (s_{12}+s_{13})}+\frac{3 s_{13} s_{23}}{s_{12}}+\frac{4 s_{23}^2}{s_{12}} \right) + \mathcal{O}(\epsilon) \, .
\end{eqnarray}
For the flavour-preserving antenna function the quark or the gluon in the final state can be identified. In the former case we find
\begin{eqnarray}
\mathcal{D}^{0, {\rm id.} q}_{3, \hat{g}}(x,z) &=&-2 \mathbf{I}_{qg}^{(1)}(\epsilon,-Q^2)\delta(1-x) \delta(1-z) \nonumber \\
&&+ \left(Q^2\right)^{-\epsilon} \bigg[ -\frac{1}{2\epsilon} \left( \delta(1-z) p^{(0)}_{gg}(x) + \delta(1-x) p^{(0)}_{qq}(z) \right) + \frac{11}{16} \delta(1-x) \delta(1-z) \nonumber \\
&&+\delta(1-z) \left(\mathcal{D}_1(x) + \left(-2 + \frac{1}{x} +x -x^2\right) \log(1-x) - \frac{(1-x+x^2)^2}{(1-x)x} \log(x)  \right) \nonumber \\
&&+\delta(1-x) \left(\frac{1}{2}-\frac{z}{2}+\mathcal{D}_1(z)-\frac{1}{2} (1+z) \log (1-z) + \frac{1+z^2}{2(1-z)} \log(z) \right) \nonumber \\
&&-\frac{3}{8} \delta(1-z) p^{(0)}_{gg}(x) -\frac{11}{24} p^{(0)}_{qq}(z) \delta(1-x)+\frac{1}{4} p^{(0)}_{gg}(x) p^{(0)}_{qq}(z) -\frac{5}{2}+\frac{1}{2 x}+\frac{x}{2} \nonumber \\
&&-\frac{x^2}{2} -z+\frac{z}{2 x}+\frac{x z}{2}-\frac{x^2 z}{2}  \bigg] + \mathcal{O}(\epsilon) 
\label{eq:D30IFgq}
\end{eqnarray}
and in the latter case
\begin{eqnarray}
\mathcal{D}^{0, {\rm id.} g}_{3, \hat{g}}(x,z) &=& \left(Q^2\right)^{-\epsilon} \bigg[ -\frac{1}{2\epsilon}\delta(1-x) p^{(0)}_{gq}(z) -\frac{7}{2}+\frac{1}{x}+x-x^2+z-\frac{z}{2 x}-\frac{x z}{2}+\frac{x^2 z}{2} \nonumber \\
&&+\left(\frac{1-2x+x^2-x^3}{2x} +\frac{1}{2} \mathcal{D}_0(x)+\frac{1}{2} \delta(1-x) \left(
\log (1-z)+\log(z)\right)\right) p^{(0)}_{gq}(z) \nonumber \\
&&+\frac{1}{2} z \delta(1-x) \bigg] + \mathcal{O}(\epsilon) \, .
\end{eqnarray}
The only integrated fragmentation antenna function of the flavour-changing $D^0_3$ antenna identifies the final-state gluon. It reads
\begin{eqnarray}
\mathcal{D}^{0, {\rm id.} g}_{3,g \to q, \hat{g}}(x,z)&=&  \left(Q^2\right)^{-\epsilon} \bigg[ -\frac{1}{2\epsilon} \delta(1-z) p^{(0)}_{qg}(x)-  \frac{3}{2}+\frac{1}{x}-\frac{z}{2 x}+\frac{1}{2} \mathcal{D}_0(z) p^{(0)}_{qg}(x) \nonumber \\
&&-\delta(1-z) \left(-\frac{1}{2}-\frac{1}{2} \log (1-x) p^{(0)}_{qg}(x)+\frac{1}{2} \log (x) p^{(0)}_{qg}(x)\right) \bigg] + \mathcal{O}(\epsilon) \, . \nonumber \\
\end{eqnarray}

There are two gluon-initiated gluon-gluon antenna functions:
\begin{eqnarray}
F^0_3(\hat{1}_{g},2_g,3_g) &=& \frac{1}{s_{123}^2} \left( \frac{2 s_{123}^2s_{23}}{s_{12} s_{13}}+\frac{2 s_{123}^2
s_{13}}{s_{12} s_{23}}+\frac{2 s_{12} s_{123}^2}{s_{13}
s_{23}} \right.  \nonumber \\
&&\left.+\frac{2 s_{13} s_{23}}{s_{12}}+\frac{2 s_{12}
s_{23}}{s_{13}}+\frac{2 s_{12} s_{13}}{s_{23}}+8 s_{123}\right) + \mathcal{O}(\epsilon) \, , \\
G^0_3(\hat{1}_g, 2_{q'} , 3_{\bar{q}'}) &=& \frac{1}{s_{123}^2} \left( \frac{(s_{12}+s_{13})^2}{s_{23}}-\frac{2 s_{12} s_{13}}{s_{23}} \right) + \mathcal{O}(\epsilon) \, .
\end{eqnarray}
Both antenna functions are symmetric under the exchange of parton 2 and 3. Therefore, for each antenna function there is only one integrated fragmentation antenna function. We find
\newpage
\begin{eqnarray}
\mathcal{F}^{0, {\rm id.} g}_{3, \hat{g}}(x,z)&=&-2 \mathbf{I}_{gg}^{(1)}(\epsilon,-Q^2)\delta(1-x) \delta(1-z) \nonumber \\
&&+ \left(Q^2\right)^{-\epsilon} \bigg[  -\frac{1}{2\epsilon} \left(\delta(1-z) p^{(0)}_{gg}(x)+ \delta(1-x) p^{(0)}_{gg}(z) \right) + \frac{121}{144}\delta(1-z)\delta(1-x) \nonumber \\
&&+\delta(1-z) \left(\mathcal{D}_1(x)+ \left(-2 + \frac{1}{x} +x -x^2 \right) \log(1-x) - \frac{(1-x+x^2)^2}{(1-x)x} \log(x)\right) \nonumber \\
&&+\delta(1-x)
\left(\mathcal{D}_1(z) + \left(-2 + \frac{1}{z} +z - z^2 \right) \log(1-z) + \frac{(1-z+z^2)^2}{(1-z)z} \log(z) \right) \nonumber \\
&&-\frac{11}{24} \left( \delta(1-z) p^{(0)}_{gg}(x)+\delta(1-x) p^{(0)}_{gg}(z) \right) +\frac{1}{4} p^{(0)}_{gg}(x)p^{(0)}_{gg}(z) \nonumber \\
&&-(2-x+x^2)(2-z+z^2) \bigg] + \mathcal{O}(\epsilon)
\end{eqnarray}
and 
\begin{eqnarray}
\mathcal{G}^{0, {\rm id.} q'}_{3, \hat{g}}(x,z) &=& \left(Q^2\right)^{-\epsilon} \bigg[ -\frac{1}{2\epsilon} \delta(1-x) p^{(0)}_{qg}(z)+ \frac{1}{2 x} p^{(0)}_{qg}(z) +\frac{1}{2} \mathcal{D}_0(x)
p^{(0)}_{qg}(z) \nonumber \\
&&+\delta(1-x) \left(\frac{1}{2}-\frac{1}{2} p^{(0)}_{qg}(z)+\frac{1}{2} \left( \log (1-z) + \log (z) \right) p^{(0)}_{qg}(z)\right) \bigg] + \mathcal{O}(\epsilon) \, . \nonumber \\
\label{eq:G30IFgqp}
\end{eqnarray}

\subsection{Final-Final Configuration}
The unintegrated $X^0_3$ antenna functions in the final-final configuration can be found in~\cite{GehrmannDeRidder:2005cm}. We recall their expressions here and give the results for their integrated form differential in the final-state momentum fraction. 

The tree-level three parton quark-anti-quark antenna function reads
\begin{equation}
A^0_3(1_{\bar{q}},3_g,2_q) =\frac{1}{s_{123}} \left(\frac{2 s_{12}^2}{s_{13} s_{23}}+\frac{2
   s_{12}}{s_{13}}+\frac{2
   s_{12}}{s_{23}}+\frac{s_{23}}{s_{13}}+\frac{s_{13}}{s_{23}} \right) + \mathcal{O}(\epsilon) \, .
\label{eq:A30FFunint}
\end{equation}
It is symmetric under the exchange of the quark pair. We find two integrated fragmentation antenna functions. Identifying the gluon, we have
\begin{eqnarray}
\mathcal{A}^{0, {\rm id.} g}_{3,\bar{q}}(z)&=& \left(q^2\right)^{-\epsilon} \bigg[ -\frac{1}{2 \epsilon } p^{(0)}_{gq}(z) +\frac{1}{4}+\frac{z}{8}+\left(-\frac{3}{8}+\frac{1}{2} \log (1-z)+\frac{\log (z)}{2}\right) p^{(0)}_{gq}(z) \bigg] \nonumber \\
&&+\mathcal{O}(\epsilon)
\end{eqnarray}
and in case the quark is identified, we find
\begin{eqnarray}
\mathcal{A}^{0, {\rm id.} q}_{3,\bar{q}}(z)&=& -2 \mathbf{I}_{q\bar{q}}^{(1)}(\epsilon ,q^2) \delta(1-z)+ \left(q^2 \right)^{-\epsilon} \bigg[ -\frac{1}{2\epsilon } p^{(0)}_{qq}(z) + \frac{3}{8}-\frac{z}{8}+\left(\frac{47}{16}+\frac{\pi ^2}{6}\right) \delta(1-z) \nonumber \\
&&+\mathcal{D}_1(z)-\frac{1}{2} (1+z) \log (1-z)+\frac{1+z^2}{2(1-z)} \log(z) -\frac{3}{8} p^{(0)}_{qq}(z) \bigg] + \mathcal{O}(\epsilon) \, .
\label{eq:A30FFqbq}
\end{eqnarray}
In the subtraction of quark-photon collinear limits the antenna function $A^0_3(1_{\bar{q}},3_{\gamma},2_q)$ is used. Its unintegrated form coincides with \eqref{eq:A30FFunint} and we have $\mathcal{A}^{0, {\rm id.} \gamma}_{3,\bar{q}} = \mathcal{A}^{0, {\rm id.} g}_{3,\bar{q}}$.

The tree-level quark-gluon antenna function can be expressed as
\begin{equation}
D^0_3(1_q, 2_g , 3_g) = d^0_3(1_q, 2_g , 3_g) + d^0_3(1_q, 3_g, 2_g)
\label{eq:D30FF}
\end{equation}
with the sub-antenna
\begin{equation}
d^0_3(1_q, 2_g, 3_g) = \frac{1}{s_{123}^2} \left( \frac{2 s_{123}^2 s_{13}}{s_{12} s_{23}}+\frac{s_{123} s_{23}}{s_{12}}+\frac{s_{12} s_{13}}{s_{23}}+\frac{s_{12}}{2}+2 s_{123}+\frac{s_{13}}{2} \right) + \mathcal{O}(\epsilon) \, .
\label{eq:d30FF}
\end{equation} 
In the sub-antenna at hand gluon 3 acts as a hard radiator while the full antenna \eqref{eq:D30FF} also contains the soft limit of gluon 3. The reference particle used in the definition of the momentum fraction has to be a hard radiator. Therefore, if we want to use the quark-gluon antenna function with the gluon as the reference particle we have to use the sub-antenna in which the reference gluon is a hard radiator. Integrating \eqref{eq:d30FF} and remaining differential in the gluon momentum fraction, we find
\begin{equation}
\mathcal{D}^{0, {\rm id.} g}_{3,g}(z)=\left(q^2\right)^{-\epsilon} \bigg[ -\frac{1}{2 \epsilon } p^{(0)}_{gq}(z) + \frac{5}{8}+\frac{z}{8}+\left(-\frac{11}{24}+\frac{1}{2} \log (1-z)+\frac{\log (z)}{2}\right) p^{(0)}_{gq}(z) \bigg] + \mathcal{O}(\epsilon)
\end{equation}
and for the case where the quark momentum is identified we have
\begin{eqnarray}
\mathcal{D}^{0, {\rm id.} q}_{3,g}(z) &=&-2 \mathbf{I}_{qg}^{(1)}(\epsilon,q^2) \delta(1-z)+ \left(q^2\right)^{-\epsilon} \bigg[ -\frac{1}{2\epsilon} p^{(0)}_{qq}(z) + \frac{3}{4}-\frac{z}{8}+\left(\frac{167}{48}+\frac{\pi ^2}{6}\right) \delta(1-z) \nonumber \\
&&+\mathcal{D}_1(z)-\frac{1}{2} (1+z)\log (1-z) + \frac{1+z^2}{2(1-z)} \log(z) -\frac{11}{24} p^{(0)}_{qq}(z) \bigg] + \mathcal{O}(\epsilon) \, .
\label{eq:D30FFgq}
\end{eqnarray}
When the quark acts as a reference particle we can integrate the full antenna function \eqref{eq:D30FF}. In this case the fragmentation antenna function reads
\begin{eqnarray}
\mathcal{D}^{0, {\rm id.} g}_{3,q}(z)&=&-2 \mathbf{I}_{qg}^{(1)}(\epsilon,q^2) \delta(1-z) +\left(q^2\right)^{-\epsilon} \bigg[ -\frac{1}{2 \epsilon} p^{(0)}_{gg}(z) + \frac{5}{3}-\frac{13 z}{12}+\frac{13 z^2}{12} \nonumber \\
&&+\left(\frac{49}{16}+\frac{\pi ^2}{6}\right) \delta(1-z)  +\mathcal{D}_1(z) + \left(-2 +\frac{1}{z} + z - z^2\right) \log (1-z) \nonumber \\
&&+\frac{(1-z+z^2)^2}{(1-z)z} \log(z) -\frac{3 }{8} p^{(0)}_{gg}(z) \bigg] + \mathcal{O}(\epsilon) \, .
\end{eqnarray}
The last quark-gluon antenna function is 
\begin{equation}
E^0_3(1_q,2_{q'},3_{\bar{q}'}) = \frac{1}{s_{123}^2} \left( \frac{(s_{12}+s_{13})^2}{s_{23}}-\frac{2 s_{12}s_{13}}{s_{23}}+ (s_{12}+s_{13}) \right) + \mathcal{O}(\epsilon) \, .
\end{equation}
It is symmetric under the exchange of particle 2 and 3. 
Phase space integration with reference particle $q$ and identified particle $q'$ yields 
\begin{equation}
\mathcal{E}^{0, {\rm id.} q'}_{3,q}(z)=  \left(q^2\right)^{-\epsilon} \bigg[ -\frac{1}{2 \epsilon } p^{(0)}_{qg}(z)  + \frac{2}{3} + \left( -\frac{17}{12}+\frac{1}{2} ( \log (1-z) +\log (z)) \right) p^{(0)}_{qg}(z) \bigg] + \mathcal{O}(\epsilon) \, .
\label{eq:E30FFqqp}
\end{equation}
In case the primary quark $q$ is identified, we have
\begin{equation}
\mathcal{E}^{0, {\rm id.} q}_{3,\bar{q}'}(z)= -4 \mathbf{I}_{qg,F}^{(1)}(\epsilon,q^2) \delta(1-z) +\left(q^2\right)^{-\epsilon} \bigg[   -\frac{1}{12}-\frac{11}{12} \delta(1-z)+\frac{1}{3} \mathcal{D}_0(z) \bigg] + \mathcal{O}(\epsilon) \, .
\end{equation}

The first gluon-gluon antenna function is 
\begin{equation}
F^0_3(1_g, 2_g , 3_g) = f^0_3(1_g , 2_g , 3_g) + f^0_3(1_g , 3_g , 2_g) 
\label{eq:F30FF}
\end{equation}
with the sub-antenna 
\begin{equation}
f^0_3(1_g, 2_g , 3_g) = \frac{1}{s_{123}^2} \left( \frac{2 s_{123}^2 s_{13}}{s_{12} s_{23}}+\frac{s_{13} s_{23}}{s_{12}}+\frac{s_{12} s_{13}}{s_{23}}+\frac{8 s_{123}}{3} \right) + \mathcal{O}(\epsilon) \, .
\end{equation}
In \eqref{eq:F30FF} we have fixed $1_g$ to be the hard radiator which is used as a reference particle in the definition of the momentum fraction. Consequently, it does not contain the sub-antenna $f^0_3(2_g, 1_g , 3_g)$. Integration of \eqref{eq:F30FF} over the phase space while remaining differential in the final-state momentum fraction yields
\begin{eqnarray}
\mathcal{F}^{0, {\rm id.} g}_{3,g}(z)&=&-2 \mathbf{I}_{gg}^{(1)}(\epsilon,q^2) \delta(1-z)+ \left(q^2\right)^{-\epsilon} \bigg[    -\frac{1}{2 \epsilon } p^{(0)}_{gg}(z) + \frac{4}{3}-\frac{11 z}{12}+\frac{11 z^2}{12} \nonumber \\
&&+\left(\frac{523}{144}+\frac{\pi ^2}{6}\right) \delta(1-z)+\mathcal{D}_1(z) + \left(-2 +\frac{1}{z} +z-z^2 \right) \log(1-z) \nonumber \\
&&+ \frac{(1-z+z^2)^2}{(1-z)z} \log(z) -\frac{11}{24} p^{(0)}_{gg}(z) \bigg] + \mathcal{O}(\epsilon) \, .
\end{eqnarray}

The second gluon-gluon antenna function is 
\begin{equation}
G^0_3(1_g, 2_{q'} , 3_{\bar{q}'}) = \frac{1}{s_{123}^2} \left( \frac{(s_{12}+s_{13})^2}{s_{23}}-\frac{2 s_{12} s_{13}}{s_{23}} \right) + \mathcal{O}(\epsilon) \, .
\end{equation}
It is symmetric under the exchange of particle 2 and 3. Using the gluon as reference particle and remaining differential in the momentum fraction of $q'$, the phase space integration gives
\begin{equation}
\mathcal{G}^{0, {\rm id.} q'}_{3,g}(z) =  \left(q^2\right)^{-\epsilon} \bigg[ -\frac{1}{2 \epsilon } p^{(0)}_{qg}(z) + \frac{1}{2} + \left(-\frac{17}{12} +\frac{1}{2} (\log (1-z) +\log (z) )\right) p^{(0)}_{qg}(z) \bigg] + \mathcal{O}(\epsilon) \, .
\label{eq:G30FFgqp}
\end{equation}
In case of an identified gluon we have
\begin{equation}
\mathcal{G}^{0, {\rm id.} g}_{3,\bar{q}'}(z)=-2 \mathbf{I}_{gg,F}^{(1)}(\epsilon,q^2) \delta(1-z) +  \left(q^2\right)^{-\epsilon} \bigg[ -\frac{1}{6}-\frac{z}{6}-\frac{11}{12} \delta(1-z)+\frac{1}{3} \mathcal{D}_0(z) \bigg]+\mathcal{O}(\epsilon) \, .
\end{equation}

\end{appendix}

\bibliographystyle{JHEP}
\bibliography{frag}

\providecommand{\href}[2]{#2}\begingroup\raggedright\begin{thebibliography}{10}

\bibitem{ATLAS:2017nah}
{\scshape ATLAS} collaboration, M.~Aaboud et~al., \emph{{Measurement of the
  cross section for inclusive isolated-photon production in $pp$ collisions at
  $\sqrt s=13$ TeV using the ATLAS detector}},
  \href{http://dx.doi.org/10.1016/j.physletb.2017.04.072}{\emph{Phys. Lett. B}
  {\bf 770} (2017) 473--493}, [\href{http://arxiv.org/abs/1701.06882}{{\tt
  1701.06882}}].

\bibitem{CMS:2018qao}
{\scshape CMS} collaboration, A.~M. Sirunyan et~al., \emph{{Measurement of
  differential cross sections for inclusive isolated-photon and photon+jets
  production in proton-proton collisions at $\sqrt{s} =$ 13 TeV}},
  \href{http://dx.doi.org/10.1140/epjc/s10052-018-6482-9}{\emph{Eur. Phys. J.
  C} {\bf 79} (2019) 20}, [\href{http://arxiv.org/abs/1807.00782}{{\tt
  1807.00782}}].

\bibitem{ATLAS:2019buk}
{\scshape ATLAS} collaboration, G.~Aad et~al., \emph{{Measurement of the
  inclusive isolated-photon cross section in $pp$ collisions at $\sqrt{s}=13$
  TeV using 36 fb$^{-1}$ of ATLAS data}},
  \href{http://dx.doi.org/10.1007/JHEP10(2019)203}{\emph{JHEP} {\bf 10} (2019)
  203}, [\href{http://arxiv.org/abs/1908.02746}{{\tt 1908.02746}}].

\bibitem{ATLAS:2019iaa}
{\scshape ATLAS} collaboration, G.~Aad et~al., \emph{{Measurement of
  isolated-photon plus two-jet production in $pp$ collisions at $\sqrt s=13$
  TeV with the ATLAS detector}},
  \href{http://dx.doi.org/10.1007/JHEP03(2020)179}{\emph{JHEP} {\bf 03} (2020)
  179}, [\href{http://arxiv.org/abs/1912.09866}{{\tt 1912.09866}}].

\bibitem{CMS:2014mvm}
{\scshape CMS} collaboration, S.~Chatrchyan et~al., \emph{{Measurement of
  differential cross sections for the production of a pair of isolated photons
  in pp collisions at $\sqrt{s}=7\,\text {TeV} $}},
  \href{http://dx.doi.org/10.1140/epjc/s10052-014-3129-3}{\emph{Eur. Phys. J.
  C} {\bf 74} (2014) 3129}, [\href{http://arxiv.org/abs/1405.7225}{{\tt
  1405.7225}}].

\bibitem{ATLAS:2017cvh}
{\scshape ATLAS} collaboration, M.~Aaboud et~al., \emph{{Measurements of
  integrated and differential cross sections for isolated photon pair
  production in $pp$ collisions at $\sqrt{s}=8$ TeV with the ATLAS detector}},
  \href{http://dx.doi.org/10.1103/PhysRevD.95.112005}{\emph{Phys. Rev. D} {\bf
  95} (2017) 112005}, [\href{http://arxiv.org/abs/1704.03839}{{\tt
  1704.03839}}].

\bibitem{ATLAS:2021mbt}
{\scshape ATLAS} collaboration, G.~Aad et~al., \emph{{Measurement of the
  production cross section of pairs of isolated photons in $pp$ collisions at
  13 TeV with the ATLAS detector}},
  \href{http://dx.doi.org/10.1007/JHEP11(2021)169}{\emph{JHEP} {\bf 11} (2021)
  169}, [\href{http://arxiv.org/abs/2107.09330}{{\tt 2107.09330}}].

\bibitem{Koller:1978kq}
K.~Koller, T.~F. Walsh and P.~M. Zerwas, \emph{{Testing {QCD}: Direct Photons
  in $e^+ e^-$ Collisions}},
  \href{http://dx.doi.org/10.1007/BF01474661}{\emph{Z. Phys. C} {\bf 2} (1979)
  197}.

\bibitem{Laermann:1982jr}
E.~Laermann, T.~F. Walsh, I.~Schmitt and P.~M. Zerwas, \emph{{Direct Photons in
  $e^+ e^-$ Annihilation}},
  \href{http://dx.doi.org/10.1016/0550-3213(82)90162-6}{\emph{Nucl. Phys.} {\bf
  B207} (1982) 205--232}.

\bibitem{Frixione:1998jh}
S.~Frixione, \emph{{Isolated photons in perturbative QCD}},
  \href{http://dx.doi.org/10.1016/S0370-2693(98)00454-7}{\emph{Phys. Lett.}
  {\bf B429} (1998) 369--374}, [\href{http://arxiv.org/abs/hep-ph/9801442}{{\tt
  hep-ph/9801442}}].

\bibitem{Aurenche:1987fs}
P.~Aurenche, R.~Baier, M.~Fontannaz and D.~Schiff, \emph{{Prompt Photon
  Production at Large $p_T$ Scheme Invariant QCD Predictions and Comparison
  with Experiment}},
  \href{http://dx.doi.org/10.1016/0550-3213(88)90553-6}{\emph{Nucl. Phys.} {\bf
  B297} (1988) 661--696}.

\bibitem{Baer:1990ra}
H.~Baer, J.~Ohnemus and J.~F. Owens, \emph{{A Next-to-leading Logarithm
  Calculation of Direct Photon Production}},
  \href{http://dx.doi.org/10.1103/PhysRevD.42.61}{\emph{Phys. Rev.} {\bf D42}
  (1990) 61--71}.

\bibitem{Aurenche:1992yc}
P.~Aurenche, P.~Chiappetta, M.~Fontannaz, J.~P. Guillet and E.~Pilon,
  \emph{{Next-to-leading order bremsstrahlung contribution to prompt photon
  production}},
  \href{http://dx.doi.org/10.1016/0550-3213(93)90615-V}{\emph{Nucl. Phys.} {\bf
  B399} (1993) 34--62}.

\bibitem{Gordon:1993qc}
L.~E. Gordon and W.~Vogelsang, \emph{{Polarized and unpolarized prompt photon
  production beyond the leading order}},
  \href{http://dx.doi.org/10.1103/PhysRevD.48.3136}{\emph{Phys. Rev.} {\bf D48}
  (1993) 3136--3159}.

\bibitem{Gluck:1994iz}
M.~Glück, L.~E. Gordon, E.~Reya and W.~Vogelsang, \emph{{High $p_T$ photon
  production at $p \bar{p}$ collider}},
  \href{http://dx.doi.org/10.1103/PhysRevLett.73.388}{\emph{Phys. Rev. Lett.}
  {\bf 73} (1994) 388--391}.

\bibitem{Catani:2002ny}
S.~Catani, M.~Fontannaz, J.~P. Guillet and E.~Pilon, \emph{{Cross-section of
  isolated prompt photons in hadron hadron collisions}},
  \href{http://dx.doi.org/10.1088/1126-6708/2002/05/028}{\emph{JHEP} {\bf 05}
  (2002) 028}, [\href{http://arxiv.org/abs/hep-ph/0204023}{{\tt
  hep-ph/0204023}}].

\bibitem{Aurenche:2006vj}
P.~Aurenche, M.~Fontannaz, J.-P. Guillet, E.~Pilon and M.~Werlen, \emph{{A New
  critical study of photon production in hadronic collisions}},
  \href{http://dx.doi.org/10.1103/PhysRevD.73.094007}{\emph{Phys. Rev. D} {\bf
  73} (2006) 094007}, [\href{http://arxiv.org/abs/hep-ph/0602133}{{\tt
  hep-ph/0602133}}].

\bibitem{Binoth:1999qq}
T.~Binoth, J.~P. Guillet, E.~Pilon and M.~Werlen, \emph{{A Full next-to-leading
  order study of direct photon pair production in hadronic collisions}},
  \href{http://dx.doi.org/10.1007/s100520050024}{\emph{Eur. Phys. J. C} {\bf
  16} (2000) 311--330}, [\href{http://arxiv.org/abs/hep-ph/9911340}{{\tt
  hep-ph/9911340}}].

\bibitem{Owens:1986mp}
J.~F. Owens, \emph{{Large Momentum Transfer Production of Direct Photons, Jets,
  and Particles}},
  \href{http://dx.doi.org/10.1103/RevModPhys.59.465}{\emph{Rev. Mod. Phys.}
  {\bf 59} (1987) 465}.

\bibitem{Gluck:1992zx}
M.~Glück, E.~Reya and A.~Vogt, \emph{{Parton fragmentation into photons beyond
  the leading order}}, \href{http://dx.doi.org/10.1103/PhysRevD.51.1427,
  10.1103/PhysRevD.48.116}{\emph{Phys. Rev.} {\bf D48} (1993) 116}. [Erratum:
  Phys. Rev.D51,1427(1995)].

\bibitem{Bourhis:1997yu}
L.~Bourhis, M.~Fontannaz and J.~P. Guillet, \emph{{Quarks and gluon
  fragmentation functions into photons}},
  \href{http://dx.doi.org/10.1007/s100520050158}{\emph{Eur. Phys. J.} {\bf C2}
  (1998) 529--537}, [\href{http://arxiv.org/abs/hep-ph/9704447}{{\tt
  hep-ph/9704447}}].

\bibitem{Buskulic:1995au}
{\scshape ALEPH} collaboration, D.~Buskulic et~al., \emph{{First measurement of
  the quark to photon fragmentation function}},
  \href{http://dx.doi.org/10.1007/BF02907417}{\emph{Z. Phys.} {\bf C69} (1996)
  365--378}.

\bibitem{Ackerstaff:1997nha}
{\scshape OPAL} collaboration, K.~Ackerstaff et~al., \emph{{Measurement of the
  quark to photon fragmentation function through the inclusive production of
  prompt photons in hadronic Z0 decays}},
  \href{http://dx.doi.org/10.1007/s100520050122}{\emph{Eur. Phys. J.} {\bf C2}
  (1998) 39--48}, [\href{http://arxiv.org/abs/hep-ex/9708020}{{\tt
  hep-ex/9708020}}].

\bibitem{GehrmannDeRidder:1997gf}
A.~Gehrmann-De~Ridder and E.~W.~N. Glover, \emph{{A Complete O ($\alpha
  \alpha_s$) calculation of the photon + 1 jet rate in $e^+ e^-$
  annihilation}},
  \href{http://dx.doi.org/10.1016/S0550-3213(97)00818-3}{\emph{Nucl. Phys.}
  {\bf B517} (1998) 269--323}, [\href{http://arxiv.org/abs/hep-ph/9707224}{{\tt
  hep-ph/9707224}}].

\bibitem{GehrmannDeRidder:1998ba}
A.~Gehrmann-De~Ridder and E.~W.~N. Glover, \emph{{Final state photon production
  at LEP}}, \href{http://dx.doi.org/10.1007/s100520050382,
  10.1007/s100529800958}{\emph{Eur. Phys. J.} {\bf C7} (1999) 29--48},
  [\href{http://arxiv.org/abs/hep-ph/9806316}{{\tt hep-ph/9806316}}].

\bibitem{Campbell_2017}
J.~M. Campbell, R.~K. Ellis and C.~Williams, \emph{{Direct Photon Production at
  Next-to\textendash{}Next-to-Leading Order}},
  \href{http://dx.doi.org/10.1103/PhysRevLett.118.222001}{\emph{Phys. Rev.
  Lett.} {\bf 118} (2017) 222001}, [\href{http://arxiv.org/abs/1612.04333}{{\tt
  1612.04333}}]. [Erratum: Phys.Rev.Lett. 124 (2020) 259901].

\bibitem{Chen_2020}
X.~Chen, T.~Gehrmann, N.~Glover, M.~H\"ofer and A.~Huss, \emph{{Isolated photon
  and photon+jet production at NNLO QCD accuracy}},
  \href{http://dx.doi.org/10.1007/JHEP04(2020)166}{\emph{JHEP} {\bf 04} (2020)
  166}, [\href{http://arxiv.org/abs/1904.01044}{{\tt 1904.01044}}].

\bibitem{Campbell_2017a}
J.~M. Campbell, R.~K. Ellis and C.~Williams, \emph{{Driving missing data at the
  LHC: NNLO predictions for the ratio of $\gamma+j$ and $Z+j$}},
  \href{http://dx.doi.org/10.1103/PhysRevD.96.014037}{\emph{Phys. Rev. D} {\bf
  96} (2017) 014037}, [\href{http://arxiv.org/abs/1703.10109}{{\tt
  1703.10109}}].

\bibitem{Catani:2011qz}
S.~Catani, L.~Cieri, D.~de~Florian, G.~Ferrera and M.~Grazzini, \emph{{Diphoton
  production at hadron colliders: a fully-differential QCD calculation at
  NNLO}}, \href{http://dx.doi.org/10.1103/PhysRevLett.108.072001}{\emph{Phys.
  Rev. Lett.} {\bf 108} (2012) 072001},
  [\href{http://arxiv.org/abs/1110.2375}{{\tt 1110.2375}}]. [Erratum:
  Phys.Rev.Lett. 117 (2016) 089901].

\bibitem{Campbell:2016yrh}
J.~M. Campbell, R.~K. Ellis, Y.~Li and C.~Williams, \emph{{Predictions for
  diphoton production at the LHC through NNLO in QCD}},
  \href{http://dx.doi.org/10.1007/JHEP07(2016)148}{\emph{JHEP} {\bf 07} (2016)
  148}, [\href{http://arxiv.org/abs/1603.02663}{{\tt 1603.02663}}].

\bibitem{Catani:2018krb}
S.~Catani, L.~Cieri, D.~de~Florian, G.~Ferrera and M.~Grazzini, \emph{{Diphoton
  production at the LHC: a QCD study up to NNLO}},
  \href{http://dx.doi.org/10.1007/JHEP04(2018)142}{\emph{JHEP} {\bf 04} (2018)
  142}, [\href{http://arxiv.org/abs/1802.02095}{{\tt 1802.02095}}].

\bibitem{Gehrmann:2020oec}
T.~Gehrmann, N.~Glover, A.~Huss and J.~Whitehead, \emph{{Scale and isolation
  sensitivity of diphoton distributions at the LHC}},
  \href{http://dx.doi.org/10.1007/JHEP01(2021)108}{\emph{JHEP} {\bf 01} (2021)
  108}, [\href{http://arxiv.org/abs/2009.11310}{{\tt 2009.11310}}].

\bibitem{Chawdhry:2021hkp}
H.~A. Chawdhry, M.~Czakon, A.~Mitov and R.~Poncelet, \emph{{NNLO QCD
  corrections to diphoton production with an additional jet at the LHC}},
  \href{http://dx.doi.org/10.1007/JHEP09(2021)093}{\emph{JHEP} {\bf 09} (2021)
  093}, [\href{http://arxiv.org/abs/2105.06940}{{\tt 2105.06940}}].

\bibitem{Badger:2021ohm}
S.~Badger, T.~Gehrmann, M.~Marcoli and R.~Moodie, \emph{{Next-to-leading order
  QCD corrections to diphoton-plus-jet production through gluon fusion at the
  LHC}}, \href{http://dx.doi.org/10.1016/j.physletb.2021.136802}{\emph{Phys.
  Lett. B} {\bf 824} (2022) 136802},
  [\href{http://arxiv.org/abs/2109.12003}{{\tt 2109.12003}}].

\bibitem{Chawdhry:2019bji}
H.~A. Chawdhry, M.~L. Czakon, A.~Mitov and R.~Poncelet, \emph{{NNLO QCD
  corrections to three-photon production at the LHC}},
  \href{http://dx.doi.org/10.1007/JHEP02(2020)057}{\emph{JHEP} {\bf 02} (2020)
  057}, [\href{http://arxiv.org/abs/1911.00479}{{\tt 1911.00479}}].

\bibitem{Kallweit:2020gcp}
S.~Kallweit, V.~Sotnikov and M.~Wiesemann, \emph{{Triphoton production at
  hadron colliders in NNLO QCD}},
  \href{http://dx.doi.org/10.1016/j.physletb.2020.136013}{\emph{Phys. Lett. B}
  {\bf 812} (2021) 136013}, [\href{http://arxiv.org/abs/2010.04681}{{\tt
  2010.04681}}].

\bibitem{Siegert:2016bre}
F.~Siegert, \emph{{A practical guide to event generation for prompt photon
  production with Sherpa}},
  \href{http://dx.doi.org/10.1088/1361-6471/aa5f29}{\emph{J. Phys.} {\bf G44}
  (2017) 044007}, [\href{http://arxiv.org/abs/1611.07226}{{\tt 1611.07226}}].

\bibitem{Czakon:2021ohs}
M.~L. Czakon, T.~Generet, A.~Mitov and R.~Poncelet, \emph{{B-hadron production
  in NNLO QCD: application to LHC t$ \overline{t} $ events with leptonic
  decays}}, \href{http://dx.doi.org/10.1007/JHEP10(2021)216}{\emph{JHEP} {\bf
  10} (2021) 216}, [\href{http://arxiv.org/abs/2102.08267}{{\tt 2102.08267}}].

\bibitem{Melnikov:2004bm}
K.~Melnikov and A.~Mitov, \emph{{Perturbative heavy quark fragmentation
  function through $\mathcal{O}(\alpha^2_s)$}},
  \href{http://dx.doi.org/10.1103/PhysRevD.70.034027}{\emph{Phys. Rev. D} {\bf
  70} (2004) 034027}, [\href{http://arxiv.org/abs/hep-ph/0404143}{{\tt
  hep-ph/0404143}}].

\bibitem{Mitov:2004du}
A.~Mitov, \emph{{Perturbative heavy quark fragmentation function through
  $\mathcal{O}(\alpha^2_s)$: Gluon initiated contribution}},
  \href{http://dx.doi.org/10.1103/PhysRevD.71.054021}{\emph{Phys. Rev. D} {\bf
  71} (2005) 054021}, [\href{http://arxiv.org/abs/hep-ph/0410205}{{\tt
  hep-ph/0410205}}].

\bibitem{GehrmannDeRidder:2005cm}
A.~Gehrmann-De~Ridder, T.~Gehrmann and E.~W.~N. Glover, \emph{{Antenna
  subtraction at NNLO}},
  \href{http://dx.doi.org/10.1088/1126-6708/2005/09/056}{\emph{JHEP} {\bf 09}
  (2005) 056}, [\href{http://arxiv.org/abs/hep-ph/0505111}{{\tt
  hep-ph/0505111}}].

\bibitem{Daleo:2006xa}
A.~Daleo, T.~Gehrmann and D.~Maitre, \emph{{Antenna subtraction with hadronic
  initial states}},
  \href{http://dx.doi.org/10.1088/1126-6708/2007/04/016}{\emph{JHEP} {\bf 04}
  (2007) 016}, [\href{http://arxiv.org/abs/hep-ph/0612257}{{\tt
  hep-ph/0612257}}].

\bibitem{Currie:2013vh}
J.~Currie, E.~W.~N. Glover and S.~Wells, \emph{{Infrared Structure at NNLO
  Using Antenna Subtraction}},
  \href{http://dx.doi.org/10.1007/JHEP04(2013)066}{\emph{JHEP} {\bf 04} (2013)
  066}, [\href{http://arxiv.org/abs/1301.4693}{{\tt 1301.4693}}].

\bibitem{Gehrmann:2021lwb}
T.~Gehrmann and R.~Sch\"urmann, \emph{{NNLO Photon Fragmentation within Antenna
  Subtraction}}, {\emph{SciPost Physics Proceedings RADCOR 2021} (2021)
  202106001}, [\href{http://arxiv.org/abs/2110.02617}{{\tt 2110.02617}}].

\bibitem{Daleo:2009yj}
A.~Daleo, A.~Gehrmann-De~Ridder, T.~Gehrmann and G.~Luisoni, \emph{{Antenna
  subtraction at NNLO with hadronic initial states: initial-final
  configurations}},
  \href{http://dx.doi.org/10.1007/JHEP01(2010)118}{\emph{JHEP} {\bf 01} (2010)
  118}, [\href{http://arxiv.org/abs/0912.0374}{{\tt 0912.0374}}].

\bibitem{vonManteuffel:2012np}
A.~von Manteuffel and C.~Studerus, \emph{{Reduze 2 - Distributed Feynman
  Integral Reduction}},  \href{http://arxiv.org/abs/1201.4330}{{\tt
  1201.4330}}.

\bibitem{Gehrmann:2011wi}
T.~Gehrmann and P.~F. Monni, \emph{{Antenna subtraction at NNLO with hadronic
  initial states: real-virtual initial-initial configurations}},
  \href{http://dx.doi.org/10.1007/JHEP12(2011)049}{\emph{JHEP} {\bf 12} (2011)
  049}, [\href{http://arxiv.org/abs/1107.4037}{{\tt 1107.4037}}].

\bibitem{bateman}
{A.\ Erd\'{e}lyi (ed.)}, \emph{{Higher Transcendental Functions, vol. 1}}.
\newblock McGraw-Hill, New York, 1953.

\bibitem{Graudenz:1993tg}
D.~Graudenz, \emph{{Next-to-leading order QCD corrections to jet cross-sections
  and jet rates in deeply inelastic electron proton scattering}},
  \href{http://dx.doi.org/10.1103/PhysRevD.49.3291}{\emph{Phys. Rev. D} {\bf
  49} (1994) 3291--3319}, [\href{http://arxiv.org/abs/hep-ph/9307311}{{\tt
  hep-ph/9307311}}].

\bibitem{Gehrmann:2002zr}
T.~Gehrmann and E.~Remiddi, \emph{{Analytic continuation of massless two loop
  four point functions}},
  \href{http://dx.doi.org/10.1016/S0550-3213(02)00569-2}{\emph{Nucl. Phys. B}
  {\bf 640} (2002) 379--411}, [\href{http://arxiv.org/abs/hep-ph/0207020}{{\tt
  hep-ph/0207020}}].

\bibitem{Currie:2017tpe}
J.~Currie, T.~Gehrmann, A.~Huss and J.~Niehues, \emph{{NNLO QCD corrections to
  jet production in deep inelastic scattering}},
  \href{http://dx.doi.org/10.1007/JHEP07(2017)018}{\emph{JHEP} {\bf 07} (2017)
  018}, [\href{http://arxiv.org/abs/1703.05977}{{\tt 1703.05977}}]. [Erratum:
  JHEP 12, 042 (2020)].

\bibitem{Glover:1993xc}
E.~W.~N. Glover and A.~G. Morgan, \emph{{Measuring the photon fragmentation
  function at LEP}}, \href{http://dx.doi.org/10.1007/BF01560245}{\emph{Z.
  Phys.} {\bf C62} (1994) 311--322}.

\bibitem{Hall:2018jub}
E.~Hall and J.~Thaler, \emph{{Photon isolation and jet substructure}},
  \href{http://dx.doi.org/10.1007/JHEP09(2018)164}{\emph{JHEP} {\bf 09} (2018)
  164}, [\href{http://arxiv.org/abs/1805.11622}{{\tt 1805.11622}}].

\bibitem{Kaufmann:2016nux}
T.~Kaufmann, A.~Mukherjee and W.~Vogelsang, \emph{{Access to Photon
  Fragmentation Functions in Hadronic Jet Production}},
  \href{http://dx.doi.org/10.1103/PhysRevD.93.114021}{\emph{Phys. Rev. D} {\bf
  93} (2016) 114021}, [\href{http://arxiv.org/abs/1604.07175}{{\tt
  1604.07175}}].

\end{thebibliography}\endgroup

\end{document}